\documentclass[11pt]{article}

\usepackage{amsmath}
\usepackage{amsfonts}
\usepackage{color}

\newcommand{\be}{\begin{equation}}
\newcommand{\ee}{\end{equation}}
\newcommand{\beqn}{\begin{eqnarray}}
\newcommand{\eeqn}{\end{eqnarray}}
\newcommand{\beqns}{\begin{eqnarray*}}
\newcommand{\eeqns}{\end{eqnarray*}}
\newcommand{\lkr}{\left(}
\newcommand{\lkv}{\left[}
\newcommand{\rkv}{\right]}
\newcommand{\rkr}{\right)}
\newcommand{\lfi}{\left\{}
\newcommand{\rfi}{\right\}}

\newcommand{\fr}[1]{(\ref{#1})}

\newcommand{\ro}{\varrho}
\newcommand{\ph}{\varphi}
\newcommand{\dd}{\delta}
\newcommand{\DD}{\Delta}

\newcommand{\af}{\alpha}
\newcommand{\eps}{\varepsilon}

\newcommand{\ga}{\gamma}

\newcommand{\lam}{\lambda}

\newcommand{\up}{\upsilon}

\newcommand{\Lam}{\Lambda}
\newcommand{\Om}{\Omega}

\newcommand{\EE}{\ensuremath{{\mathbb E}}}

\newcommand{\II}{\ensuremath{{\mathbb I}}}
\newcommand{\ZZ}{\ensuremath{{\mathbb Z}}}
\newcommand{\PP}{\ensuremath{{\mathbb P}}}

\newcommand{\Var}{\mbox{Var}}
\newcommand{\Cov}{\mbox{Cov}}
\newcommand{\diag}{\mbox{diag}}
\newcommand{\supp}{\mbox{supp}}
\newcommand{\etal}{ {\it et. al }}

\newtheorem{theorem}{Theorem}
\newtheorem{lemma}{Lemma}

\newtheorem{remark}{Remark}

\newcommand{\bA}{\mathbf{A}}
\newcommand{\bB}{\mathbf{B}}
\newcommand{\bD}{\mathbf{D}}
\newcommand{\bI}{\mathbf{I}}
\newcommand{\bQ}{\mathbf{Q}}
\newcommand{\bH}{\mathbf{H}}

\newcommand{\bc}{\mathbf{c}}
\newcommand{\bu}{\mathbf{u}}
\newcommand{\bv}{\mathbf{v}}

\newcommand{\beps}{\mbox{\mathversion{bold}$\eps$}}

\newcommand{\bAm}{\mathbf{A}^{(m)}}
\newcommand{\bBm}{\mathbf{B}^{(m)}}

\newcommand{\bcm}{\mathbf{c}^{(m)}}
\newcommand{\bum}{\mathbf{u}^{(m)}}
\newcommand{\bvm}{\mathbf{v}^{(m)}}

\newcommand{\bepsm}{\mbox{\mathversion{bold}$\eps$}^{(m)}}

\newcommand{\bAmkl}{A^{(m)}_{kl}}
\newcommand{\bAmkk}{A^{(m)}_{kk}}
\newcommand{\bAmll}{A^{(m)}_{ll}}
\newcommand{\bBmkl}{B^{(m)}_{kl}}
\newcommand{\bcml}{c^{(m)}_l}
\newcommand{\bcmk}{c^{(m)}_k}

\newcommand{\bAs}{\bA^*}
\newcommand{\bBs}{\bB^*}

\newcommand{\mo}{m_0}

\newcommand{\hbu}{\hat{\bu}}
\newcommand{\hbv}{\hat{\bv}}

\newcommand{\hbc}{\hat{\bc}}

\newcommand{\hbum}{\hat{\bu}^{(m)}}

\newcommand{\hbvm}{\hat{\bv}^{(m)}}

\newcommand{\hbcm}{\hat{\bc}^{(m)}}

\newcommand{\hbcml}{\hat{c}^{(m)}_l}
\newcommand{\hbcmk}{\hat{c}^{(m)}_k}
\newcommand{\hbumk}{\hat{u}^{(m)}_{k}}
\newcommand{\hbvmk}{\hat{v}^{(m)}_{k}}

\newcommand{\Bpqsa}{B_{p,q}^s (A) }

\newcommand{\phimk}{\ph_{mk}}
\newcommand{\phimks}{\ph^*_{mk}}

\newcommand{\phiml}{\ph_{ml}}
\newcommand{\psijk}{\psi_{jk}}
\newcommand{\psijl}{\psi_{jl}}
\newcommand{\wjk}{w_{jk}}
\newcommand{\hfn}{\hat{f}_n} 

\newcommand{\amk}{a_{mk}}
\newcommand{\bjk}{b_{jk}}
\newcommand{\bjl}{b_{jl}}
\newcommand{\betajk}{\beta_{jk}}
\newcommand{\hamk}{\hat{a}_{mk}}
\newcommand{\haml}{\hat{a}_{ml}}
\newcommand{\hbjk}{\hat{b}_{jk}}
\newcommand{\tbjk}{\tilde{b}_{jk}}
\newcommand{\hbetajk}{\hat{\beta}_{jk}}
\newcommand{\bjkp}{b_{j'k}}
\newcommand{\hbjkp}{\hat{b}_{j'k}}

\newcommand{\fjk}{f_{jk}}
\newcommand{\gaj}{\ga_j}

\newcommand{\Jmkl}{J_{mkl}}
\newcommand{\Ijmkl}{I_{jmkl}}
\newcommand{\Lmjk}{{\cal L}_{mjk}}
\newcommand{\Lmjks}{{\cal L}^*_{mjk}}
\newcommand{\Lmjkc}{{\cal L}^c_{mjk}}
\newcommand{\hjmkl}{h_{jmkl}}

\newcommand{\Lh}{{L_\ph}}
\newcommand{\Ls}{{L_\psi}}
\newcommand{\Uh}{{U_\ph}}
\newcommand{\Us}{{U_\psi}}
\newcommand{\koj}{k_{0j}}
\newcommand{\kom}{k_{0m}}
\newcommand{\Komh}{{K_{0m}^\ph}}
\newcommand{\Kojs}{{K_{0j}^\psi}}  
\newcommand{\Komhc}{{K_{0mc}^\ph}}
\newcommand{\Kojsc}{{K_{0jc}^\psi}}  
\newcommand{\Koms}{K_{0m}^*}
\newcommand{\Kmmo}{\tilde{K}_{m, m_0}}
\newcommand{\Kjmo}{\tilde{K}_{j, m_0}}
\newcommand{\Kjmop}{\tilde{K}_{j', m_0}}

\newcommand{\sumJ}{\sum_{j=0}^{J-1}}
\newcommand{\summJ}{\sum_{j=m}^{J-1}}
\newcommand{\sumjk}{\sum_{k=0}^{2^j-1}}
\newcommand{\sumkomh}{\sum_{k \in \Komh}}
\newcommand{\sumkomhc}{\sum_{k \in \Komhc}}

\newcommand{\sumkojsc}{\sum_{k \in \Kojsc}}

\newcommand{\hfo}{\hat{f}_0}

\newcommand{\hf}{\hat{f}}

\newcommand{\hm}{\hat{m}}

\newcommand{\hfom}{\hf_{0,m}}
\newcommand{\fom}{f_{0,m}}
\newcommand{\hfcm}{\hat{f}_{c,m}}
\newcommand{\fcm}{f_{c,m}}
\newcommand{\hfm}{\hf_{m}}
\newcommand{\hfoj}{\hf_{0,j}}
\newcommand{\foj}{f_{0,j}}
\newcommand{\hfcj}{\hat{f}_{c,j}}
\newcommand{\fcj}{f_{c,j}}
\newcommand{\hfj}{\hf_{j}}
\newcommand{\hfomo}{\hf_{0,m_0}}
\newcommand{\fomo}{f_{0,m_0}}
\newcommand{\hfcmo}{\hat{f}_{c,m_0}}
\newcommand{\fcmo}{f_{c,m_0}}
\newcommand{\hfmo}{\hf_{m_0}}
\newcommand{\hfohm}{\hf_{0,\hm}}

\newcommand{\hfchm}{\hat{f}_{c,\hm}}

\newcommand{\ron}{\ro_n}

\newcommand{\zmax}{z_{\max}}
\newcommand{\zmaxlk}{z_{\max}^{(l,k, \dd)} }
\newcommand{\zmaxkk}{z_{\max}^{(k,k, \dd)} }
\newcommand{\zmaxll}{z_{\max}^{(l,l, \dd)} }

\newcommand{\zmaxkko}{z_{\max}^{(k,k,0)} }

\newcommand{\tmaxkl}{t_{\max}^{(k,l)} }

\newcommand{\ints}{\ensuremath{{\mathbb Z}}}

\newcommand{\RR}{\ensuremath{{\mathbb R}}}

\begin{document}

\title{\bf {Nonparametric Regression Estimation Based on Spatially Inhomogeneous Data: 
Minimax Global Convergence Rates   and Adaptivity}}

\author{{\em Anestis ~Antoniadis}, \\
                Laboratoire Jean Kuntzmann,
                Universite Joseph Fourier,\\
                38041 Grenoble CEDEX 9, France.\\
        {\em Email}:~\texttt{Anestis.Antoniadis@imag.fr}
               \\ \\
        {\em Marianna ~Pensky}, \\
         Department of Mathematics,
         University of Central Florida,\\
         Orlando, FL 32816-1364, USA.\\
        {\em Email}:~\texttt{Marianna.Pensky@ucf.edu}
          \\ \\
        and
          \\ \\
        {\em Theofanis ~Sapatinas},\\
        Department of Mathematics and Statistics,
        University of Cyprus,\\
        P.O. Box 20537,
        CY 1678 Nicosia,
        Cyprus.\\
        {\em Email}:~\texttt{fanis@ucy.ac.cy}}

\date{}

\bibliographystyle{plain}
\maketitle

\begin{abstract}
We  consider the nonparametric regression estimation problem of recovering an unknown response function $f$ on the basis 
of spatially inhomogeneous data when  the design points follow a known density $g$ with a finite number of well-separated zeros.  
In particular, we  consider two different cases: when $g$ has zeros of a polynomial order and when $g$ has zeros of an exponential order. 
These two cases correspond to moderate and severe data losses, respectively. We obtain  asymptotic (as the sample size increases)  minimax lower bounds for the $L^2$-risk when $f$ is assumed to belong to a Besov ball, and construct adaptive wavelet thresholding estimators of $f$ that are asymptotically optimal (in the minimax sense) or near-optimal within a logarithmic factor (in the case of a zero of a polynomial order), over a wide range of Besov balls. 

The spatially inhomogeneous ill-posed problem that we investigate is inherently more difficult 
than spatially homogeneous ill-posed problems like, e.g., deconvolution. In particular, due to spatial irregularity, 
assessment of asymptotic minimax global convergence rates is a much harder task than the derivation of  
asymptotic minimax local convergence rates studied recently in the literature. Furthermore, the resulting estimators exhibit 
very different behavior and asymptotic minimax global convergence rates in comparison 
with the solution of spatially homogeneous ill-posed problems.
For example, unlike in the deconvolution problem, the asymptotic minimax global convergence rates  are greatly influenced not only by the 
extent of data loss but also by the degree of spatial homogeneity of   $f$.
Specifically,  even if $1/g$ is non-integrable, one can recover $f$ as well as in the case of an equispaced 
design (in terms of asymptotic minimax global convergence rates) when it is homogeneous enough since 
 the estimator is ``borrowing strength'' in the areas where $f$ is adequately sampled. 

\vspace{2mm} {\bf  Keywords}: { Adaptivity, Besov spaces, inhomogeneous data, minimax estimation, 
nonparametric regression, thresholding, wavelet estimation.}

\vspace{2mm}{\bf AMS (2000) Subject Classification}: {Primary: 62G08, Secondary: 62G05, 62G20}
\end{abstract}

\section{Introduction} 
\label{introduction}
\setcounter{equation}{0}

Applicability of majority of techniques for estimation in the nonparametric regression model 
rests on the assumption that data is equispaced and complete.  These assumptions were mainly
adopted by signal processing community where the signal is assumed to be recorded 
at equal intervals in time. However, in reality, due to unexpected losses of data or
limitations of data sampling techniques, data may fail to be equispaced and complete. 
To this end, we consider the problem of recovering an unknown response function $f \in L^2([0,1])$ 
on the basis of irregularly spaced observations, i.e., when one observes $y_i$ governed by
\begin{equation} 
\label{main_eq}
y_i = f(x_i) + \sigma \, \xi_i, \quad i=1,2, \ldots, n,
\end{equation}
where $x_i \in [0,1]$, $i=1,2, \ldots, n$, are fixed (non-equidistant) or random points, 
$\xi_i$, $i=1,2, \ldots, n$, are independent standard Gaussian random variables 
and $\sigma^2>0$ (the noise level)  is assumed to be known and finite.
Model \fr{main_eq} can be viewed as a problem of recovering a signal 
when part of data is lost (e.g., in cell phone use) or unavailable 
(e.g., in military applications). 
Model \fr{main_eq} is also intimately connected to the problem of missing
data since points $x_i$, $i=1,2, \ldots, n$, can be viewed as the
remainder of $N$ equidistant points $j/N$, $j=1,2, \ldots, N$, after
observations at $(N-n)$ points have been lost. However, there is a
great advantage in treating  the missing data problem  as a
particular case of a nonparametric regression problem: with the last two decades seeing
tremendous advancement in the field of nonparametric statistics, a
nonparametric regression approach to incomplete data brings along all the modern
tools in this field such as asymptotic minimax convergence rates, Besov spaces, wavelets and adaptive estimators.
 
The problem of estimating an unknown response function in the context of wavelet thresholding 
in the nonparametric regression setting with irregular design has been now addressed by many authors, 
see, e.g.,  Hall  and  Turlach  (1997), Antoniadis  and Pham  (1998), Cai  and Brown  (1998), 
Sardy {\it et al.}  (1999), Kovac   and Silverman  (2000),   Pensky  and  Vidakovic  (2001), Brown  {\em et al.} (2002), 
Zhang {\em et al.} (2002), Kohler  (2003) and Amato {\it et al.}  (2006). 
Several tools were suggested for attacking the problem; here, we shall review only few of them. 
For instance, the procedure of  Kovac and Silverman (2000) relies upon  a linear
interpolation transformation $R$ to the observed data vector $y=(y_1,y_2,\ldots,y_n)$
that maps it to a new vector of size $2^J$ ($2^{J-1} < n \leq 2^J$), 
corresponding to a new design with equispaced points. After
the transformation, the new vector is multivariate normal with
mean $Rf$ and covariance matrix which is assumed to have a finite
bandwidth, so that the computational complexity of their algorithm
is of order $n$. Cai and Brown  (1998) attacked the problem by using multiresolution analysis, 
projection  and wavelet nonlinear thresholding while Sardy {\it et al.}  (1999) applied an isometric method.
Pensky and Vidakovic  (2001) estimated the conditional expectation $\EE(Y|X)$ directly by constructing 
its wavelet expansion, while Amato  {\it et al.}  (2006) applied a reproducing kernel Hilbert space (RKHS)
approach in the spirit of Wahba (1990). However, until very recently, all studies have been carried out 
under the assumption that the nonequispaced design still possesses some regularity, namely, 
the density function $g$ of the design points $x_i$, $i=1,2,\ldots,n$, is uniformly bounded from below, 
i.e., $\inf_{x \in [0,1]}g(x) \geq c$ for some constant $c>0$. In this case, asymptotically, model \fr{main_eq}
is equivalent to the case of the standard (equispaced) nonparametric regression model, as long as the 
design density function $g$ is known (see, e.g., Brown {\it et al.} (2002)).

Recently, an attempt  has been  made of  more advanced 
investigations of the problem.  Kerkyacharian and Picard (2004)   introduced
warped wavelets to construct estimators of the unknown response function $f$ under model \fr{main_eq} 
when the design density function $g$ has zeros of polynomial order. They, however,  measured the error 
of their suggested estimator in the warped Besov spaces which is, practically, equivalent to measuring 
the error of the estimator at the design points only. For this reason, the derived estimators posses  
the usual asymptotic (as the sample size increases) minimax global convergence rates which do  not depend 
on the order of the zeros of the design density function $g$. This line of investigation was continued 
by Chesneau (2007) who constructed  asymptotic  minimax lower bounds over a wide range of Besov 
balls, under the assumption that the design density function $g$ is known and  that $1/g$ is integrable, 
and, furthermore, suggested adaptive wavelet thresholding estimators for the unknown response function $f$. 
However,  in Kerkyacharian and Picard (2004) and Chesneau (2007), the assumptions on the design 
density function $g$ are restrictive enough so that the asymptotic minimax global convergence rates of any 
estimator coincide with the asymptotic minimax global convergence rates under the assumption that $g$ is 
bounded from below, i.e., the corresponding nonparametric estimation problem is a {\em well-posed} problem.

Ga\"{i}ffas  (2005, 2007)  was the first  author who considered nonparametric regression estimation on the basis of spatially inhomogeneous data as an {\em ill-posed} problem. 
In particular, he constructed pointwise adaptive estimators of $f$ on the basis of local polynomials when $1/g$ is non-integrable and showed that the asymptotic minimax local convergence rates of the suggested estimators are slower than in the case when $g$ is bounded from below, hence, demonstrating that  the aforementioned estimation problem is an ill-posed problem. 
Since his techniques  are intended for local reconstruction and depend on cross-validation at each point, they become too involved when one tries to adapt them to the whole domain of $f$. Furthermore, Ga\"{i}ffas (2006, 2009) studied asymptotic minimax uniform convergence rates. However, these rates are expressed in a very complex form which is very hard to obtain for $f$ belonging to standard functional classes (see Remark \ref{Gaiffas_and_us}). Note also that some of his results were recently extended to the multivariate case by Guillou  and Klutchnikoff (2011).

Our objective is to study how the zeros of the design density function $g$ affect the asymptotic minimax global 
convergence rates of $f$ in model (\ref{main_eq}), and to construct adaptive wavelet thresholding 
estimators of $f$ which attain these rates, over a wide range of Besov balls. As we show below 
(see Remark \ref{local-global}), assessing asymptotic minimax global convergence rates 
is a much harder task than assessing asymptotic minimax  local convergence rates. Model \fr{main_eq} can be viewed 
as a spatially inhomogeneous ill-posed problem which is inherently more difficult than spatially homogeneous ill-posed problems like, 
e.g., deconvolution, especially in the case when the unknown response function is spatially homogeneous.
To the best of our knowledge, so far, there are no results for asymptotic minimax global convergence rates
in the case of spatially inhomogeneous ill-posed problems when its solution is  spatially homogeneous
since this problem is usually avoided by restricting attention to the case when the estimated function
is spatially inhomogeneous, or, at most, belongs to a Sobolev ball (see, e.g.,  Hoffmann and  Reiss  (2008)).

In what follows, we  address these issues. In particular, we mainly consider two different cases:
when $g$ has zeros of a polynomial order and when $g$ has zeros of an exponential order.  We obtain  asymptotic (as the sample size increases)  minimax lower bounds for the $L^2$-risk when $f$ is assumed to belong to a Besov ball, and construct adaptive wavelet thresholding estimators of $f$ that are asymptotically optimal (in the minimax sense) or near-optimal within a logarithmic factor (in the case of a zero of a polynomial order), over a wide range of Besov balls. Due to spatial irregularity, the suggested estimators exhibit very different behavior  and asymptotic minimax global convergence rates in comparison 
with the solution of spatially homogeneous ill-posed problems (see Remark~\ref{conv_rates}).
Specifically,  even if $1/g$ is non-integrable, one can recover $f$ as well as in the case of an equispaced 
design (in terms of asymptotic minimax global convergence rates) when the function is homogeneous enough since 
 the estimator is ``borrowing strength'' in the areas where $f$ is adequately sampled. 
These features lead to a  different structure of  estimators of $f$ described in Section \ref{estimation_strategies}.
The complementary case when $1/g$ is integrable  has been partially handled by Chesneau (2007)
who showed that the problem is well-posed  (i.e., data loss does not affect the asymptotic minimax global convergence rates) when $f$ is spatially homogeneous. A complete study of the case when $1/g$ is integrable is considered in 
Section \ref{sec:small_alpha}. In depth discussion of the differences of the spatial features in the spatially inhomogeneous 
ill-posed problem considered in this paper is presented in Section   \ref{discussion}.

To address spatial irregularity of the design in the case  when the design density  function $g$ has a zero of a polynomial order, 
we develop a novel, two-stage, adaptive wavelet thresholding estimator. 
This  estimator  consists of a linear part which is taken at a resolution level that is chosen adaptively by Lepski's method and which estimates $f$ in the neighborhood of the zero of $g$. We refer to this as the {\em zero-affected}
part of the estimator. The second part is nonlinear (thresholding) and is used 
outside the immediate neighborhood of the zero of $g$. We refer to this as the 
{\em zero-free} part of the estimator. The lowest resolution level of the nonlinear part 
coincides with the resolution level of the linear part of the estimator, so that the sum of the two parts represents
$f$ correctly. If $1/g$ is integrable, then the zero-affected portion of the estimator vanishes and $f$ 
can be estimated by an adaptive wavelet thresholding estimator in the spirit of Chesneau (2007).

We limit our attention only to the   $L^2$-risk since the consideration of a wider class 
of risk functions will make the exposition of the present work even longer; all results, however, obtained can be extended 
to the case of $L^u$-risks, $1 \leq u < \infty$. Moreover, we consider only the univariate case, leaving generalizations to the multivariate case for future investigation.

The rest of the paper  is organized as follows.
Section \ref{formulation} discusses the formulation of the nonparametric regression estimation problem 
of the unknown response function $f$ on the basis 
of spatially inhomogeneous data, in particular when the design density function $g$ has either a zero of a polynomial order or a zero of an exponential order. Section \ref{lowbounds}  contains  the asymptotic 
minimax lower bounds for the $L^2$-risk when $f$ is assumed to belong to a Besov ball.   Section \ref{estimation_strategies}
talks about estimation strategies when $1/g$ is non-integrable, in particular, about partitioning $f$ and its  
estimator into the zero-affected and zero-free parts. Section \ref{est_fc_f0} elaborates on 
the estimation of the zero-affected and the zero-free parts, and is followed by Section \ref{minimax_risk} 
which discusses  the  choice of  adaptive resolution level and 
derives the asymptotic minimax upper bounds for the $L^2$-risk in the case when $1/g$ is non-integrable. 
Section \ref{sec:small_alpha} studies the complementary case when $g$ has   zeros but $1/g$ is still integrable.
Section \ref{discussion} concludes the paper  with a   discussion. 
Finally, Section \ref{append}  contains the proofs of the statements in the earlier sections.


\section{Formulation of the  problem}
\label{formulation}
\setcounter{equation}{0}

Consider the nonparametric regression model \fr{main_eq}. Since the noise level is assumed to be known and finite, 
without loss of generality, we set $\sigma =1$. 
Therefore, from now onwards, we work with observations 
$y_i$ governed by equation \fr{main_eq}, where $f \in L^2([0,1])$ is the unknown response function to be recovered, $x_i \in [0,1]$, $i=1,2,\ldots,n$, 
are random design points with the underlying density function $g$, and $\xi_i$, $i=1,2,\ldots,n$, 
are independent standard Gaussian random variables, independent of $x_i$, $i=1,2,\ldots,n$.
Furthermore,  we assume that  the design density  function $g$  is known and has  a finite number of zeros  which are well-separated,
i.e., there exist a constant $\delta>0$ such that the distance between two consecutive  zeros is at least $\delta$. 
The last  assumption is motivated by the following considerations. 
If $g$ vanishes on an interval $[a,b] \subset [0,1]$, $a < b$,  then consistent estimation of $f(x)$, for  $x \in [a,b]$,  is impossible. 
Also,  $g$ has an infinite number of zeros on $[0,1]$  only in the case when  $g$ is highly oscillatory, which is not 
a very likely scenario. Finally, the assumption that $g$ has low values on a part of its domain but is still separated from zero 
is not an interesting case to consider, since the lower bound on $g$ will appear in the constant of 
the  well-known expressions for the asymptotic minimax convergence rates (see, e.g.,  Tsybakov  (2009),  Chapters 1-2).

\medskip
Note that the above assumptions are not restrictive.  If the noise level $\sigma$ is unknown, it can be easily estimated with 
parametric precision using observations in the region where $g$ is separated from zero.
The assumption that the design points $x_i$, $i=1,2,\ldots,n$,  are random is not confining  either. 
In fact, with small modifications of the theory below, one can consider fixed points  $0\leq x_1 < x_2 < \cdots < x_n\leq 1$, 
generated by an increasing and continuously differentiable function $G$ such that $G(0)=0$, $G(1) =1$ and 
$G(x_i) = i/n$, $i=1,2,\ldots,n$. Then, the function $G$ plays the role of a ``surrogate'' distribution function
with density function $g$; the design points $x_i$, $i=1,2,\ldots,n$, can be then obtained as  
$x_i = G^{-1} (i/n)$, $i=1,2,\ldots,n$.

Moreover, since  the design density function $g$ is assumed to be known with a finite number of zeros that are also well-separated,
one can partition the interval $[0,1]$ into subintervals in such a manner that each subinterval  
contains only one zero of $g$.  
For this reason, in what follows, without loss of generality, we assume that $g$ has only one zero $x_0  \in [0,1]$,
and that the following condition holds.

\medskip

\noindent{\bf Assumption A.} Let the design density $g $ be a continuous function on the interval $[0,1]$ with $g(x_0)=0$,  $x_0 \in [0,1]$. 
Then, there exists   constants $\alpha \in \RR$,   $b \geq 0$ ($\alpha >0$ if $b=0$),  $\beta>0$  and $C_g>0$ such that, for   any $x$, with $x, x+x_0 \in [0,1]$,
\be \label{zeros}
\lim_{x \rightarrow 0} \, g(x_0 + x) |x|^{-\alpha} \exp(b|x|^{-\beta}) = C_g.
\ee 

If $b=0$, we shall say that $x_0$ is a  zero of {\it polynomial order}. If $b>0$, we shall say that $x_0$ is a  zero of {\it exponential order}. Observe  that \fr{zeros} implies that there exist some constants $0<C_{g1} < C_g < C_{g2}$
 such that for any $x$, with $x, x+x_0 \in [0,1]$ and $x_0 \in [0,1]$, one has
\be \label{gineq}
 g(x_0 + x) \leq C_{g2}  |x|^{ \alpha} \exp(-b|x|^{-\beta}),\ \ \ 
 g(x_0 + x) \geq C_{g1}  |x|^{ \alpha} \exp(-b|x|^{-\beta}). 
\ee

Note that the two cases in Assumption A correspond to the situations of  moderate ($b=0$) and severe ($b>0$) 
data losses, respectively. Chesneau (2007)  showed that in the case  of a moderate loss ($b=0$) with $0<\af <1$ 
(i.e., $1/g$ is integrable),  and for a response function $f$ that is spatially homogeneous, 
$f$ can be estimated with the same asymptotic minimax  global convergence rates as in the case of $b=0$ with $\af=0$ (i.e., $g$ is uniformly bounded from below); hence, in this case, the nonparametric regression estimation problem turns out 
to be a {\em well-posed} problem.

\medskip

Therefore, we shall be  mainly interested only in the complementary situation when $1/g$ is non-integrable: (i) moderate losses (i.e., $b=0$)   
with $\af \geq 1$ and  (ii) severe losses (i.e., $b>0$) with $\af \in \RR$ and $\beta >0$. As we shall see below, 
usually in those cases, the asymptotically  optimal  (in the minimax sense) estimation procedures yield  estimators with lower convergence rates   than in the case of equispaced observations, so that the corresponding nonparametric regression estimation problem under model \fr{main_eq}  becomes {\em ill-posed} 
(see Remark \ref{rem-ill-p}), with  the degree of  ill-posedeness growing as $\af \geq 1$ increases when $b=0$ or as $\beta >0$ increases when $b>0$.

\medskip

In what follows, we use the symbol $C$ for a generic positive constant, independent
of the sample size $n$, which may take different values at different places.


\begin{remark}
\label{rem-ill-p} 
{\bf  (Risk functions and design)}.  {\em    
As indicated above, we shall measure the precision of any estimator $\hfn$ of $f$ by its $L^2$-risk, i.e.,
$$
\Delta(\hfn) =  \EE \|\hfn - f\|^2.
$$
If the design points $x_i \in [0, 1]$, $i=1,2 \ldots,n$, in model (\ref{main_eq}) are treated as fixed (i.e., non-random), then, the above risk, 
evaluated  at  the equispaced design $\{i/n\}$, $i=1,2,\ldots,n$,  
corresponds to  
$$
\Delta^d(\hfn) =\frac{1}{n} \sum_{i=1}^n \EE [\hfn (i/n) - f (i/n)]^2,
$$
and leads to an {\em ill-posed}~ nonparametric regression estimation problem. 
However, it is instructive to note that if one measures the  precision of an estimator $\hfn$ at the design points $x_i \in [0,1]$, $i=1,2,\ldots,n$, only,
by calculating 
$$
\Delta^d_{fixed}(\hfn, x_i) =  \frac{1}{n}  \sum_{l=1}^n \EE [\hfn (x_i) - f (x_i)]^2,
$$ 
as it was done in, e.g., Amato \etal (2006), 
then the problem ceases to be ill-posed. Moreover, in this case, no special treatment 
is necessary to account for the irregular design. To confirm that, note that 
model \fr{main_eq} can be re-written as 
\be \label{F_function}
y_i = F(i/n) + \xi_i,  \quad i=1,2, \ldots, n,
\ee  
where $F(x) = f(G^{-1} (x))$, $x \in [0,1]$, and $G$ is the ``surrogate'' distribution function mentioned earlier. Construct now an estimator  $\hat{F}_n $ of $F$  using, e.g.,
any of the standard wavelet thresholding techniques, and set $\hfn (x) = \hat{F}_n (G(x))$, $x \in [0,1]$.
Then, 
$$
\hat{F}_n (x) = \hfn (G^{-1} (x)), \quad x \in [0,1],
$$ 
and  $\Delta^d_{fixed}(\hfn, x_i)$ takes the form
$$
\Delta^d_{fixed}(\hfn, x_i) =  \frac{1}{n} \sum_{i=1}^n \EE [\hat{F}_n (i/n) - F(i/n)]^2.
$$
Therefore, if  the observed data vector $y=(y_1,y_2,\ldots,y_n)$ is treated  as if the measurements were carried out at
equispaced design points, then, by using, e.g., available wavelet denoising algorithms, the resulting 
estimator $\hat{F}_n$ of function $F$  will be adaptive and it will lead to the smallest possible risk $\Delta^d_{fixed}(\hfn, x_i) $.
This phenomenon was noticed earlier by Cai and Brown (1998), Sardy {\it et al.} (1999) and Brown {\it et al.} (2002). }
\end{remark}

\begin{remark}
\label{local-global} 
{\em   {\bf (Local versus global convergence rates). } The nonparametric regression estimation problem of recovering $f$ globally,
on the basis of spatially inhomogeneous data, is a much more difficult task
than the corresponding problem of estimating $f$ locally, say at a given point $a$. Indeed, if $G$, the distribution function associated with the design density function $g$, is known, then $F(G(a))= f(a)$ and, hence, one can estimate $F$ 
at the point $G(a)$ instead of  estimating $f$ at the point $a$, where $F(x) = f(G^{-1} (x))$, $x\in [0,1]$, and $F$ is equispaced sampled, as in \fr{F_function}. 
Hence, local estimation can be reduced  to a well-addressed  pointwise regression estimation problem.
If $g(a) \neq 0$,  then the problem is well-posed and has been extensively studied
before. If, instead, $a =x_0$ is a zero of $g$, then one can deduce asymptotic minimax pointwise convergence rates directly from considerations of Remark \ref{rem-ill-p}
and straightforward calculus. Let, for simplicity, $x_0=0$ and 
$g(x) = (\alpha+1) x^{\af}$, so that   $G(x) = x^{\af +1}$ and $G^{-1} (x) = x^{1/(\af+1)}$,  $x\in [0,1]$.
Let $f$ satisfy a H\"older condition of order $s$ at $x_0$, i.e.,  $|f(x) - f(x_0)| \leq C |x-x_0|^s$.
Then, since $x_0=0$,   $F(x) = f(G^{-1} (x))$, $x \in [0,1]$,  satisfies a H\"older condition of order $s' = s/(\af +1)$ at $0$, i.e., for $x_0=0$,
\beqns
|F(x) - F(x_0)| & = & |f(G^{-1} (x)) - f(G^{-1} (x_0))| \leq C |G^{-1} (x) - G^{-1} (x_0)|^s
   =  C |x- x_0|^{s/(\af+1)}.
\eeqns
Since, for $x_0=0$,  $f(x_0) = F(0)$, one can set $\hat{f} (x_0) = \hat{F}(0)$ and obtain asymptotic minimax pointwise convergence rates for $\hat{f}(x_0)$, on noting that
$$
\EE \|\hat{f} (x_0) - f(x_0)\|^2 = \EE \|\hat{F}(0) - F(0)\|^2 \leq C\, n^{-\frac{2s'}{2s' +1}} = 
O \lkr n^{-\frac{2s}{2s + \alpha +1}} \rkr,
$$ 
which coincides with the asymptotic minimax pointwise convergence rates obtained by Gaiffas (2005).
The whole argument here rests on the fact that $f(x_0) = F(G(x_0))$, $x_0 \in [0,1]$, so one can estimate $F$ at $G(x_0)$ instead of estimating $f$ at the $x_0$. This, however, cannot be accomplished when a global estimation procedure is required since, in such a case, a  Taylor expansion is needed, that
can be applied only locally.}
\end{remark}


\section {Minimax lower bounds for the $L^2$-risk over Besov balls}
\label{lowbounds}
\setcounter{equation}{0}

 Before constructing an adaptive estimator of the unknown response function $f$ under model \fr{main_eq}, 
we  first derive the asymptotic minimax lower bounds for the $L^2$-risk over a wide range of Besov balls. 
 
Among the various characterizations of Besov spaces $B_{p,q}^s$  in terms of wavelet bases, we recall that for an $r$-regular multiresolution analysis (see, e.g., Meyer, 1992, Chapter 2, pp 21--25),  with $0<s<r$,  and for a Besov ball $B_{p,q}^s (A)$ defined as
$$
B_{p,q}^s (A) =\{f  \in L^p([0,1]):~f \in
B_{p,q}^s,\; \|f\|_{B_{p,q}^s} \leq A \},
$$  
of radius $A>0$ with $1\leq p,q \leq \infty$, one has, with $s' = s+1/2-1/p$,
\begin{equation}
\label{bpqs}
B_{p,q}^s (A)   = \lfi f  \in L^p([0,1]):  \lkr
\sum_{k=0}^{2^m -1}  |\amk|^p \rkr^{1/p}  
 + \lkr \sum_{j=m}^{\infty} 2^{js'q} \lkr \sum_{k=0}^{2^j -1} 
|\bjk|^p \rkr^{q/p} \rkr^{1/q} \leq A \rfi,
\end{equation}
with respective sum(s) replaced by maximum if  $p=\infty$ and/or $q=\infty$, where $s' = s +1/2-1/p$ (see, e.g., Johnstone \etal (2004)). 
We study below the $L^2$-risk
over Besov balls  $B_{p,q}^s (A)$ defined as
$$
R_n (B_{p,q}^s (A)) = \inf_{\tilde{f}_n} \, \sup_{f \in B_{p,q}^s (A)} 
\EE  \|  \tilde{f}_n - f \|^2,
$$
where $\|h\|$ is the $L^2$-norm of a function $h$ defined on $[0,1]$, and the
infimum is taken over all possible square-integrable estimators (i.e., measurable functions) $\tilde{f}_n$
of $f$ based on observations $y_i$ from model \fr{main_eq}.  

\medskip

The following statement provides the asymptotic minimax lower bounds for the
$L^2$-risk. 
\begin{theorem}  \label{th:lower}
 Let  $1 \leq p, q \leq \infty$  and $\max(1/p, 1/2) \leq s <r$, and let Assumption A 
(with $\af > 0 $ if  $b=0$, and $\af \in \RR$ and $\beta >0$ if $b>0$) hold. Then, as $n \rightarrow \infty$, 
\be \label{low1} 
R_n (B_{p,q}^s (A)) \geq  \lfi
\begin{array}{ll}
C\, n^{-\frac{2s}{2s  + 1}} & {\rm if}\ \  b=0,\, \af s < s',\\
C\, n^{-\frac{2s'}{2s' + \alpha}} & {\rm if}\ \  b=0,\, \af s \geq s',\\
C\,  (\ln n)^{-\frac{2s'}{\beta}} &  {\rm if}\ \   b>0.
\end{array} \right.
\ee
\end{theorem}

Note that the asymptotic minimax lower bound for the $L^2$-risk in the first part of \fr{low1} is obtained by the arguments in Theorem 3.1 of Chesneau  (2007).

\begin{remark}
\label{conv_rates} 
{\em   {\bf  (Global convergence rates).  }  As we shall show below, the asymptotic  minimax lower bounds for the $L^2$-risk obtained in Theorem \ref{th:lower}
are attainable for $b>0$ and are attainable up to a logarithmic factor for $b=0$. If $\af s = s'$, the asymptotic minimax global  
convergence rates in the first and second parts of \fr{low1} coincide.  Hence, 
whenever $\af s \leq s'$, the aforementioned nonparametric regression estimation problem is not ill-posed but well-posed, in the sense that the asymptotic minimax global convergence rates are the same as in the case of an equispaced design.
For $\af \geq 1$, this relation can take place only if $2 \leq p \leq \infty$, i.e., when the function is 
spatially homogeneous. In particular, $\af s \leq s'$ holds true for any $\af$ such that
$1 \leq \af \leq 1  + (1/2 - 1/p)/s$, i.e., when $f$ is very spatially homogeneous 
($p$ is large, in particular, when $p > 2/(1 - (\af -1)s)$ provided that $1 < \af < 1 + 1/s$), so that
even a relatively severe data loss does not lead to the reduction of asymptotic minimax global convergence rates.
If $0 < \af < 1$, then the considered nonparametric regression estimation problem  is always well-posed whenever $f$ is spatially homogeneous ($p \geq 2$) and also 
when $f$ is spatially inhomogeneous ($1 \leq p < 2$) and  $0 < \af < 1- (1/p-1/2)/s$.
Therefore, even if $f$ is spatially inhomogeneous, the aforementioned nonparametric regression estimation problem is well-posed whenever data loss is 
very limited ($0 < \af < 1- (1/p-1/2)/s$).
}
\end{remark}


\section{Estimation  strategies when $1/g$ is non-integrable}
\label{estimation_strategies}
\setcounter{equation}{0}

We consider a scaling function $\ph^*$  and a mother wavelet $\psi^*$ that generate an orthonormal wavelet basis
 in $L^2(\RR)$, as  those obtained from, e.g., an $r$-regular multiresolution 
analysis of $L^2(\RR)$, for some $r>0$. We shall also assume that $\ph^*$ and $\psi^*$ are both 
compactly supported, with integer bounds on their supports so that, for some 
$L_{\ph^*},\ U_{\ph^*},\ L_{\psi^*},\ U_{\psi^*} \in \ZZ$, with 
$L_{\ph^*} < U_{\ph^*}$,  $L_{\psi^*} < U_{\psi^*}$,
$$
\supp(\ph^*) = [L_{\ph^*}, U_{\ph^*}], \ \ \ \supp(\psi^*) = [L_{\psi^*}, U_{\psi^*}],\ \  
L_{\ph^*} \leq 0,\  U_{\ph^*} \geq 0,\ U_{\ph^*} - L_{\ph^*} \geq 4.
$$
(For instance, the Daubechies or Symmlets scaling functions $\ph^*$ and mother wavelets $\psi^*$, 
with filter number (number of vanishing moments) $N \geq 3$, satisfy \fr{support} with $L_{\ph^*} = 0$, 
$U_{\ph^*}=2N-1$, $L_{\psi^*}=1-N$ and $U_{\psi^*}=N$, see, e.g., Mallat (1999),  Section 7.2.)

We then obtain a periodized version of the wavelet basis on the unit interval, i.e., 
for $j \geq 0$ and $k=0,1,\ldots,2^j-1$, as
$$
\ph_{jk}(x) = \sum_{i \in \ints} 2^{j/2} \ph^*(2^j (x +i) -  k),
\quad \psi_{jk}(x) = \sum_{i \in \ints} 2^{j/2} \psi^*(2^j (x +i) - k), \quad x \in [0,1],
$$
so that, for any $m \geq 0$,  the set
$$
\{\phimk,\ \psijk:\ j \geq m, \, k=0,1,\ldots, 2^{j}-1\},
$$ 
where
$$
\phimk (x) = 2^{m/2}  \ph(2^m x -k), \quad
\psijk(x) = 2^{j/2}  \psi(2^j x -k), \quad x \in [0,1],
$$
forms an orthonormal wavelet basis   for $L^2([0,1])$
(see, e.g., Mallat (1999), Theorem 7.16).  Hence, for any $m \geq 0$, any $f\in L^2([0,1])$, can be expanded as
\be 
f(x) = \sum_{k=0}^{2^{m}-1}a_{mk} \phimk (x) + \sum_{j=m}^\infty \sum_{k=0}^{2^{j}-1} b_{jk} \psijk (x), \quad  x \in [0,1],
\label{funf} 
\ee
where
$$
a_{mk} = \int^1_0 f(x)\phimk(x)\,dx, \quad k=0,1,\ldots,2^{m}-1,
$$
$$
b_{jk}=\int^1_0 f(x)\psijk(x)\,dx,\quad j \geq m, \quad k=0,1,\ldots,2^{j}-1.
$$

Denote by $L_{\ph}, U_{\ph}, L_{\psi}$ and $U_{\psi}$ the support bounds of the periodic 
scaling function $\ph$ and mother wavelet $\psi$.
Note that the supports of $\ph^*_{mk}$  and $\ph_{mk}$ coincide if and only if  $2^m > U_{\ph^*} - L_{\ph^*}$, and, similarly, 
the supports of $\psi^*_{jk}$ and $\psi_{jk}$ coincide if and only if  $2^m > U_{\psi^*} - L_{\psi^*}$.
Choose the lowest resolution level $m_1$  such that 
$2^{m_1} > \max \lkr U_{\ph^*} - L_{\ph^*},  U_{\psi^*} - L_{\psi^*} \rkr$,
so that supports of periodic and non-periodic wavelets coincide.
In this case, we obtain that
\be  \label{support}
 L_{\ph^*}= L_{\ph},\  U_{\ph^*}= U_{\ph},\   L_{\psi^*}=L_{\psi},\ U_{\psi^*}=U_{\psi}, \  L_{\ph} \leq 0,\  U_{\ph} \geq  0,
\ U_{\ph } - L_{\ph } \geq 4.
\ee

For any integer $l \geq 1$, denote $k_{0l} = 2^l x_0$. (Note that $k_{0l}$ is  not necessarily a rational quantity
and can take any value.) At each resolution level, we partition the set of all indices into the indices which are 
{\it zero--affected   }   and {\it zero--free}. In particular, let 
 $K_{0m}^\ph$ and $K_{0j}^\psi$ be 
the sets such that, for any integer $m\geq m_1$ and $j=m,m+1,\ldots$,
\beqns
K_{0m}^\ph & = & \lfi k: 0 \leq k \leq 2^m -1, \Lh -1 < \kom - k < \Uh+1 \rfi,\\ 
K_{0j}^\psi & = & \lfi k: 0 \leq k \leq 2^j -1, \Ls -1 < \koj - k < \Us+1 \rfi
\eeqns
and let 
$$
\Komhc = \lfi k: 0 \leq k \leq 2^m -1, k \notin \Komh \rfi, \quad
\Kojsc = \lfi k: 0 \leq k \leq 2^m -1, k \notin \Kojs \rfi.
$$
Simple calculations yield that   $k \in \Komhc$ and $k \in \Kojsc$ imply that   $x_0 \not\in \supp\ \phimk$
and     $x_0 \not\in \supp\ \psijk$, respectively, so that the sets $\Komhc$ and $\Kojsc$ are zero--free
while the sets $\Komh$ and $\Kojs$ are zero--affected.

With the above notation it is easy to see that, for any $m \geq m_1$ and $j=m,m+1,\ldots$, $f$ can be partitioned as 
the sum of zero--affected and zero--free parts, i.e.,
$$
f(x) = \fom (x) + \fcm (x), \quad x \in [0,1],
$$
where   
\beqn
\fom (x) & = & \sum_{k \in \Komh} a_{mk} \phimk (x) + \sum_{j=m}^{\infty}  \sum_{k \in \Kojs} b_{jk} \psijk (x), \quad x \in [0,1],
\label{fo}\\
\fcm (x) & = & \sum_{k \in \Komhc} a_{mk} \phimk (x) + \sum_{j=m}^{\infty} \sum_{k \in \Kojsc} b_{jk} \psijk (x), \quad x \in [0,1].
\label{fc}
\eeqn 
We then construct estimators $\hfom$ and $\hfcm$ of $\fom$ and $\fcm$, respectively,
and estimate $f$ by
\be \label{total_est}
\hfm  (x) = \hfom  (x) + \hfcm  (x), \quad x \in [0,1].
\ee
(We emphasize the unusual feature in the construction of $\hfm$: as we shall see below, $\hfom$  is a linear wavelet estimator   while 
$\hfcm$ is a nonlinear (thresholding) wavelet estimator with the lowest resolution level $m$ determined by the linear part.) 

\medskip
By observing that, for any function  $u \in L^2[0,1]$,  we have 
$$
 \int_0^1 u(x) f(x) dx = \EE \left( \frac{f(X) u(X)}{g(X)} \right),
$$
when the random variable $X \sim g$, and setting, for any $m \geq m_1$ and $j=m,m+1,\ldots$, $u(x) = \phimk (x)$ and $u(x) = \psijk (x)$, $x \in [0,1]$,  in turn, similarly to (3.3) in 
Chesneau (2007), we estimate $a_{mk}$, $k \in \Komhc$, and  $b_{jk}$, $k \in \Kojsc$, respectively, by 
 \be  \label{hamk}
\hamk = \frac{1}{n} \sum_{i=1}^n \frac{\phimk(x_i) y_i}{g(x_i)},\ \  k \in \Komhc,\ \ \ \ \ 
\tbjk = \frac{1}{n} \sum_{i=1}^n \frac{\psijk (x_i) y_i}{g(x_i)},\ \   k \in \Kojsc.
\ee
Hence, we can construct an estimator $\hfcm$ of $\fcm$ by estimating $a_{mk}$, $k \in \Komhc$, and  $b_{jk}$, $k \in \Kojsc$, by $\hamk$, $k \in \Komhc$, and $\tbjk$, $k \in \Kojsc$, respectively,  given in (\ref{hamk}), along with a thresholding step (see below).

Note that since $1/g$ is non-integrable, the estimators given in \fr{hamk} would have infinite variances if 
$k \in \Komh$ or  $k \in \Kojs$,   so that one cannot construct an analogous estimator $\hfom$ of $\fom$
by direct estimation of the appropriate scaling and wavelet coefficients. Instead, in this case, we shall use a linear estimator with the lowest resolution level $m$ estimated from the data.
In what follows, we shall consider the estimation of $\fom$ and $\fcm$ separately.

\section{Estimation of  the  zero-free and the  zero--affected parts. } 
\label{est_fc_f0}
\setcounter{equation}{0}

Consider first the estimation of the zero-free part.
 In order to estimate  $\fcm$, we construct a wavelet thresholding  estimator $\hfcm$   as 
\be \label{hfcmfixed}
\hfcm  (x)   =   \sum_{k \in \Komhc } \hamk \phimk (x) + \sum_{j=m}^{J-1} \sum_{k \in \Kojsc} \hbjk  \psijk (x), \quad m_1 \leq m \leq J-1,\  
x \in [0,1], 
\ee
where   $\hamk$  are given in \fr{hamk}, $J$ is defined below in \fr{Jvalueo},   while the  coefficients    $\hbjk$ are 
thresholded estimators of the wavelet coefficients $\bjk$ defined as
\be  \label{hbjk}
\hbjk  = \lfi \begin{array}{ll}
\tbjk \, \II( \tbjk^2 >  d^2 n^{-1}\,  \ln n \, 2^{j\af}\, |k - \koj|^{-\af} ) & \mbox{if}\ \ b=0,\\
\tbjk \, \II( |k - \koj| > 2^{j-m} ) &    \mbox{if}\ \ b>0.
\end{array} \right.
\ee 
Here, $d>0$  is a constant,   $\tbjk$ are defined by \fr{hamk} and  $m$ is such that $m_1 \leq m \leq J-1$, where  
\be \label{Jvalueo}
2^{m_1} = \max \lkr U_{\ph^*} - L_{\ph^*},  U_{\psi^*} - L_{\psi^*} \rkr + 1,\ \ \ \ \ 
2^J = \lfi \begin{array}{ll}
(n/\ln  n)^{1/(\af+1)}& \mbox{if}\ \ b=0,\\
(\ln n)^{2/\beta} & \mbox{if}\ \ b>0.
\end{array} \right.
\ee

Consider now the estimation of the zero-affected part. 
Since   the estimators $\hamk$ of $\amk$, given in \fr{hamk}, have infinite variances when $k \in \Komh$, we estimate those coefficients by  
solving a system of linear equations. Note that there is a finite known number of indices in  $\Komh$, at most, 
$w_\phi=\Uh - \Lh$  indices. For any given $m$, such that $m_1 \leq m \leq J-1$, denote
\be \label{fmepsm}
f_m(x) = \sum_{k=0}^{2^m-1} a_{mk} \phimk (x), \ \ \ 
\eps_m(x) = \sum_{j=m}^{\infty} \sum_{k=0}^{2^j-1} b_{jk} \psijk (x), \quad x \in [0,1],
\ee
and observe that $f(x) = f_m(x) + \eps_m(x)$, so that
\be  \label{eq_sum}
 \sum_{k \in \Komh} a_{mk} \phimk (x) = f_m (x) - \eps_m (x) - \sum_{k \in \Komhc} a_{mk} \phimk (x), \quad x \in [0,1].
\ee
Denote  $\Om_\delta = [\Lh + \delta_b, \Uh - \delta_b]$, and choose $\delta_b$ such that 
\be \label{delta}
\delta_b = \lfi \begin{array}{ll}
0 < \delta_b < 1/2,\  \ph(\Lh + \delta_b) \neq 0, \ \ph(\Uh - \delta_b) \neq 0, & 
 \mbox{if} \ \ b> 0, \\
0, & \mbox{if} \ \ b=0.
\end{array} \right.
\ee
Introduce also a finite set of indices
\be \label{Komstar}  
K_{0m}^* =  \lfi k: 0 \leq k \leq 2^m -1,\ 2 \Lh - \Uh \leq \kom - k < \Lh \ \ 
\mbox{or}\ \  \Uh < \kom - k  \leq 2 \Uh - \Lh \rfi.
\ee
Now, multiply both sides of formula \fr{eq_sum} by $g(x)\, \ph_{ml} (x)\, \II(2^m x -l \in \Om_\delta)$, $l \in \Komh$, 
where $\II(x \in \Om)$ is the indicator of set $\Om$, and integrate.
As a result,   obtain the following system of linear equations
\be  \label{system}
\bAm \bum = \bcm - \bepsm - \bBm \bvm.
\ee
Here, matrices $\bAm$  and $\bBm$ and vectors $\bcm$, $\bepsm$, $\bum$ and $\bvm$   have, respectively,  elements
\begin{align}
A^{(m)}_{lk} & = \int_0^1 \phimk (x) \phiml (x) g(x)\, \II(2^m x -l \in \Om_\delta) dx,\ \ k,l \in \Komh, \label{matrA}\\
B^{(m)}_{lk} & = \int_0^1 \phimk (x) \phiml (x) g(x)\, \II(2^m x -l \in \Om_\delta) dx,\ \ l \in \Komh, \ k \in K_{0m}^*,  \label{matrB}\\
c^{(m)}_l & = \int_0^1  f(x) \phiml (x) g(x) \, \II(2^m x -l \in \Om_\delta) dx,\ \  l \in \Komh, \label{vecc} \\
\eps^{(m)}_l & = \int_0^1  \eps_m (x) \phiml (x) g(x) \, \II(2^m x -l \in \Om_\delta) dx,\ \  l \in \Komh, \label{veceps}\\
u^{(m)}_{k} & = \amk,\  \ k \in \Komh,\ \ \ \  v^{(m)}_{k}   = \amk,\  \ k \in K_{0m}^*. \label{vecuv} 
\end{align}
(Note   that the matrices $\bAm$ and $\bBm$ are completely known, and also observe   
that $B^{(m)}_{lk} \neq 0$ only if $k \in K_{0m}^*$, since, 
for $k \not\in K_{0m}^*$, one has $\phimk (x) \phiml (x) = 0$.)

Since $K_{0m}^* \subset \Komhc$, it follows from \fr{vecuv}  that components $v_{k}^{(m)}$ of vector $\bv^{(m)}$  can be estimated 
by 
$$
\hbvmk = \hamk, \quad k \in \Komhc,
$$ 
using \fr{hamk}. We also estimate $c^{(m)}_l$ by 
\be \label{hbcl}
\hbcml =  \frac{1}{n} \sum_{i=1}^n  y_i \, \phiml (x_i)\, \II(2^m x_i -l \in \Om_\delta),\ \ \ l \in \Komh, 
\ee
and ignore vector $\beps$ in \fr{system}, thus, replacing \fr{system} by the following system of linear equations 
\be  \label{estsys}
\bAm \hbum = \hbcm   - \bBm \hbvm.
\ee
Since matrix $\bAm$ is a positive definite matrix of non-asymptotic size,  ${\rm det}(\bAm ) \neq 0$ and  we   obtain the solution 
$$
\hbum = (\bAm)^{-1} (\hbcm   - \bBm \hbvm)
$$ 
of the system of linear equations \fr{estsys}. 

Finally, for any given $m$, such that $m_1 \leq m \leq J-1$, we  set $\hamk = \hbumk,$ $k \in \Komh$, and estimate $\fom$ by the following linear wavelet estimator 
 \be  \label{hfom}
\hfom (x)   =   \sum_{k \in \Komh} \hamk \phimk (x), \quad x \in [0,1].
\ee

The following statement provides the asymptotic upper bounds for the bias and 
the variance of the estimator $\hfom$ given in  \fr{hfom}.

 \begin{lemma} \label{lem:error_sys_sol}
Denote $\fom (x) =   \sum_{k \in \Komh} \amk \phimk (x)$ and let $m=m(n)$ 
be a non-random, non-negative integer, quantity such that  $m(n) \rightarrow \infty$ as $n \rightarrow \infty$.  Let the estimator $\hfom$ be defined by \fr{hfom}. Then, as $n \rightarrow \infty$,  
\be   \label{linest_error}
\| \EE \hfom  - \fom \|^2 = O \lkr  2^{-2ms'} \rkr,\ \ \  
\EE \| \hfom  - \EE \hfom \|^2 = O \lkr  n^{-1} 2^{m \af}\ \exp( b\, 2^{m \beta} [2^{\beta +1} + 1] ) \rkr.
\ee
Moreover, if $b=0$, then, as $n \rightarrow \infty$, $\EE \| \hfom - \EE \hfom \|^4 = o(1)$.
\end{lemma}

Define $m_0$ to be such that
\be \label{mvalue}
2^{m_0} = \lfi
\begin{array}{ll}
n^{^\frac{1}{2 s' + \af}} & \mbox{if}\ \ \ b=0,   \\
\lkr b^{-1} 2^{- (\beta + 2)} \ln n \rkr^{\frac{1}{\beta}} & \mbox{if}\ \ \ b>0. 
\end{array} \right.
\ee 
It follows from Lemma \ref{lem:error_sys_sol} that, if $m=m_0$, the error $\EE \| \hfom - \fom \|^2$
of the estimator $\hfom$ attains the asymptotic minimax lower bounds for the $L^2$-risk obtained in Theorem \ref{th:lower}. 
Since $\af$, $b$ and  $\beta$ in   \fr{mvalue}  are known, the value of $m_0$ is also known in the case of $b>0$.
Therefore,  one can select $m_0$ as the lowest resolution level in the estimator of the zero-free part \fr{hfcmfixed}.
 
 \medskip
On the other hand, the following lemma demonstrates that the wavelet thresholding estimator $\hfcm$, defined in \fr{hfcmfixed} with $m=m_0$ given in (\ref{mvalue}), attains 
the asymptotic minimax lower bounds for the $L^2$-risk obtained in Theorem \ref{th:lower}, in the case of $b>0$.


\begin{lemma}  \label{lem:exponential_zero}
Let  $1 \leq p,q \leq \infty $  and $\max{(1/2,  1/p)} \leq s <r$, and let Assumption A (with  $b>0$, $\beta >0$  and  $\af  \in \RR$) hold.   
Let the estimator $\hfcm$ be defined by 
\fr{hfcmfixed} with $m=m_0$ given in (\ref{mvalue}),
Then,   as $n \rightarrow \infty$, 
 \be \label{expon_zero} 
 \sup_{f \in B_{p,q}^s (A)} \EE  \| \hfcmo  - \fcmo \|^2  \leq  C\,  (\ln n)^{-\frac{2s'}{\beta}}. 
\ee
\end{lemma}

Unfortunately, this idea cannot be implemented in the case of $b=0$. Indeed, 
 though $\af$  in   \fr{mvalue} is known, the value of $s'$ is unknown  and, 
therefore, the estimator $\hfom$, defined  in 
\fr{hfcmfixed} with $m=m_0$ given in (\ref{mvalue}), is not realizable  if $b=0$.
 In this case, we need to adequately choose a  resolution level, say $\hm$, which approximates  $m_0$ in some sense, and then estimate $f$ by 
$\hf  (x) =   \hf_{0,\hm}  (x)  + \hf_{c,\hm}  (x)$. The choice of such resolution level is a rather difficult task. 
On the one hand, $\hm$ should not be too small since, otherwise, the linear portion of the estimator would have bias that will be too large. On the other hand, since $\hfom$ is the linear estimator, 
in order to represent $f = f_{0,\hm} + f_{c,\hm}$ adequately, $\hm$ has to be used as the lowest resolution level in 
$\hf_{c,m}$. 

\medskip
The following lemma provides the asymptotic minimax upper bounds for the $L^2$-risk of the wavelet thresholding estimator  $\hfcm$, defined in \fr{hfcmfixed},  in the case of $b=0$. In particular, it shows that this risk contains the component $n^{-1}\ 2^{m \af}$, so that  in order to attain the asymptotic minimax lower bounds for the $L^2$-risk in the case of $b=0$, obtained in Theorem \ref{th:lower} (up to a logarithmic factor),
one needs $\hm \leq m_0$ with high probability.


\begin{lemma}  \label{lem:polynomial_zero}
Let  $1 \leq p,q \leq \infty$  and $\max{(1/2,  1/p)} \leq s <r$, and let Assumption A  (with $b=0$ and $\af \geq 1$) hold. Let the estimator $\hfcm$  be defined by  \fr{hfcmfixed},
 where  $m$ is such that $m_1 \leq m \leq J-1$, with $m_1$ and $J$   defined in \fr{Jvalueo}.
Let $\hbjk$ be given by \fr{hbjk} with $d > 4 C_d$, where $C_d$ is given by 
\be \label{Cdvalue}
C_d = 8 C_\psi C_{g1}^{-1} \max \lkr 2, 2 \|f\|_\infty^2,  \|f\|_\infty  \|\psi\|_\infty/3, 
\|\psi \|_\infty  \rkr\ \ \ \mbox{with}\ \ \ C_\psi = [ 2 \max ( |\Ls|, |\Us| ) ]^\af.
\ee
Then,  as $n \rightarrow \infty$,
\be \label{risk_hfc}
 \sup_{f \in B_{p,q}^s (A)}  \EE \| \hfcm - \fcm \|^2 \leq  \lfi
\begin{array}{ll} 
C\,  (n^{-1}\ 2^{m \af}\, (\ln n)^{\II(\af =1)} + 
n^{-\frac{2s}{2s + 1}}\ (\ln n)^{\mu_1}) 
& {\rm if}\ \ \  b=0,\ \af s < s',\\ 
C\,  (n^{-1}\ 2^{m \af}\, (\ln n)^{\II(\af =1)} + 
 n^{-\frac{2s'}{2s' + \af}}\ (\ln n)^{\mu_2} )
& {\rm if}\ \ \  b=0,\  \af s \geq s',
\end{array} \right. 
\ee
where, 
$$
\mu_1 =  \frac{2s (1 +\II(\af =1))}{2s + 1} \quad \text{and} \quad
\mu_2 = \frac{2s'(1 + \II(\af =1))}{2s' + \af}  + \II \lkr \frac{s'}{s} =\af >1 \rkr.
$$
Moreover,  as $n \rightarrow \infty$,
\be \label{risk_hfc_4}
\EE \| \hfcm - \fcm \|^4 = o \lkr  1 \rkr.  
\ee
\end{lemma}

\begin{remark} \label{unknown_g}
{\rm  {\bf  (The case of an unknown design density).  }
So far, we  have made the assumption that the design density function $g$ is known.  In many practical situations, however, this may not be true.  
Nevertheless,  the suggested method can be applied to the case of an unknown $g$.
%
In particular, one should start with the construction of lower and upper confidence limits $\hat{g}_L$ and $\hat{g}_U$, respectively, 
for the  unknown  $g$.  This can be accomplished by using a variety of nonparametric methodologies for  constructing  simultaneous confidence intervals 
of a probability density function (see, e.g., Tribouley  (2004), Bissantz {\it et al.} (2007) and  Gin\'e and Nickl (2010)). 
The lower estimator confidence limit $\hat{g}_L$   allows to   assess  the areas where $g$  vanishes.
If there are several distinct areas like that, we partition the interval $[0,1]$ into subintervals,
so that each of the intervals contains only one zero of $g$. After that, we can estimate the location of  the  zero of $g$ 
as the middle of the interval where the lower confidence bound for $g$ is equal  to  zero.
From this point onwards, without loss of generality,   we   assume that $g$ vanishes at only one point 
of the interval $[0,1]$. We shall also limit our attention to  the case of   zero of  a  polynomial order, 
since,  in the case of exponential zero, 
data loss around zero is so severe  that in practice   $f$ cannot be adequately estimated. 
In order to implement our estimators, 
we need to assess the value  of $\alpha$ and the constants 
$C_{g1}$ and $C_{g2}$ in \fr{gineq}. For this purpose, note that whenever $z$ is small,  one has the following relation for the  
distribution function $G$ of $g$
$$
G(x_0 + z) - G(x_0 -z) \approx 2 C_g (\af +1)^{-1} z^{\af +1}.
$$
Therefore, $\alpha + 1$ and $C_g$ can be recovered using linear regression of $\log [\hat{G}(x_0 + z) - \hat{G}(x_0 -z)]$
onto $\log z$   for small values of $z$ (i.e., using observations surrounding $x_0$), where $\hat{G}$ is the empirical distribution function  of  
$G$ based on  $ x_1,x_2, \ldots, x_n$. 
As the value of $\alpha$ has been  estimated  by $\hat{\alpha}$, the constants $C_{g1}$ and $C_{g2}$ can be  estimated by 
$$
{\widehat C}_{g1}= \min_{ 1\leq i\leq n}  \ \frac{| {\hat G} (x_i )  -  {\hat G}(x_0)|
({\hat \af}  +1)}{|x_i-x_0|^{ {\hat \af}  +1}},\ \ \ \ 
 {\widehat C}_{g2}= \max_{ 1\leq i\leq n}  \ \frac{| {\hat G} (x_i )  -  {\hat G} (x_0)|( {\hat \af}  +1)}{|x_i-x_0|^{ {\hat \af}  +1}}. 
$$
Note that the estimated values  $\hat{\af}$, $\widehat{C}_{g1}$ and $\widehat{C}_{g2}$ of  $\af$, $C_{g1}$ and $C_{g2}$, 
respectively,   are necessary for finding the highest resolution level $J$
and for  the construction of the involved thresholds. 
Once the above estimates haven been obtained, we then estimate the zero-affected part  of $f$. This procedure is relatively easy to generalize to the case of an unknown $g$: 
one just needs to replace  the elements $A^{(m)}_{lk}$ and 
$B^{(m)}_{lk}$ of the matrices  $\bAm$  and $\bBm$, given by (\ref{matrA}) and (\ref{matrB}), respectively, by their corresponding unbiased estimators
\beqns
\hat{A}^{(m)}_{lk} & = &  n^{-1} \sum_{i=1}^n  \phimk (x_i) \phiml (x_i),\ \ k,l \in \Komh,\\
\hat{B}^{(m)}_{lk} & = &  n^{-1} \sum_{i=1}^n  \phimk (x_i) \phiml (x_i),\ \ l \in \Komh, \ k \in K_{0m}^*,  
\eeqns 
to solve the  corresponding system of linear equations for various values of $m$ and  
to  carry out Lepski's procedure (see Section \ref{minimax_risk}) to choose a suitable value of $\hat{m}$.
Subsequently, we estimate  the  wavelet coefficients using an estimator $\hat{g}$ of $g$  in  \fr{hamk}.
Note that we  only  need to evaluate $\hat{g}$ at the  points where $g$  cannot vanish. Moreover, since 
we need to use $\hat{g}$ only for the ``zero-free'' part, we need estimators of $g$ away from its zero 
where the density of observations is reasonably high.}
\end{remark}


\section {Adaptive estimation and the minimax upper bounds for the $L^2$-risk   when $1/g$ is non-integrable }
\label{minimax_risk}
\setcounter{equation}{0}

In order to construct an adaptive wavelet estimator of $f$ in the case of  $b=0$, we shall use the 
technique of optimal tuning parameter selection pioneered by Lepski (1990, 1991) and further exploited 
in Lepski and Spokoiny (1997) and  Lepski {\em et al.} (1997). The idea behind this technique 
is to construct estimators for various values of the tuning parameter in question ($m$, in our case),
and then choose an optimal value of the tuning parameter by regulating the differences between the 
estimators constructed with different values of the parameter.

In particular, if $b=0$, for various values of $m$, we construct versions of the system of equations
\fr{estsys},  where the estimators $\hbv^{(m)}$ are constructed as before, 
solve those systems and obtain the estimators \fr{hfom}, where $\hamk = \hbumk,$ $k \in \Komh$.
We then construct an estimator $\hfm$ of $f$ using formula \fr{total_est}, where $\hfom$ and $\hfcm$ are given by 
\fr{hfom} and \fr{hfcmfixed}, respectively, where $m$ is the lowest resolution level of $\hfcm$.  
The choice of the optimal resolution level   is driven 
by the zero-affected part of $f$ rather than the zero-free part. For this reason, for any resolution level $m >0$,
we define a neighborhood $\Xi_m$ of $x_0$ as
\be \label{Xi_neighborhood}
\Xi_m = \lfi x:\ 2^{-m}[\min(\Lh,\Ls)- \Uh] < x-x_0 < 2^{-m}[\max(\Uh,\Us)- \Lh] \rfi
\ee
and observe that $\Xi_m$ is designed so that $\supp (\fom ) \subseteq \Xi_m$, $\supp (\hfom ) \subseteq \Xi_m$
and $\Xi_j \subset \Xi_m$ if $j > m$.

For $b=0$, choose $m  = \hm$ such that $m_1 \leq m \leq J-1$, where $m_1$ and $J$ are defined in \fr{Jvalueo} and 
\be \label{mopt}
\hm = \min \ \lfi m: \ \   \| (\hfm   - \hfj) \II(\Xi_m) \|^2 \leq \lam^2\, 2^{j\af}\, n^{-1} \ln n 
\ \ \mbox{for all}\ \ j,\   m  < j \leq J-1 \rfi,
\ee
where $\lam >0$ is a constant to be defined below. For completeness, define
\be \label{moptlargeb}
\hm= \mo\ \ \ \mbox{if}\ \ b>0,
\ee
where $\mo$ is defined in \fr{mvalue}.

The construction of $\hm$ for $b=0$ is based on the following idea.
Note that, when $\hm  \leq m_0$,    then, for $m=\hm$, one has 
\be \label{Lepskii-expl}
\EE \| \hfm - f  \|^2 \leq 2 \lkv \EE \| \hfm -  \hfmo \|^2 +   \EE \| \hfmo - f  \|^2 \rkv.
\ee
The first component in \fr{Lepskii-expl} is small due to definition of the resolution level $\hm$
while the second component is calculated at the optimal resolution level $m_0$  and, 
hence,  tends to zero at the optimal (in the minimax sense) global convergence rate (up to a logarithmic factor). 
On the other hand, if $m=\hm  > m_0$, then there exists $j > m_0$ such that 
$\| (\hat{f}_{m_0}   - \hfj) \II(\Xi_{m_0}) \|^2 > \lam^2\, 2^{j\af}\, n^{-1} \ln n. $

\medskip
The following lemma shows that,  if $\lam$ is large enough, the probability of the above-mentioned event is infinitesimally small. (Here, $||h||_{\infty}$ is the   uniform norm  of a bounded function $h$ defined on $[0,1]$.)


\begin{lemma} \label{lem:mopt_dev}
Let $b=0$ and let $\mo$ and $\hm$ be given by \fr{mvalue} and \fr{mopt}, respectively.
Denote 
\be \label{Clam1lam2}
C_{\lam 0} = 4 \sqrt{2 (U_\ph - L_\ph+1)},\ \ 
C_{\lam 1} = C_{\lam 0} (\sqrt{2}\, C_{g2})^{-1} \, \|(\bAs)^{-1}\|, 
\quad C_{\lam 2} = C_{\lam 0} \,  \|(\bAs)^{-1} \bBs \|.
\ee
Let  $C_d$ be given by   \fr{Cdvalue}  and 
\be \label{Clambda}
C_\lam = \max \lkr 2 C_u, C_\tau C_{\lam 0} \rkr,
\ee
where 
\beqn
 C_u & = &  \max \lkr C_{\lam 1} C_\kappa,  C_{\lam 2} C_\tau   \rkr,
 \label{Cu}
\\
C_\tau & = &  8 C_\ph C_{g1}^{-1} \max \lkr 2, 2 \|f\|_\infty^2,  \|f\|_\infty  \|\ph \|_\infty/3,
\| \ph \|_\infty  \rkr, 
 \label{Ctauph}
\\
C_\kappa & = & \min_{a>0} \max \lkr 16 C_\ph C_{g2} \|f\|_\infty, 16 a, \frac{8 \|f\|_\infty \|\ph\|_\infty}{3},
16 C_\ph C_{g2}, \frac{4 C_\ph C_{g2} \|\ph\|_\infty}{a^2}, \frac{4 \|\ph\|_\infty^2}{3a} \rkr,\ \ \ \ \ 
 \label{Ckappa}
\eeqn  
$C_{g1}$ and $C_{g2}$ are defined by \fr{gineq} and   $C_\ph = [ 2 \max ( |L_\ph|, |U_\ph| ) ]^\af$.
If 
$ \lam  \geq   \max \lkr  C_{\lam 1}, C_{\lam 2} \rkr$,
then,  as $n \rightarrow \infty$,     
 \be  \label{large_devm} 
\PP \lkr \hm > \mo \rkr = O \lkr n^{- \frac{\lam}{C_\lam} } \rkr + O \lkr n^{ \frac{1}{\alpha+1} - \frac{d}{2 C_d}   } \rkr.
\ee
\end{lemma}


Lemma \ref{lem:mopt_dev} confirms that indeed $m= \hm$ can be chosen as the lowest resolution level in the 
nonlinear part of the estimator, so that we estimate $f$ by
\be  \label{nonlin}
\hf  (x) =   \hfohm (x)  + \hfchm (x) , \quad x \in [0,1],
\ee
where $\hfom (x)$  and $\hfcm (x)$ are defined in \fr{hfom}  and \fr{hfcmfixed}, respectively. 

\medskip

The following statement confirms that, when $1/g$ is non-integrable,  the adaptive wavelet thresholding estimator $\hf$ defined by \fr{nonlin}
attains (up to a logarithmic factor if $b=0$) 
the asymptotic minimax  lower bounds for the $L^2$-risk obtained in Theorem \ref{th:lower}.


\begin{theorem}  \label{cor:upper_bound}
Let  $1 \leq p,q  \leq \infty $  and $\max(1/2,  1/p) \leq s < r$, and let Assumption A  
(with   $\af \geq 1$ if $b=0$ and $\af \in \RR$ if $b>0$) hold.  Let $\hf$ be the estimator
defined by \fr{nonlin} with $\lam > \max  \lkr 2 C_\lam, C_{\lam 1}, C_{\lam 2} \rkr$ in \fr{mopt} and 
$d > 2 (\alpha +1)^{-1} (2 \alpha +3)  C_d $,    where $C_\lam$ is defined in \fr{Clambda}, 
$C_{\lam 1}, C_{\lam 2}$ are defined in \fr{Clam1lam2},  and  $C_d$  
is defined in  \fr{Cdvalue}.  Then, as $n \rightarrow \infty$, 
$$
 \sup_{f \in B_{p,q}^s (A)} \EE  \| \hf   - f \|^2  \leq \lfi
\begin{array}{ll} 
C\,    n^{-\frac{2s}{2s  + 1}}\ (\ln n)^{ \frac{2s (1 +\II(\af =1))}{2s + 1}} & {\rm if}\;\;\; b=0,\ \af s < s',\\ 
C\,  n^{-\frac{2s'}{2s' + \af}}\ (\ln n)^{ \frac{2s'(1 + \II(\af =1))}{2s' + \af}  + \II \lkr \frac{s'}{s} =\af >1 \rkr } 
& {\rm if}\;\;\; b=0,\ \af s \geq s',\\
C\,  (\ln n)^{-\frac{2s'}{\beta}}  &  {\rm if}\,\;\; b>0.
\end{array} \right. 
$$
\end{theorem}

\medskip


 \begin{remark}
\label{adapt} 
{\em   {\bf (Adaptivity)  }  Theorems \ref{th:lower} and   \ref{cor:upper_bound} demonstrate that, for severe data losses ($b>0$),  
the adaptive wavelet thresholding estimator $\hf$ given by \fr{nonlin} 
attains the asymptotically optimal (in the minimax sense) global convergence rates. 
 For moderate data losses ($b=0$  with $\af \geq 1$), however, the adaptive wavelet thresholding estimator $\hf$ given by \fr{nonlin}  is  
asymptotically near-optimal (up to a logarithmic factor). Moreover, if $p$ is large and
$\af >1$ is relatively small ($1 < \af < (1/2 - 1/p)/s$), then data loss does not affect the asymptotic minimax  global convergence 
rates and they coincide with the asymptotic minimax global convergence rates obtained in the absence of data losses. 
}
\end{remark}


\begin{remark}
\label{Gaiffas_and_us} 
{\em {\bf (Relation to local and uniform convergence rates). }
The suggested estimation of the zero-affected part of $f$ is 
somewhat similar to the procedure of Ga\"{i}ffas (2005, 2007), with the difference that he used 
local polynomials while we are using wavelets. However, the significant difference 
is that  we use this estimator only for the zero-affected part and not for the whole function $f$.
Another difference between our and Ga\"{i}ffas' studies   is that, first,   we are able to formulate the 
asymptotic minimax  convergence rates  explicitly, in a simple meaningful way, and, due to the fact that we are using thresholding of wavelet 
coefficients rather than solution of the system of equations as in Ga\"{i}ffas, our estimator can adapt to the case 
when the estimated function is spatially inhomogeneous. Moreover, we should point out that our asymptotic minimax convergence rates 
are global and over a wide range of Besov balls compare to Gaiffas that are local or uniform and only for H\"older spaces. 
In particular,  Ga\"{i}ffas (2005, 2007) deals  only with estimation of $f$ at $x_0$, the zero of the design density function $g$.
The asymptotic minimax local convergence rates of his estimator can be expressed explicitly via 
$\alpha$ and the parameters of the H\"{o}lder ball that $f$ belongs. However, as we pointed out in Remark \ref{local-global}, 
this problem is much easier than the global estimation problem we considered and can be solved by straightforward calculus. 
Furthermore, Ga\"{i}ffas (2006, 2009) studied asymptotic minimax uniform convergence rates.
The derived convergence rates  are formulated in terms of
a solution of a nonlinear equation, and there are no explicit expressions for these rates in a general situation.
For instance,   the only example, which appears in Ga\"{i}ffas (2009),
is produced  for the  simplest situation when $\sigma=1$, $f$ belongs to a 
H\"{o}lder class with parameters $s=L=1$ and the design density $g$ is of the form $g(x) =4|x-1/2|$,
i.e., $\af =1$.  In this case, the  asymptotic minimax uniform convergence rates  are given by
$$
r_n(x) = (\log n/n)^{\alpha_n(x)},
$$
where
$$
\alpha_n(x) = \lfi \begin{array}{ll}
\frac{1	}{3} \lkr 1 - \frac{1 - 2x)}{\log(\log n/n)}\rkr, &   x \in [0, 0.5 - (\log n/2n)^{1/4}),\\
\frac{\lkr [ (x-0.5)^4 + 4  \log n/n]^{1/2} - (x-0.5)^2 \rkr - \log 2}{2 \log (\log n/n) }, &
x \in [0.5 - (\log n/2n)^{1/4}, 0.5 - (\log n/2n)^{1/4}],\\
\frac{1	}{3} \lkr 1 - \frac{2x - 1)}{\log(\log n/n)}\rkr, &   x \in (0.5 + (\log n/2n)^{1/4}, 1]\\
\end{array} \right.
$$
On the other hand, in the case when $\alpha>1$ or in the case when $\alpha$ is not an integer, a solution of the corresponding equation which produces the asymptotic minimax uniform convergence rates, as well as 
derivation of the explicit expression for these rates, require  very nontrivial investigation. 
}
\end{remark}

 
\section {Adaptive estimation and the minimax upper bounds for the $L^2$-risk  when $1/g$ is   integrable }
\label{sec:small_alpha}
\setcounter{equation}{0}

The case when $1/g$ is integrable, i.e., when $g$ has a zero of a polynomial order $\alpha$, $0 < \alpha <1$, has been considered by Chesneau (2007)
who demonstrated that the problem is well-posed when   $f$ is spatially homogeneous, i.e., when $p \geq 2$. However, the  lower bounds 
in Theorem \ref{th:lower} show that the problem becomes ill-posed when $\af s > s'$, i.e., when $1 \leq p < (s (1-\af) +1/2)^{-1}$.
Hence, by considering only spatially homogeneous regression functions ($p \geq 2$), Chesneau (2007) missed 
the ``elbow rate" when $f$ is spatially inhomogeneous and the fact that the problem becomes ill-posed in this case.
However, since the estimators \fr{hamk} (of the scaling and wavelet coefficients) have finite variances for $0 < \af < 1$, one can construct an adaptive estimator, similar to the one considered in
Chesneau (2007), by simply  thresholding wavelet coefficients. In particular, set
\be \label{fest_small_af}
\hf (x)   = \sum_{k=0}^{2^{m_1}-1}  \hat{a}_{m_1 k} \ph_{m_1 k} +     \sum_{j=m_1}^{J-1} \sum_{k=0}^{2^j-1}   \hbjk  \psijk (x),
\ee
where $\hamk$ and  $\hbjk$ are defined in \fr{hbjk},  and $m_1$ and $J$ are defined in \fr{Jvalueo} with $b=0$.

\medskip
The following statement confirms that estimator \fr{fest_small_af} attains (up to a logarithmic factor) 
the asymptotic minimax  lower bounds for the $L^2$-risk obtained in Theorem \ref{th:lower}.

\begin{theorem}  \label{th:frisk_af_small}
Let $1 \leq p < \infty, 1 \leq q \leq \infty$, $\max(1/2,  1/p) \leq s <r$, and let $d$ in   \fr{hbjk}
satisfy 
\be \label{d_value}
d > \frac{2 C_d (3 \af + 5)}{(1-\af)(1+\af)},
\ee
where $C_d$ is given by \fr{Cdvalue}.
Let Assumption A (with $b=0$ and $0 < \af < 1$) hold, and let $\hf$ be the estimator
defined by \fr{fest_small_af}. Then, as $n \rightarrow \infty$, 
 \be \label{upper_f_af_small} 
 \sup_{f \in B_{p,q}^s (A)} \EE  \| \hf  - f \|^2  \leq\lfi
\begin{array}{ll} 
C\,  n^{-\frac{2s}{2s  + 1}}\ (\ln n)^{ \frac{2s  }{2s + 1}}  & {\rm if}\;\;\;   \af s < s',\\ 
C\, n^{-\frac{2s'}{2s' + \af}}\ (\ln n)^{ \frac{2s'}{2s' + \af} + \II(\af s = s') }  & {\rm if}\;\;\;   \af s \geq s'. 
\end{array} \right.
\ee
\end{theorem}

\medskip

Theorem \ref{th:frisk_af_small} shows that, for $b=0$ and $0 <\af<1$,  the aforementioned estimation problem is well-posed as long as 
$p >  (s (1-\af) +1/2)^{-1}$ and it becomes ill-posed when $p < (s (1-\af) +1/2)^{-1}$.
Therefore, even when data loss is very moderate ($b=0$ and $0 < \af < 1$), the estimation problem becomes ill-posed 
whenever $f$ is rather spatially inhomogeneous  ($p < (s (1-\af) +1/2)^{-1}$).

\begin{remark} \label{counter-intuitive}
{\rm  {\bf  (Integrable and non-integrable design density).  }
For $b = 0$, the asymptotic minimax global convergence rates in Theorems 2 and 3 are the same,
except for  $\alpha  = 1$.  The reason for this lies in the fact that these rates are not driven by the fact whether 
$1/g$ is integrable ($b=0$ and $0<\alpha < 1$) or non-integrable ($b=0$ and $\alpha \geq 1$) but by the relation between $\alpha s$ and $s'$. 
This   clearly follows  from  Theorem \ref{th:lower} which establishes the asymptotic minimax lower bounds for the $L^2$-risk.
}
\end{remark}

\section{Discussion}
\label{discussion}
\setcounter{equation}{0}

 We considered the nonparametric regression estimation problem of recovering an unknown 
response function $f$ on the unit interval $[0,1]$ on the basis of incomplete 
data when the design density function $g$ is known and has a zero $x_0 \in [0,1]$  
of a polynomial or an exponential order. We investigated the asymptotic (as the sample size increases) global estimation (in the minimax sense and for an $L^2$-risk) of $f$ over a wide range of Besov balls $B_{p,q}^{s}(A)$ of radius $A>0$, where $1 \leq p, q \leq \infty$  and $\max(1/p, 1/2) \leq s <r$, where $r>0$ is the regularity parameter associated with the wavelet system.
The aforementioned global nonparametric regression estimation problem is a much harder 
problem than the local nonparametric regression estimation problem studied by Gaiffas (2005, 2007),  since it cannot 
be reduced to the estimation of a related regularly-sampled function (see Remarks \ref{rem-ill-p}  and  \ref{local-global}).  
As a spatially inhomogeneous ill-posed problem, the resulting 
estimators demonstrate completely different patterns of behavior
in comparison with spatially homogeneous ill-posed problems like, e.g.,  deconvolution.

We studied various regimes of data loss, ranging from relatively minor data losses (when $g$ has 
a zero of polynomial order $0<\af <1$, so that $1/g$ is integrable) to moderate data losses 
(when $g$ has a zero of polynomial order $\af \geq 1$ and, hence,  $1/g$ is non-integrable) 
and, last, to severe data losses (when $g$ has a zero of exponential order, so that   $1/g$ is non-integrable).

Asymptotic minimax global convergence rates in the case of minor data losses  ($0<\af <1$) were studied by  Chesneau  (2007) who showed that  
the problem is well-posed (the asymptotic minimax global convergence rates are the same as in the absence of data loss) whenever 
the regression function $f$ is spatially homogeneous ($p \geq 2$). As our study shows, the problem remains well-posed even if $f$ is
spatially inhomogeneous as long as the data loss is very minor ($0<\af < 1 - (1/p - 1/2)/s$) or the function is relatively 
smooth ($p > (1/2 - s(\af -1))^{-1}$). When $\af \geq 1 - (1/p - 1/2)/s$ ($p \leq (1/2 - s(\af -1))^{-1}$),
the problem becomes ill-posed.

Now, consider the situation when data loss is moderate ($b=0$ and the zero of $g$ is of a polynomial order $\af \geq 1$).
The problem is now ill-posed if  $\af \geq (1/2 - 1/p)/s$, i.e., it is always ill-posed when $f$ is spatially inhomogeneous 
($1 \leq p < 2$). However, as Remark \ref{conv_rates} points out, when $f$ is very spatially homogeneous ($p$ is rather large)
and data loss is relatively moderate ($1 < \af < (1/2 - 1/p)/s$), the estimation problem of $f$  ceases to be ill-posed and exhibits asymptotic minimax global convergence rates observed when $g$ is bounded from below. Thus, in the case when $f$ is very spatially homogeneous, the estimator of $f$ is ``borrowing strength'' in the areas where $f$ is adequately sampled and exhibits asymptotic minimax global convergence rates common for regularly spaced regression estimation problems.
This is very dissimilar to spatially homogeneous ill-posed problems (e.g., deconvolution) where
there is a change point in the asymptotic minimax global convergence rates (the, so-called, elbow effect) when $f$ is spatially inhomogeneous ($1 \leq p <2$) and they are independent of $p$ when it is spatially homogeneous ($2 \leq p \leq \infty$). On the contrary, in the case of spatially inhomogeneous ill-posed problems, 
like the one considered herein, the asymptotic minimax global convergence rates depend on $p$ even when the function is spatially homogeneous ($2 < p \leq \infty$) as long as $\af \geq (1/2 - 1/p)/s$.   Thus, the elbow effect occurs when $p >2$,
in particular, when $p > 2/(1 - (\af -1)s)$ provided that $1 < \af < 1 + 1/s$.

 In the case when data loss is severe
($b>0$ and the zero of $g$ is of an exponential order $\beta>0$),
the asymptotic minimax global convergence rates grow with $p$, i.e., the more spatially homogeneous  $f$  is, the better it can be estimated.
This is unlike spatially homogeneous ill-posed problems (e.g., deconvolution) where the minimax global convergence rates improve  when $p$ is growing when $1<p <2$ and 
are independent of $p$ when $f$ is spatially homogeneous ($2 \leq p \leq \infty$) (see Pensky and Sapatinas (2009, 2010).

 The unusual behavior of the asymptotic minimax global convergence rates in the case of the  spatially inhomogeneous ill-posed problem considered above
calls for different adaptive estimation strategies. In particular,  whenever data loss is moderate or severe, we partition $f$ 
into zero-affected and zero-free parts. First, we construct a linear wavelet estimator of the zero-affected 
part where the lowest resolution level $m=\hm$ is independent
of the unknown parameters of the Besov ball that $f$ is assumed to belong and, therefore, known when $b>0$, and is chosen using Lepski's method 
when  $b=0$. After that, we construct a nonlinear (thresholding) wavelet estimator of the zero-free part of $f$ starting from the lowest resolution level $m=\hm$. Note that the nonlinear estimator is required even if $g$ has a zero of exponential order ($b>0$). 
This is very different from the case of spatially homogeneous ill-posed problems (e.g., deconvolution),
where in the case of exponentially growing eigenvalues, a linear estimator usually attains asymptotically optimal 
(in the minimax sense) global convergence rates (see Pensky and Sapatinas (2009, 2010).)

We should also mention that there is a significant difference between asymptotic minimax local and asymptotic minimax global convergence rates.
Note that asymptotic minimax local convergence rates at a zero of $g$ are always affected by loss of data, even for moderate data losses.
The asymptotic minimax global convergence rates, however,  are not affected when data loss is limited and the regression function is 
very spatially homogeneous ($1 < \af < 1 + 1/s$ and $p > 2/(1 - (\af -1)s)$.  

Finally, we point out that some of the logarithmic factors which appear in Theorems 1-3 could be possibly removed by using 
block thresholding rather than the considered term-by-term thresholding of wavelet coefficients. 
Furthermore, due to its construction, the suggested adaptive wavelet thresholding estimator is not easily computable, so it is of limited practical use. Therefore, it is desirable to construct an alternative, more computational feasible, adaptive estimator which attains the 
asymptotic minimax global convergence rates, that was the aim of this work. This is the  project  for future work 
that we hope to address elsewhere.

\section*{Acknowledgements}

Marianna Pensky was supported in part by National Science Foundation
(NSF), grant  DMS-1106564. The authors want to thank Yuri Golubev 
for valuable discussion of results and methodologies used in this paper. 
We would like also to thank the referees for constructive comments that led to improvements.


\section {Proofs}
\label{append}
\setcounter{equation}{0}

Since the paper contains a large number of statements, below we give a road map of Section \ref{append}.
Section \ref{pr:lower_bounds} contains the proof of the asymptotic minimax lower  bounds for the $L^2$-risk.
Theorem \ref{cor:upper_bound}, which provides the asymptotic minimax upper bounds for the $L^2$-risk of an adaptive estimator of $f$
 in the case of $\alpha \geq 1$ if $b=0$ and $\alpha \in \RR$ if $b>0$, is proved in Section \ref{sec:minimax_proof}. The proof of  Theorem \ref{cor:upper_bound} is based on 
Lemmas \ref{lem:error_sys_sol}-\ref{lem:mopt_dev}.
In particular,  Lemma \ref{lem:error_sys_sol}, which is proved in Section  \ref{proofs_f0}, gives an asymptotic minimax upper bound for the $L^2$-risk 
of the zero-affected portion of the estimator at a fixed resolution level $m$. 
Lemmas  \ref{lem:exponential_zero}  and  \ref{lem:polynomial_zero} provide asymptotic minimax upper bounds for the $L^2$-risk 
of the zero-free part of the estimator when estimation  is carried out 
(in the case of an exponential zero)  or is started (in the case of a polynomial zero) at a fixed
resolution level chosen  in advance. Last, Lemma \ref{lem:mopt_dev} proves that,  with high probability,  the resolution level chosen 
by Lepski's procedure is not higher than the optimal resolution level. The proof of Lemma \ref{lem:error_sys_sol} is included 
in Section \ref{proofs_f0}  while the proofs of Lemmas  \ref{lem:exponential_zero} and \ref{lem:polynomial_zero} are given in
Section \ref{sec:zero-free}. Section \ref{sec:large_dev} contains the proof of Lemma \ref{lem:mopt_dev} 
as well as large deviation results for a wavelet or scaling coefficient (Lemma \ref{lem:large_dev1})
or the right-hand side of the system of linear equations (Lemma \ref{lem:large_dev2}).

Sections  \ref{pr:wavelet_coefs} and \ref{proof_linear_service} contain supplementary statements 
which are used in the proofs of Lemmas \ref{lem:error_sys_sol}-\ref{lem:polynomial_zero}.
In particular,  Lemma \ref{lem:coef_moments}, proved in Section \ref{pr:wavelet_coefs},
 provides upper bounds for moments and covariances of wavelet and scaling coefficients.  
In Section \ref{proof_linear_service},  Lemma  \ref{auxiliary} contains a purely technical auxiliary result, while 
Lemma  \ref{lem:sys_matr}  provides upper and lower bounds for the entries of the matrices 
which appear in the system of linear equations which is used for the construction of the zero-affected part
of the estimator.

Finally, Theorem \ref{th:frisk_af_small}, which delivers the asymptotic minimax upper bounds for the $L^2$-risk of an adaptive estimator of $f$
 in the case of $b=0$ and $0<\alpha < 1$, is proved in Section \ref{sec:proofs_small_alpha}. The proof of this theorem requires 
a technical result provided by Lemma \ref{lem:moments} which precedes  Theorem \ref{th:frisk_af_small} in 
 Section \ref{sec:proofs_small_alpha}.


\subsection{Proof of the asymptotic minimax lower bounds for the $L^2$-risk}
\label{pr:lower_bounds}
 
{\bf Proof of Theorem \ref{th:lower}}.  
On noting that the asymptotic minimax lower bounds for the $L^2$-risk  in Theorem 3.1 of Chesneau  (2007a) 
is also true when $b=0$ and $1/g$ is non-integrable ($\af \geq 1$), 
the asymptotic minimax lower bounds for the $L^2$-risk  in the first part of \fr{low1} can be obtain by 
the arguments of Chesneau  (2007a) and, hence,  we need to prove only the 
asymptotic minimax lower bounds for the $L^2$-risk in the second and third parts of  \fr{low1}.  
For this purpose, we consider   functions $f_{jk}$ be of the form $\fjk= \gaj \psijk$ and let
$f_0 \equiv 0$. Note that by \fr{bpqs}, in order $\fjk \in \Bpqsa$,
we need $\gaj \leq A 2^{-js'}$. Set $\gaj = c 2^{-j s'}$, where $c$
is a positive constant such that $c<A$, and apply the following
classical lemma on lower bounds:

\begin{lemma} \label{korost}
{\bf (H\"{a}rdle, Kerkyacharian, Picard \& Tsybakov (1998), Lemma
10.1).\ } Let $V$ be a functional space, and let $d(\cdot, \cdot)$
be a distance on $V$. For $f, g \in V$, denote by $\varLambda_n
(f,g)$ the likelihood ratio $\varLambda_n (f,g) = d\PP_{X_n^{(f)}}/
d\PP_{X_n^{(g)}}$, where $d\PP_{X_n^{(h)}}$ is the probability
distribution of the process $X_n$ when $h$ is true. Let $V$ contains
the functions $f_0, f_1, \ldots, f_\aleph$ such that
\begin{itemize}
\item[$(a)$] $d(f_k, f_{k'}) \geq \delta >0$ for $k=0,1,\ldots,\aleph$,\; $k \neq k'$,

\item[$(b)$] $\aleph \geq \exp(\lam_n)$ for some $\lam_n >0$,

\item[$(c)$] $\ln \varLambda_n (f_0, f_k) = u_{nk} - v_{nk}$, where 
$v_{nk}$ are constants and $u_{nk}$ is a
random variable such that there exists $\pi_0>0$ with
$\PP _{f_k}(u_{nk} >0) \geq \pi_0$,

\item[$(d)$] $\sup_k v_{nk} \leq \lam_n$.\\
\end{itemize}
\vspace{-0.5cm} Then, for an arbitrary estimator $\hf$,
$$
\sup_{f \in V} \PP_{X_n^{(f)}} \big(d(\hf, f) \geq \delta/2\big)
\geq \pi_0/2.
$$
\end{lemma}

Let now $V=\lfi \fjk: |k-k_{0j}| \leq K/2 \rfi$, 
where $K>2$ is a fixed positive constant, 
so that $\aleph = K$. Choose $d(f,g) = \| f-g\|$, where, as before, $\| \cdot \|$ denotes the
$L^2$-norm on the   interval $[0,1]$. Then, $d(\fjk, f_0) = \gaj = \delta$. 
Let $v_{nk} = \lam_n =  \ln K$ and 
$u_{nk} = \ln \varLambda_n (f_0, \fjk) + \ln K$.
 Now, in order to apply Lemma \ref{korost}, we need to show that for some $\pi_0 >0$, uniformly
for all $\fjk$, we have
$$
\PP_{\fjk} (u_{nk} > 0) =  \PP_{\fjk} \lkr \ln \varLambda_n (f_0, \fjk) > - \ln K \rkr \geq \pi_0 >0.
$$ 
Since, by Chebychev's inequality,
$$
\PP_{\fjk} \lkr \ln \varLambda_n (f_0, \fjk) > - \ln K \rkr \geq 1 - 
\frac{\EE_{\fjk} \big|\ln \Lam_n (f_0, \fjk)\big|}{\ln K},
$$ 
we need to find a uniform upper bound for
$\EE_{\fjk} |\ln \varLambda_n (f_0, \fjk)|$.

Note that  
$$
- 2 \ln \varLambda_n (f_0, \fjk)   =    
\sum_{i=1}^n  \ga_j^2 \psijk^2 (x_i) + 2 \sum_{i=1}^n  \ga_j \psijk (x_i) \xi_i 
$$
where $\xi_i $, $i=1,2,\ldots,n$, are independent standard Gaussian random variables. Thus, 
$$
\EE |- 2 \ln \varLambda_n (f_0, \fjk) | \leq A_n + 2 B_n, 
$$
where 
$$
A_n = \EE |\sum_{i=1}^n  \ga_j^2 \psijk^2 (x_i)| = n \ga_j^2 \int_0^1 \psijk^2 (x) g(x) dx, \quad
B_n = \EE | \sum_{i=1}^n  \ga_j \psijk (x_i) \xi_i|.
$$ 
Note that by Jensen's inequality,
\beqns
B_n & = & \EE  \lfi  \EE  \lkv \Big| \sum_{i=1}^n  \ga_j \psijk (x_i) \xi_i \Big|\  \Bigg| x_1,x_2, \ldots, x_n \rkv \rfi \\
& \leq & \EE \lfi \EE  \lkv \lkr \sum_{i=1}^n  \ga_j \psijk (x_i) \xi_i\rkr^2 \Bigg| x_1,x_2, \ldots, x_n \rkv \rfi^{1/2}
= \sqrt{A_n},  
\eeqns
so that one needs uniform upper bounds for $A_n$ only. 

If $j$ is large enough, $A_n$ can be presented as 
$$
A_n = n \ga_j^2 \int_{L_\psi}^{U_\psi} \psi^2 (z) g(x_0 + 2^{-j}(z+k-k_{0j}))  dz,
$$
where $k_{0j}=2^j x_0$. 
Observe that condition \fr{zeros} implies that one has  
$$
g(x_0 + x) \leq C |x|^{\alpha} \exp( -b|x|^{-\beta}).
$$ 
Let $M_\psi = \max(|L_\psi|, |U_\psi|)$.  Then, for a finite value of $K$,
one has 
$$ 
A_n \leq C n \ga_j^2 2^{-j \af}\ (M_\psi^\af + K^\af) \exp \lkr  -b 2^{j \beta} (M_\psi  + K)^{-\beta} \rkr. 
$$
Now,  recall that  $\gaj = c 2^{-j s'}$ and choose the smallest possible value of $j$ such that $A_n$ are uniformly bounded.
Simple calculation yield that 
$$
A_n = O \lkr 2^{-j(2s' + \af)} \exp \bigg(-b 2^{j \beta} [M_\psi  + K]^{-\beta} \bigg) \rkr,
$$ 
so that
$2^j = O \lkr n^{1/(2s' +\af)} \rkr$ if $b=0$ and $2^j = O \lkr (\ln n)^{1/\beta} \rkr$ if $b>0$.

Now, applying Lemma \ref{korost} and Chebyshev inequality, we finally obtain 
$$
\inf_{\tilde{f}_n} \, \sup_{f \in B_{p,q}^s (A)} \EE  \|  \tilde{f}_n - f \|^2 \geq
\inf_{\tilde{f}_n} \, \sup_{f \in V} (\gaj^2/4)\ \PP (\| \tilde{f}_n - f \| > \gaj/2) \geq \pi_0 \gaj^2/8,
$$
which, on noting that 
\be \label{main_rel}
\frac{2s'}{2s'+\af} < \frac{2s}{2s+1} \quad \text{if and only if}  \quad     s' < \af s,
\ee
completes the proof of the theorem.


\subsection{Properties of the estimators of scaling and wavelet coefficients}
\label{pr:wavelet_coefs}

Consider the quantity
\be  \label{Jkml}
\Jmkl = \int  2^m |\ph(2^m x - k) \ph(2^m x - l)|\, g^{-1} (x)\, dx.
\ee

\begin{lemma} \label{lem:coef_moments}
Let $m=m(n)$ be a non-random, non-negative integer, quantity, with $m(n) \rightarrow \infty$ as $n \rightarrow \infty$,  and let $\hamk$ be defined by \fr{hamk}.
Then, for $k,l \in \Komhc$, as $ n \rightarrow \infty$, 
\be  \label{covar_a}
|\Cov (\hamk,\haml)| = O \lkr n^{-1}\ (\Jmkl + 1) \rkr, 
\ee
where
\be \label{upperJkml}
\Jmkl = O \lkr n^{-1}\, 2^{m \af} |k - \kom|^{-\af}  \exp(b 2^{m \beta} |k - \kom|^{-\beta}) \rkr\ \ \mbox{if}\  \ 
|k-l| \leq \Uh-\Lh, 
\ee
and $\Jmkl=0$ otherwise. 
Moreover, if $b=0$, then, as $ n \rightarrow \infty$, 
\beqn 
\Var (\hamk) & = & O \lkr n^{-1}\,  2^{m  \af} |k - \kom|^{-\af} \rkr,  \nonumber \\
\EE  (\hamk - \amk)^4 & = & O \lkr n^{-3}\,  2^{m (3\af+1)} |k - \kom|^{-3\af} \rkr 
+ O \lkr n^{-2}\,  2^{2 m \af} |k - \kom|^{-2\af} \rkr. \label{4thmoment}
\eeqn 
Similarly, if $k,l \in \Kojsc$ and $b=0$, then $\tbjk$,    defined in \fr{hamk}, satisfy, as $n \rightarrow \infty$,
\beqn 
\Var (\tbjk) & = & O \lkr n^{-1}\,  2^{j  \af}\ |k - \koj|^{-\af} \rkr,  \label{lemmom1}\\
\EE  (\tbjk - \bjk)^4 & = & O \lkr n^{-3}\,  2^{j (3\af+1)} |k - \koj|^{-3\af} \rkr
+ O \lkr n^{-2}\,  2^{2 j \af} |k - \koj|^{-2\af} \rkr, \label{lemmom2}\\
\EE  (\tbjk - \bjk)^6 & = & O \lkr n^{-5}\,  2^{j (5\af+2)} |k - \koj|^{-5\af} \rkr 
+ O \lkr n^{-4}\,  2^{j (4\af+1)} |k - \koj|^{-4\af} \rkr \label{lemmom3}\\
& & + \, \, O \lkr n^{-3}\,  2^{3 j \af} |k - \koj|^{-3\af} \rkr. \nonumber
\eeqn

\end{lemma}

\noindent
{\bf Proof of Lemma  \ref{lem:coef_moments}}.  
Let us first prove formula \fr{upperJkml}. 
Changing variables $z = 2^m (x - x_0)$ in the   integral in \fr{Jkml}, and using
inequality  \fr{gineq}, derive that 
$$
\Jmkl  \leq \frac{2^{m \af}}{C_{g1}}\ 
\int_\Lh^\Uh \frac{|\ph (z)| |\ph(z + k-l)|\ dz }{|z + k - \kom|^\af \, 
\exp \lkr - b 2^{m  \beta} |z + k - \kom|^{-\beta} \rkr }.
$$
It is easy to note that $\Jmkl = 0$ if $|k-l| > \Uh-\Lh$. Also,   
$k \in \Kojsc$ implies   that $\kom - k \leq \Lh -1$  or  $\kom - k \geq \Uh+1$,
so that one has  $|z + k - \kom| \geq 1$ and, hence, 
$ |z + k - \kom| \propto |k - \kom|$ which proves \fr{upperJkml}. 
Now, by direct calculations we obtain that
$$
\Cov(\hamk,\haml) = n^{-1} \lfi \int  [\sigma^2 + f^2(x)]\ \phimk (x) \phiml(x) g^{-1} (x) dx  
-  \amk a_{ml} \rfi,
$$
so that \fr{covar_a} is valid.

Since the proofs for the scaling and the wavelet coefficients in Lemma \ref{lem:coef_moments}
are similar, we shall prove only formulae \fr{lemmom1}-\fr{lemmom3}. 
Observe that, due to \fr{gineq} and the fact that $k  \in \Kojsc$ implies
$ \koj - k \leq \Ls -1$ or $\koj - k \geq \Us+1$,  by considerations similar to
the ones provided above, for integers $r_1, r_2 >0$, one has 
\be  \label{serv_ineq}
\int  (g(x))^{-r_2}\, (\psijk (x))^{2 r_1}\,   dx \leq  
C\, 2^{j (r_1 -1)}\  2^{j r_2 \af}\ |k - \koj|^{- r_2 \af}.
\ee
Now, to complete the proof of \fr{lemmom1}--\fr{lemmom3},  as $n \rightarrow \infty$,
apply \fr{serv_ineq} to the following formulae
\beqns 
\Var(\tbjk) & = & O \lkr n^{-1}\ \int   g^{-1}(x)   \psijk^2 (x)   dx \rkr, \\
\EE  (\tbjk - \bjk)^4 & = &   O \lkr n^{-3}\ \int  g^{-3}(x)   \psijk^4 (x)  dx 
+    n^{-2} \lkv \int  g^{-1}(x)  \psijk^2 (x)  dx \rkv^2 \rkr, \\
\EE  (\tbjk - \bjk)^6 & = & O \lkr n^{-5}\ \int  g^{-5}(x)  \psijk^6 (x)  dx
+  n^{-3} \lkv \int  g^{-1}(x)  \psijk^2 (x)  dx \rkv^3 \right. \\  
& & + \,\, \left. n^{-4}\ \int  g^{-3}(x)  \psijk^4 (x)  dx\  \int  g^{-1}(x)  \psijk^2 (x)  dx \ \rkr.
\eeqns


\subsection{Proofs of the supplementary statements used in the proof of  Lemma \ref{lem:error_sys_sol}. }
\label{proof_linear_service}


\begin{lemma} \label{auxiliary} 
Let $\dd_0 = 0.5\ \, 3^{\beta+1}\ (2\ 3^{\beta +1} + (2M)^{\beta + 1} )^{-1}$, where $\beta >0$ and $M>0$, and
let  $0 < \dd < \dd_0$ and  $a, b \in [2 - \dd, M]$.
 Let $c>0$ be such that $c \leq \min(a,b) + \dd$ and  $c \leq \max(a,b) -(1 - 2 \dd)$. 
Then, 
$$
a^{-\beta} + b^{-\beta} - 2 c^{-\beta} \leq - 0.5\, \beta M^{-(\beta+1)}.
$$
\end{lemma}

\noindent
{\bf Proof of Lemma  \ref{auxiliary}. }
Note that $\dd_0 <1/2$ and that $\dd < \dd_0$ implies $\dd  <1/2$.
Let, without loss of generality, $a \leq b$. Then, $c \leq a+ \dd$, 
$c \leq b -(1-2  \dd)$ and 
\beqns
a^{-\beta} - c^{-\beta} & \leq &  a^{-(\beta+1)}  \beta \dd \leq (2-\dd)^{-(\beta+1)}  \beta \dd \\
b^{-\beta} - c^{-\beta} & \leq & b^{-\beta} - (b -1+ 2\dd)^{-\beta} \leq - \beta M^{-(\beta+1)}\, (1 - 2\dd).
\eeqns
Therefore, taking into account that $2-\dd > 3/2$ and $0<\dd < \dd_0$, we obtain
\begin{align*}
a^{-\beta} & + b^{-\beta} - 2c^{-\beta} \leq (2-\dd)^{-(\beta+1)}  \beta \dd - \beta M^{-(\beta+1)}\, (1 - 2\dd) \\
& < - \beta M^{-(\beta+1)} (3/2)^{-(\beta+1)} [ (3/2)^{\beta+1} - \dd_0 (M^{\beta+1} + 2 (3/2)^{\beta+1})]
= - 0.5 \beta M^{-(\beta+1)}, 
\end{align*}
which proves  the lemma.
\\

\begin{lemma} \label{lem:sys_matr}
Let $\bAmkl$,  $\bBmkl$, $\bcml$ and $\hbcml$ be  given by 
\fr{matrA}, \fr{matrB}, \fr{vecc} and \fr{hbcl}, respectively.
Then, $\Var (\hbcmk) = O \lkr n^{-1}\, \bAmkk \rkr$,  
and, for some constants $C_1 >0$  and $C_2 >0$, one has
\be \label{Amkk}
C_1 2^{-m \af} \exp(-b 2^{\beta(m+1)})   \leq \bAmkk \leq C_2 2^{-m \af} 
\exp ( -b \, M_\ph^{-\beta}\ 2^{m \beta} ),
\ee
where $M_\ph = \Uh - \Lh + \max(|\Uh|, |\Lh|)$.
Moreover, if $b >0$ and  $0 < \delta_b < \delta_0$ for 
$ 
\delta_0 = 0.5\ \, 3^{\beta+1}\ [2\ 3^{\beta +1} + (\Uh + \Lh)^{\beta + 1} ]^{-1}, 
$ 
then 
\be \label{fraction}
\frac{|\bAmkl|}{\sqrt{\bAmkk } \sqrt{A_{ll}^{(m)}}} \leq
C \exp\lkr - 0.25\ b \, (\Uh + \Lh)^{-(\beta + 1)}   \ 2^{m\beta} \rkr.
\ee
In addition, if $b=0$ and $m_1 \leq m \leq J-1$, then, as $n \rightarrow \infty$,  
\be \label{Bc_bounnds}
 \|\bBm \|= O \lkr 2^{- m\af/2} \rkr,\ \ \ \ 
\EE (\hbcmk - \bcmk)^4 = O \lkr n^{-2} 2^{-2m\af} \rkr,
\ee
where $\|\bBm \|$ is the spectral norm of matrix $\bBm$.
\end{lemma}

\noindent
{\bf Proof of Lemma  \ref{lem:sys_matr}.}
First, note that, by \fr{matrA}, one has 
$$\Var  (\hbcmk) = n^{-1} \int \phimk^2 (x) (f^2 (x) + \sigma^2) g(x) \, \II(2^m x -l \in \Om_\delta) dx
= O \lkr n^{-1}\, \bAmkk \rkr.$$ If $b=0$, then  
\beqns
\EE (\hbcmk - \bcmk)^4 & = & O \lkr n^{-3}\, \int   \phimk^4 (x) g(x) dx + 
n^{-2} \lkv \int   \phimk^2 (x) g(x) dx \rkv^2 \rkr \\
& = &  O \lkr n^{-3}\, 2^m 2^{-m \af} + n^{-2} 2^{-2 m \af} \rkr =  O \lkr n^{-2} 2^{-2m\af} \rkr,
\eeqns
since $n^{-1} 2^{m(1 + \af)} < 1$ for $m_1 \leq m \leq J-1$, which completes the proof of the second half of \fr{Bc_bounnds}.
Now, observe that, as $n \rightarrow \infty$, 
\beqn \label{Amkl}
A^{(m)}_{kl} & = & \int \ph(z + \kom - k) \ph(z + \kom - l) g(x_0 + 2^{-m} z) \II (z + \kom - l \in  \Om_\delta) dz\\
& \sim & C_g \ 2^{-m \af} \int \ph(z + \kom - k) \ph(z + \kom - l) |z|^\af \exp(- b 2{m\beta} |z|^{- \beta})\,  dz,
 \ \ \ k,l \in  \Komh,
\eeqn
and $B^{(m)}_{kl}$ has a similar expression, just with $k \in \Koms$ and $l \in \Komh$,
where $\Koms$ is defined in \fr{Komstar}.
Recalling that $b=0$ and the quantities  $|k - \kom|$ and $|l - \kom|$ are uniformly bounded 
for $k \in \Komhc$ and $l \in \Komh$, obtain  (for $b=0$) that  $|B^{(m)}_{kl}|  = O(2^{-m\af})$,
so that the first statement in \fr{Bc_bounnds} is true due to the fact that matrix $\bB^{(m)}$ is finite dimensional.

Now, let $b>0$ and let us prove \fr{Amkk}. Observe that $\Lh \leq z + \kom - k \leq \Uh$ and $k \in \Komh$
imply $|z| \leq M_\ph$. Hence, the upper bound in \fr{Amkk} follows from \fr{gineq} and \fr{Amkl}.
In order to prove the lower bound in \fr{Amkk}, note that 
$$
\bAmkk \geq C_{g1} 2^{-m\af}   \int_{\Om^*_\delta} \ph^2 (z) |z - (\kom -k)|^\af 
\exp(- b 2{m\beta} |z - (\kom -k)|^{- \beta})\,  dz
$$
where  $\Om^*_\dd =  (\Lh + \dd_b, (\Lh + \Uh - 1)/2) \cup ((\Lh + \Uh + 1)/2, \Uh - \dd_b)$ 
and $\dd_b$ is defined in \fr{delta}. Since  $|z - (\kom -k)| \geq 1/2$  for $z \in \Om^*_\dd$,
and  by \fr{support}, $(\Lh + \Uh - 1)/2 - (\Lh + \dd_b) \geq 1$  and 
$(\Uh - \dd_b) - (\Lh + \Uh + 1)/2 \geq 1$, one has 
$$
\bAmkk \geq C_{g1} 2^{-\af (m+1)} \ \exp \lkr - b 2^{\beta(m+1)} \rkr \ 
\min \lkr \int_{\Lh + \dd_b}^{(\Lh + \Uh - 1)/2} \ph^2 (z) dz,\  \ 
\int_{(\Lh + \Uh + 1)/2}^{\Uh - \dd_b} \ph^2 (z) dz \rkr,
$$
which completes the proof of  \fr{Amkk}.

Finally, let us prove \fr{fraction}. Note that asymptotic value of the integral in 
\fr{Amkl} is defined by the value at a point which maximizes the argument of the exponential function.
Recall that (see, e.g.,  Dingle  (1973)) if $F(\lam) = \int_a^b h(x) \exp(\lam S(x)) dx$, where
$\max S(x)$ is achieved   at $x=a$ and $S(x)$ is a decreasing function of $x$, and if the functions $f(x)$ and $S(x)$ are continuous for $x \in [a,b]$ and 
infinitely differentiable in the neighborhood of $x=a$ with $S'(a) \neq 0$, then,
as $\lam \rightarrow \infty$,  $F(\lam)$ has the following asymptotic expression 
\be \label{gen_asymp}
F(\lam) \sim \exp(\lam S(a))\ \sum_{k=o}^\infty c_k \lam^{-(k+1)}\ \ \ 
\mbox{with} \ \ \ c_k = - D^k (h(x)/S'(x)),
\ee
where $D$ is the differential operator of the form $D= - \frac{1}{S'(x)}\, \frac{d}{dx}$.

It is easy to calculate  that $\exp(- b 2{m\beta} |z|^{- \beta})$ takes its maximum value at 
$z =   \zmax^{(l,k,\dd)} = \max (u_\dd, v_\dd)$, where 
$u_\dd^{(l,k)} = \max(\Lh + k -\kom, \Lh + \dd_b + l -\kom)$, 
$v_\dd^{(l,k)} =\min(\Uh + k -\kom, \Uh - \dd_b + l -\kom)$ and 
$\Lh \leq k-\kom, l-\kom \leq \Uh$. In what follows, we shall drop the superscripts
whenever it does not cause confusion.

 First, consider the case of $k=l$. Then, by examining the cases $\kom - l \leq (\Lh + \Uh)/2$ and 
$\kom - l > (\Lh + \Uh)/2$ separately, one can easily conclude that
\be \label{zmaxll}
\zmaxll = \lfi \begin{array}{ll}
|\Lh + \dd_b + l -\kom|, & \mbox{if}\ \ \  \kom - l > (\Lh + \Uh)/2, \\
|\Uh - \dd_b + l -\kom|, & \mbox{if}\ \ \  \kom - l \leq (\Lh + \Uh)/2,
\end{array} \right.
\ee
where,  by \fr{support}, $\zmaxll \geq (\Uh - \Lh - 2 \dd_b)/2 \geq 2-\dd_b >1$  in both cases. 
Hence, since $\ph(\zmaxll) \neq 0$ by definition of $\dd_b$,  formula \fr{gen_asymp} yields
\beqn \label{AmllAsymp}
\bAmll & \sim & C_g (b \beta)^{-1} \ph^2 (\zmaxll + \kom -l)\ |\zmaxll|^\af \  2^{-m (\af + \beta)}\,   
\exp(- b 2^{m\beta} |\zmaxll|^{- \beta}) \nonumber\\
& \geq & K_1  2^{-m (\af + \beta)}\, \exp(- b 2^{m\beta} |\zmaxll|^{- \beta}).   
\eeqn

If $k \neq l$, then  $|k-l| \geq 1$ and one has
four cases, depending on whether $\kom -k$ and $\kom -l$ are smaller or greater than $(\Lh + \Uh)/2$.
 We shall consider two of those since the other two cases are similar. In what follows, we denote 
by $\zmaxkko$ the value of $\zmaxkk$ obtained if $\dd=\dd_b =0$.

If $\kom - l \leq  (\Lh + \Uh)/2$ and $\kom - k \leq  (\Lh + \Uh)/2$
then $|\zmaxll|= \Uh - \dd_b +l - \kom$, $|\zmaxkko|  = \Uh +k - \kom$ and, since $\dd_b < 1/2$,  
$$
|\zmaxlk| = \lfi \begin{array}{ll}
\Uh - \dd_b + l -\kom, & \mbox{if} \ \ \ l>k\\
\Uh + k - \kom, & \mbox{if} \ \ \ l < k. 
\end{array} \right.
$$
Therefore, taking into account that $|\zmaxkk| = |\zmaxkko| - \dd_b$, one derives
that 
\be \label{zmax_ineq}
\max \lkr |\zmaxll|, |\zmaxkk| \rkr - |\zmaxlk| \geq 1 - 2 \dd_b. 
\ee
 Now, consider the case when $\kom - l \leq  (\Lh + \Uh)/2$ and $\kom - k >  (\Lh + \Uh)/2$.
In this situation, $|\zmaxll|= \Uh - \dd_b +l - \kom$, $|\zmaxkko|  = \kom- k - \Lh$
and $|\zmaxlk| = \max(|\Uh + k - \kom|, |\Lh + \dd_b + l - \kom|)$, so that relation 
\fr{zmax_ineq} is again true. Cases when $\kom - l >  (\Lh + \Uh)/2$  can be examined in a similar 
manner and it can be shown that \fr{zmax_ineq}  is valid. 

 The asymptotic expression for $\bAmkl$ as $m \rightarrow \infty$ can be obtained using formula  \fr{gen_asymp}
\beqn \label{AmklAsymp}
\bAmkl & \sim & C_g K(\ph,b, \beta, \zmax)\  \ 2^{-m (\af + \beta)} \, 2^{-m \beta r^*}\,  
\exp(- b 2{m\beta} |\zmax|^{- \beta}), 
\eeqn
where $K(\ph, b, \beta, \zmax)$ depends on $\ph, b, \beta$ and $\zmax$ only and, hence, uniformly bounded,
$r^* =0$ if $\zmax$ does not coincide with $\Lh$ or $\Uh$ and $r^* = r_0 +1$ if it does.   (Here, $r_0$ is 
the number of continuous derivatives of $\ph$.)

We are now ready to complete the proof of the lemma.
Recall that
$$
|\zmaxlk | \leq \min(|\zmaxll |, |\zmaxkko |) \leq \min(|\zmaxll |, |\zmaxkk  |) + \dd_b,
$$ 
and, by \fr{zmax_ineq},   that
$$
|\zmaxlk | \leq \max(|\zmaxll |, |\zmaxkk  |) - (1 - 2 \dd_b).
$$
Since $|\zmaxll |> 2 - \dd_b$ and $|\zmaxkk  |> 2 - \dd_b$,  an  application of  Lemma \ref{auxiliary}, with 
$\delta=\delta_b$, $a = |\zmaxll |$, $b =  |\zmaxkk  |$, $c = |\zmaxlk |$ and $M = (\Lh + \Uh)/2$,  completes the proof  of the lemma. 
\\

\begin{lemma} \label{lem:inv_matr}
 Let $\bA$ be the matrix with the entries  given by \fr{matrA} with $g$ satisfying Assumption A 
and let $\bD$ be  the diagonal matrix $\bD   = \sqrt{\diag(\bA)}$. Denote $\bQ = \bD^{-1} \bA \bD^{-1}$. 
Then, for any $b \geq 0$, one has $\| \bQ^{-1} \| = O(1)$ as $m \rightarrow \infty$, where $\| \cdot \|$ denotes the spectral norm.
Moreover, if $b >0$, then $\bQ^{-1} = \bI + \bH$, where 
$$
\| \bH \| = O \lkr \exp (- 0.125\ b \delta_0^2 2^{m \beta}  ) \rkr,\ \ \ 
m \rightarrow \infty,
$$
and $\delta_0$  is defined in Lemma \ref{lem:sys_matr},
i.e., $\bQ^{-1} = \bI\, (1 + o(1))$ as $m \rightarrow \infty$.
\end{lemma}

\noindent
{\bf Proof of Lemma  \ref{lem:inv_matr}.  }   
Note that matrix $\bQ$ is an $(\Uh-\Lh +1)$-dimensional positive definite matrix with 
a unit main diagonal and smaller off-diagonal entries, so that, it has a non-asymptotic bounded 
inverse $\bQ^{-1}$ with $\| \bQ^{-1} \| = O(1)$.

If $b>0$, then $Q_{kk} =1$, so that $\bQ = \bI + \bH$.  Here,  by Lemma \ref{lem:sys_matr},
$\bH$ is a finite dimensional matrix with elements
$H_{lk} =  O \lkr \exp \lfi - 0.25\ b \, (\Uh + \Lh)^{-(\beta + 1)}   \ 2^{m\beta} \rfi \rkr$,  as $m \rightarrow  \infty$.
Hence,  $\|\bH\| \leq C_H \exp \lfi - 0.25\ b \, (\Uh + \Lh)^{-(\beta + 1)}   \ 2^{m\beta} \rfi$ for some $C_H >0$, so that
 $\| \bH \| \to 0$ as $m \to \infty$.   
To complete the proof  of the lemma, it suffices to  note that 
$$
\bQ^{-1} = \bI + \sum_{k=1}^\infty (-1)^k \bH^k,  
\quad
\mbox{ where} \quad 
\left \| \sum_{k=1}^\infty (-1)^k \bH^k \right \| \leq \sum_{k=1}^\infty \|\bH\|^k =  O\lkr \| \bH \| \rkr \to 0 \ \ \ (m\to \infty). 
$$
 

\subsection{Proofs of the large deviation results   }
\label{sec:large_dev}


Denote
\be \label{ron}
\ro_n = n^{-1/2}\, \sqrt{\ln n}.
\ee
In order to prove Lemma \ref{lem:mopt_dev}, we need the following three large deviation results (Lemmas \ref{lem:large_dev1}-\ref{lem:system_sol_dev}).
(We note that the slightly unusual formulation of Lemma \ref{lem:large_dev1} is due to the fact that we 
are planning to use it with both $w= \ph$ and $w= \psi$.)
\\


\begin{lemma} \label{lem:large_dev1}
Let $b=0$.
Let $w$ be a bounded function with a compact support $[W_1,W_2]$ 
and a unit $L^2$-norm. Denote $\wjk (x) = 2^{j/2} w (2^j x - k)$ and set
$$
\betajk = \int \wjk(x) f(x) dx,\ \ \ \ 
\hbetajk = n^{-1} \sum_{l=1}^n \frac{\wjk(x_i) y_i}{g(x_i)},
$$ 
where  $f$ is the unknown  response function in model \fr{main_eq}. 
Let $C_{g1}$ be defined in \fr{gineq},  and let  
\be \label{Ctau}
C_w = \lkv 2 \max ( |W_1|, |W_2| ) \rkv]^\af,  \quad C_\tau =  8 C_w
 C_{g1}^{-1} \max \Big( 2,  2 \|f\|_\infty^2,  \|f\|_\infty  \|w\|_\infty/3,
\|w\|_\infty  \Big).
\ee
Let $m_1$ and $J$ be defined by \fr{Jvalueo}, let $\ron$  be defined by \fr{ron},  and  let
$$ 
K_{0jc}^w = \lfi k: \ 0 \leq k \leq 2^j -1,\ x_0 \notin \supp\ \wjk \rfi.
$$
Then, for $m_1 \leq j  \leq J-1$, $k \in K_{0jc}^w$ and $\tau \geq 1$,  as $n \rightarrow \infty$, 
\be  \label{large_dev1} 
\PP \lkr |\hbetajk  - \betajk | >  \tau\,  \ron\, 
2^{j\af/2} |k - k_{0j}|^{-\af/2} \rkr = O \lkr n ^{- \frac{\tau}{C_\tau} }  \rkr.
\ee
\end{lemma}

\noindent
{\bf Proof of Lemma  \ref{lem:large_dev1}}.  
The proof of the  lemma is based on ideas presented in Chesneau (2007a).
Observe that 
$$
\PP \lkr |\hbetajk  - \betajk | >  \tau\,  \ron\, 
2^{j\af/2} |k - k_{0j}|^{-\af/2} \rkr   \leq P_1 + P_2, 
$$
where 
\beqns
P_1 & = & \PP \lkr \left| n^{-1} \ \sum_{i=1}^n  [g(x_i)]^{-1}\, \wjk(x_i) f(x_i)  - \betajk \right| > 
0.5\, \tau\,  \ron\, 2^{j\af/2} |k - k_{0j}|^{-\af/2} \rkr ,\\
P_2 & = & \PP \lkr \left| n^{-1} \ \sum_{i=1}^n  [g(x_i)]^{-1}\, \wjk(x_i) \xi_i  \right| > 
0.5\, \tau\,  \ron\, 2^{j\af/2} |k - k_{0j}|^{-\af/2} \rkr .
\eeqns

The proof of the statement is now based on Bernstein's inequality, that we recall for completeness, 
\be  \label{bernstein}
 \PP  \lkr \left| n^{-1} \sum_{i=1}^n \eta_i \right| > z \rkr \leq 
2\ \exp \lkr-\frac{nz^2}{2(\sigma^2 + \|\eta\|_\infty\, z/3)}\rkr, 
\ee
where  $\eta_i,\   i=1,2,\ldots,n, $ are independent and identically distributed random variables with ${\EE} \eta_i =0$, 
 ${\EE} \eta_i^2 = \sigma^2$ and $\|\eta_i\| \leq \|\eta\|_\infty < \infty$.
\\

First, let us construct an upper bound for $P_1$. Note that, for $k \in K_{0jc}^w$, and for $i=1,2,\ldots,n$, one has, for  $x_i \in \supp\, w_{jk}$,
\be \label{gxi_ineq}
g(x_i) \geq C_w^{-1}\, C_{g1} 2^{-j\af} |k - \koj|^\af.
\ee
Let $\eta_i =  [g(x_i)]^{-1}\, \wjk(x_i) f(x_i) - \betajk$. Then,  
$\EE  \eta_i = 0$,  and, by \fr{gxi_ineq}, we derive
$\|\eta\|_\infty \leq  C_w C_{g1}^{-1}\, |k -\koj|^{-\af}\, 2^{j(\af + 1/2)}\,  \|w\|_\infty \|f\|_\infty$, 
so that 
\beqns 
\EE\eta_i^2 = \int \frac{\wjk^2(x) f^2(x)}{g(x)}\, dx \leq 
\frac{\|f\|^2_\infty}{C_{g1}}\ \int_{W_1}^{W_2} \frac{w^2(t) 2^{j \af}}{|t+ k -\koj|^\af} dt
\leq \frac{\|f\|^2_\infty\,  2^{j \af} C_w}{C_{g1} \, |k -\koj|^\af}. 
\eeqns 
Now, applying Bernstein's inequality and recalling that $m_1 \leq j \leq J-1$, $2^{J(\af +1)} = n/\ln n$
and  $|k- \koj| \geq 1$, we obtain 
\beqns  
P_1 & \leq & 2 \exp \lkr - \frac{C_{g1}\ \tau^2 \ln n}{8 C_w\, \|f\|_\infty 
(\|f\|_\infty  +  \|w\|_\infty\, \tau/6)} \rkr.
\eeqns 
Using  the inequality $a/(b+c)  \geq \min \lkr a/(2b), a/(2c) \rkr$, where
$a,b,c>0$,  and taking into account that $\tau^2 \geq \tau$ for $\tau \geq 1$, we obtain 
\be  \label{p1}
P_1 \leq 2 \exp (- \tau \ln n/D_1 )\ \ \ \mbox{with}\ \ \ 
D_1 = 8\, C_w C_{g1}^{-1} \, \max ( 2 \|f\|_\infty^2, \|f\|_\infty  \|w\|_\infty/3).   
\ee 

In order to construct an upper  bound for $P_2$, note that, conditionally on 
$(x_1,x_2, \ldots, x_n)$, one has, for  $x_i \in \supp\, w_{jk}$, 
$$
n^{-1} \ \sum_{i=1}^n  (g(x_i))^{-1}\, \wjk(x_i) \xi_i \sim {\cal N} (0, s_{jk}^2),
$$
where, by \fr{gxi_ineq} and $\sigma=1$, 
$$
s_{jk}^2 = \frac{1}{n^2} \sum_{i=1}^n \frac{\wjk^2 (x_i)}{g^2 (x_i)} 
\leq \frac{C_w  2^{j \af}}{C_{g1} |k - \koj|^\af \ n^2}\ \sum_{i=1}^n \frac{\wjk^2 (x_i)}{g  (x_i)}.  
$$
Hence, conditionally on $ (x_1,x_2, \ldots, x_n)$,  
$$
\PP \lkr \left| n^{-1} \ \sum_{i=1}^n  [g(x_i)]^{-1}\, \wjk(x_i)\, \xi_i  \right| > 
\frac{  \tau\,  \ron\, 2^{j\af/2}}{2\, |k - k_{0j}|^{\af/2}} \Bigg|  x_1,x_2, \ldots, x_n \rkr  
\leq \exp \lkr - \frac{\tau^2 2^{j \af} \, \ln n}{8 n |k - \koj|^\af s_{jk}^2} \rkr.
$$
Now,  consider  the following two sets:  
$$
\Omega_\up   (x_1,x_2, \ldots, x_n)  = \lfi  (x_1,x_2, \ldots, x_n) : 
\left| n^{-1}\ \sum_{i=1}^n \frac{\wjk^2 (x_i)}{g  (x_i)} - 1 \right| \geq \up \rfi, 
$$
and its complementary, $\Omega_\up^c  (x_1,x_2, \ldots, x_n)$. Then,
$P_2 \leq P_{21} + P_{22}$, where
\beqns
P_{21} & = & \EE \lkv \PP \lkr \left| n^{-1} \ \sum_{i=1}^n  (g(x_i))^{-1}\, \wjk(x_i) \xi_i \right| > 
\frac{  \tau\,  \ron\, 2^{j\af/2}}{2\, |k - k_{0j}|^{\af/2}} \Bigg|  x_1,x_2, \ldots, x_n  \rkr  
\II (\Omega_\up^c  (x_1,x_2, \ldots, x_n)) \rkv,  \\
P_{22} & = & \EE \lkv\, \II (\Omega_\up   (x_1,x_2, \ldots, x_n))\, \rkv, 
\eeqns
and $\II (\Om)$ is the indicator of the set $\Om$.
Since, for $\Omega_\up^c  (x_1,x_2, \ldots, x_n)$, we have
$$
n^{-1}\,  \sum_{i=1}^n [g  (x_i)]^{-1}\, \wjk^2 (x_i) \leq  
\lkr 1 + \left|n^{-1} \sum_{i=1}^n   [g  (x_i)]^{-1}\, \wjk^2 (x_i)   -1 \right| \rkr 
\leq \up + 1, 
$$
it is easy to check that 
\be \label{p21}
P_{21}  \leq \exp (- \tau^2 \ln n/D_2 )\ \ \ \mbox{with}\ \ \ 
D_2 = 8\, C_w C_{g1}^{-1} \, (\up +1).     
\ee 

In order to find an upper bound for $P_{22}$, we apply  Bernstein's inequality 
with $Z_i = [g  (x_i)]^{-1}\, \wjk^2 (x_i)-1$. Note that $\EE Z_i = 0$, 
$\EE Z_i^2 \leq C_w C_{g1}^{-1} \, \|w\|_\infty 2^{j (\af +1)} |k - \koj|^{-\af}$,
 $\|Z\|_\infty \leq 2 C_w C_{g1}^{-1} \, \|w\|_\infty 2^{j (\af +1)} |k - \koj|^{-\af}$.
Application of \fr{bernstein} with $z = \up$, yields
\be \label{p22}
P_{22}  \leq 2 \exp (- \up^2 \ln n/D_3 )\ \ \ \mbox{with}\ \ \ 
D_3 = 2\, \|w\|^2_\infty\, C_w C_{g1}^{-1} \, (1+2 \up/3).    
\ee 
Now,  set   $\up = 0.5\, \tau \| w \|_\infty$ and observe that, for 
$\tau \geq 1$,  one has 
$$
4  \|w\|^{-2}_\infty\, (1+2 \up/3)^{-1}\, \up^2 \geq \tau^2 /(\up +1)  \geq 
\tau \cdot \min(1/2, \| w \|_\infty^{-1}). 
$$
To  complete the proof,  we only need to combine  \fr{p1}, \fr{p21} and \fr{p22}.\\


\begin{lemma} \label{lem:large_dev2}
Let $b=0$,  $C_\ph = [ 2 \max ( |L_\ph|, |U_\ph| ) ]^\af$
and $C_{g2}$ be defined in \fr{gineq}.
Let $m$ be an integer such that  $m_1 \leq m  \leq J-1$,   and   let $k \in \Komh$. 
Let $\ron$ be defined ny  \fr{ron} and let $C_\kappa$ be given by \fr{Ckappa}.
Then, for $ \bcml$ and $\hbcml$ given by \fr{vecc} and \fr{hbcl}, respectively, 
 and an arbitrary constant   $\kappa \geq 1$,   
\be  \label{large_dev2} 
\PP \lkr |\hbcml  - \bcml | >  \kappa\, \ron\,   
2^{-\frac{m\af}{2}}   \rkr = O \lkr n^{- \frac{\kappa}{C_\kappa}} \rkr, \ \  n \rightarrow \infty.
\ee
\end{lemma}

\noindent
{\bf Proof of Lemma  \ref{lem:large_dev2}}. 
The proof is very similar to the proof of Lemma \ref{lem:large_dev1}, therefore, we shall 
just provide its outline. Partition the probability in \fr{large_dev2} into $P_1$ and $P_2$
with 
\beqns
P_1 & = & \PP \lkr \left| n^{-1} \ \sum_{i=1}^n    \phimk (x_i) f(x_i)  - \bcmk \right| > 
0.5\, \kappa\,  \ron\, 2^{-\frac{m\af}{2}} \rkr ,\\
P_2 & = & \PP \lkr \left| n^{-1} \ \sum_{i=1}^n  \phimk (x_i) \xi_i  \right| > 
0.5\, \kappa\,  \ron\, 2^{-\frac{m\af}{2}} \rkr .
\eeqns
An upper bound for $P_1$,  obtained by applying  Bernstein's  inequality,  is of the form 
\be  \label{pp1}
P_1 \leq 2 \exp (- \kappa \ln n/D_4 )\ \ \ \mbox{with}\ \ \ 
D_4 = 8\,  \|f \|_\infty\,   \max ( 2 C_{g2}  C_\ph,  \|\ph \|_\infty/3).   
\ee 

In order to derive an upper bound for $P_2$, introduce a set 
$$
\Theta_v  (x_1,x_2, \ldots, x_n)  = \lfi  (x_1,x_2, \ldots, x_n): 
\left| n^{-1}\ \sum_{i=1}^n  \phimk^2 (x_i)  - \int \phimk^2 (x) g(x) dx \right| \geq v 2^{-m \af} \rfi, 
$$
and its complementary, $\Theta_v^c   (x_1,x_2, \ldots, x_n)$. 
Then, similarly to the proof of Lemma~\ref{lem:large_dev1}, obtain 
$P_2 \leq P_{21} + P_{22},$ where
\beqns
P_{21} & = & \EE \lkv \PP \lkr \left| n^{-1} \ \sum_{i=1}^n    \phimk(x_i) \xi_i \right| > 
0.5\, \kappa \,  \ron\, 2^{-m\af/2} \Bigg|  x_1,x_2, \ldots, x_n  \rkr   
\II (\Theta_v^c  (x_1,x_2, \ldots, x_n) ) \rkv \\
& \leq & \exp \lkr - \frac{\kappa^2 \ln n}{8 (v + C_\ph C_{g2})} \rkr.
\eeqns
Also, application of   \fr{bernstein} with $\eta_i = \lkr \phimk^2(x_i) - \int \phimk^2 (x) g(x) dx \rkr$, 
yields
\beqns
P_{22} & = & \EE \lkv \II (\Theta_v  (x_1,x_2, \ldots, x_n)) \rkv 
\leq 2\, \exp \lkr - \frac{n v^2\, 2^{-m(1+\af)}}{2  \|\ph\|^2_\infty (C_\ph C_{g2} + v/3)} \rkr. 
\eeqns 
Setting $v = a \kappa$,    noting that for any $A, B, C>0$ one has 
$A/(B+C) \geq \min(A/(2B), A/(2C))$, and recalling that $\kappa \geq 1$ and $n 2^{-m(1+\af)} \geq \ln n$ by \fr{Jvalueo},   we derive
\be  \label{pp2}
P_2 \leq 2 \exp (- \kappa \ln n/D_5 )\ \ \ \mbox{with}\ \ \ 
D_5 = \max \lkr 16a,  16 C_\ph C_{g2}, \frac{4 C_\ph C_{g2} \|\ph\|_\infty}{a}, \frac{4 \|\ph\|_\infty^2}{3a} \rkr.  
\ee 
To complete the proof,  it suffices to  note that $a>0$ is arbitrary.


\begin{lemma} \label{lem:system_sol_dev}
Let $b=0$, let $\mo$ and $\hm$ be given by   \fr{mvalue} and \fr{mopt}, respectively.
Consider the non-asymptotic finite dimension matrices $\bAs$ and $\bBs$ 
with elements 
\beqn  
 A^{*}_{kl} & = & \int \ph(z + \kom -k) \ph(z + \kom-l) |z|^\af dz,\ \ \ k,l \in \Komh, \label{elementAs}\\
 B^{*}_{lk} & = & \int \ph(z + \kom -k) \ph(z + \kom-l) |z|^\af dz,\ \ \ l \in \Komh,\ k \in K_{0m}^*. \label{elementBs} 
\eeqn 
Let $\hbum$ be the solution of the system of equations \fr{estsys}.
Let $C_{\lam 1}$ and $C_{\lam 2}$ be defined by \fr{Clam1lam2} and  let $C_u$  be defined by \fr{Cu}.
If $ \lam  \geq   \max \lkr  C_{\lam 1}, C_{\lam 2} \rkr$,
then,  as $n \rightarrow \infty$,   
 \be  \label{large_dev_u} 
   \PP \lkr \| \hbu^{(m)} - \EE \hbu^{(m)} \|  >   \lam \,  \ron\,  2^{m \af/2}  \rkr   = O \lkr n^{-   \frac{2\lam}{C_u}  } \rkr.
\ee
\end{lemma}

\noindent
{\bf Proof of Lemma  \ref{lem:system_sol_dev}}. 
Observe that for any $m$, by \fr{estsys}, one has
$$
\| \hbum - \EE \hbum\| \leq \|(\bAm)^{-1} (\hbcm - \bcm) \| + \|(\bAm)^{-1} \bBm (\hbvm - \bvm)\|,
$$ 
so that  
\begin{align*}
& \PP \lkr \| \hbu^{(m)} - \EE \hbu^{(m)} \|  >   \lam \,  \ron\,  2^{m \af/2}  \rkr 
 \leq  \PP \lkr \|(\bAm)^{-1} (\hbcm - \bcm) \| > 0.5\, \lam \,  \ron\,  2^{m \af/2}  \rkr \\
& + \PP \lkr  \|(\bAm)^{-1} \bBm (\hbvm - \bvm)\| > 0.5\, \lam \,  \ron\,  2^{m \af/2}  \rkr
\equiv P_1 + P_2.
\end{align*}

Now note that, by assumption \fr{zeros} and the dominated convergence theorem, as $n \rightarrow \infty$, one has 
$\bAm = C_g 2^{-m\af} \bAs (1 + o(1))$ and $\bBm = C_g 2^{-m\af} \bBs (1 + o(1))$,
where  the matrices $\bAs$ and $\bBs$, defined in \fr{elementAs} and \fr{elementBs},  are independent of  $m$, since
the  sets $\Komh$ and $K_{0m}^*$ are defined in  terms of $k-\kom$ and $l-\kom$.
Therefore,  $\|(\bAm)^{-1}\| = C_g^{-1}\, 2^{m \af}\, \| (\bAs)^{-1}\| (1+o(1))$
and  $\|(\bAm)^{-1} \bBm \| = \| (\bAs)^{-1}  \, \bBs \| (1+o(1))$. 
Hence, setting $\kappa= C_{\lam 1}^{-1}\ \lam$ in Lemma \ref{lem:large_dev2}, where $C_{\lam 1}$ is defined 
in \fr{Clam1lam2},  and taking into account that the set 
$\Komh$ contains no more than $\Uh - \Lh+1$ indices,  we  obtain 
\beqns
P_1 & \leq & \PP \lkr \|\hbcm - \bcm \|>  \frac{ C_{g2}\, \lam \,  \ron\,  2^{-m \af/2}}{2\,  \| (\bAs)^{-1}\| } \rkr
\leq \sumkomh \PP \lkr |\hbcmk - \bcmk| > \frac{ C_{g2}\, \lam \,  \ron\,  2^{-m \af/2}}{2\, \sqrt{\Uh -\Lh +1}\ \| (\bAs)^{-1}\| } \rkr  \\
& = & \sumkomh \PP \lkr |\hbcmk - \bcmk| > 2 C_{\lam 1}^{-1}\ \lam \,  \ron\,  2^{-m \af/2} \rkr 
= O \lkr n^{- 2(C_\kappa\, C_{\lam 1})^{-1}\ \lam }   \rkr.
\eeqns
Similarly, using Lemma \ref{lem:large_dev1} with $w=\ph$ and $C_\tau$ given by \fr{Ctauph},  and recalling the  definitions of 
$\hbvm$ and $\bvm$, one can derive an upper bound for $P_2$ as 
\beqns
P_2 & \leq & \PP \lkr \|\hbvm - \bvm \|>  \frac{ \lam \,  \ron\, 2^{m \af/2} }{2\, \| (\bAs)^{-1}  \, \bBs \|} \rkr 
\leq \sum_{k \in K_{0m}^*} \PP \lkr |\hamk - \amk| > 
\frac{ \lam \,  \ron\, 2^{m \af/2} }{2\, \sqrt{\Uh -\Lh +1}\  \| (\bAs)^{-1}  \, \bBs \|} \rkr  \\
& = & O \lkr n^{- 2 (C_\tau\, C_{\lam 2})^{-1}\ \lam }  \rkr,
\eeqns
which completes the proof  of the lemma.\\


\noindent  
{\bf Proof of Lemma  \ref{lem:mopt_dev}}.  
Note that by definition of $\hm$, whenever $\hm > \mo$, there exists $j > \mo$ 
such that $\| (\hfmo   - \hfj) \II(\Xi_{m_0}) \|^2 > \lam^2\, 2^{j\af}\, \rho_n^2$,
where $\ron$ is defined in \fr{ron}.  Therefore, 
\beqn \label{eq:Pj}
\PP (\hm > \mo) & \leq & \sum_{j=\mo}^{J-1}  {\cal P}_j \ \ \mbox{with}\ \   
{\cal P}_j = \PP \lkr \| (\hfmo   - \hfj) \II(\Xi_{m_0}) \|^2 > \lam^2\, 2^{j\af}\, \rho_n^2 \rkr.
\eeqn 
Observe  that since
\beqns
\| (\hfmo   - \hfj) \II(\Xi_{m_0}) \| & \leq & 
\| (\hfoj   - \foj) \II(\Xi_{m_0}) \|  + \| (\hfcj   - \fcj) \II(\Xi_{m_0}) \| \\
& + & \| (\hfomo   - \fomo) \II(\Xi_{m_0}) \|  + \| (\hfcmo   - \fcmo) \II(\Xi_{m_0}) \|,
\eeqns 
one has the following upper bound for ${\cal P}_j$ defined in \fr{eq:Pj}:
\beqns 
{\cal P}_j   & \leq &  {\cal P}_{0,j,\mo} + {\cal P}_{0,j,j} + {\cal P}_{c,j,\mo} + {\cal P}_{c,j,j},
\eeqns
where, for  any   $\mo \leq m \leq j$,
\beqns 
{\cal P}_{0,j,m} & = & \PP \lkr \| (\hfom    - \fom ) \II(\Xi_{m_0}) \|  > 0.25\, \lam \, 2^{j\af/2}\, \rho_n  \rkr,\\
{\cal P}_{c,j,m} & = & \PP \lkr \| (\hfcm    - \fcm ) \II(\Xi_{m_0}) \|  > 0.25\, \lam \, 2^{j\af/2}\, \rho_n  \rkr.
\eeqns
Since $\supp (\fom) \subseteq \Xi_{m} \in \Xi_{m_0}$    for $m \geq m_0$, one has
\beqn \label{eq:part1} 
\| (\hfom   - \fom) \II(\Xi_{m_0}) \|^2  & = & \| (\hfom   - \fom) \II(\Xi_{m}) \|^2 = \|  \hfom   - \fom \|^2  \\
& \leq & \| \hbum - \bum \|^2 + 2 (\Uh - \Lh +1)A^2 2^{-2ms'}. \nonumber
\eeqn
Hence, by \fr{eq:part1}  and Lemma \ref{lem:system_sol_dev}, since $\mo \leq m \leq j$, one derives
\beqns 
{\cal P}_{0,j,m} & \leq &  \PP \lkr \| \hbum - \bum \|  > 0.25\, \lam \, 2^{j\af/2}\, \rho_n - A \sqrt{2(\Uh - \Lh +1)} 2^{- js'} \rkr
= O \lkr n^{-   \frac{\lam}{2C_u}  } \rkr.
\eeqns

Now, let us consider the second term,  ${\cal P}_{c,j,m}$.
Note that $\supp (\phimk)$ and $\Xi_{m_0}$ have non-empty intersection if and only if
$k \in \Kmmo$, where
\beqns
\Kmmo = \lfi k:\ 2^{m-m_0} [\min(\Lh, \Ls) - \Uh]  - \Uh < k - \kom <  2^{m-m_0} [\max(\Uh, \Us) - \Lh]  - \Lh \rfi.
\eeqns
Hence, for $m \geq m_0$,
\beqns 
\| (\hfcm   - \fcm) \II(\Xi_{m_0}) \|^2  & \leq & \| \hbvm - \bvm \|^2 + 
\sum_{j'=m}^{J-1} \sum_{k \in \Kjmop} (\hbjkp  - \bjkp)^2 + \sum_{j=J}^\infty \sum_{k \in \Kjmo} \bjk^2.
\eeqns
Here, by \fr{R2}, we have 
$$
\sum_{j=J}^\infty \sum_{k \in \Kjmo} \bjk^2 \leq A^2 2^{-2J s^*}  =  
O \lkr n^{-\frac{2{s'}}{2{s'} + \af}}\ (\ln n)^{ \frac{2{s'}}{2{s'} + \af}} \rkr,
$$ 
 where $s^*$ is defined in \fr{besball}. Also,
\beqns
(\hbjkp  - \bjkp)^2 & \leq &  (\hbjkp  - \bjkp )^2 \II(|\hbjkp  - \bjkp |> 0.5 d 2^{j \alpha/2} |k - k_{0j'}|^{-\alpha/2}) 
 + \bjkp^2
\eeqns
since 
$\II (|\hbjkp|>   d 2^{j \alpha/2} |k - k_{0j'}|^{-\alpha/2})  \leq \II (|\bjkp|>  0.5 d 2^{j \alpha/2} |k - k_{0j'}|^{-\alpha/2}) 
+ \II(|\hbjkp - \bjkp|> 0.5 d 2^{j \alpha/2} |k - k_{0j'}|^{-\alpha/2}) $ and, for $j \geq m_0$ and  $n$ large enough, 
$\II (|\bjkp|>  0.5 d 2^{j \alpha/2} |k - k_{0j'}|^{-\alpha/2})  =0$.
Denote 
$ C_{LU} = \max \lkr | \min(\Lh, \Ls) - 2 \Uh|, |\max(\Uh, \Us) -   2\Lh| \rkr$
and observe that 
$\Kjmop \subset \lfi k:\ |k - k_{0j'}| <  2^{j'-m_0} C_{LU} \rfi$.
Hence, using the Cauchy-Schwarz inequality and \fr{besball}, one obtains   
$$
\sum_{j'=m}^{J-1} \sum_{k \in \Kjmop} \bjkp^2 \leq A^2 (2 C_{LU})^{(1-2/p)_+ }\  2^{-2\mo s'} 2^{-2s^*(m-\mo)}.
$$
Combining all inequalities above, we derive that for any  $m \geq m_0$,  
\beqns 
\| (\hfcm   - \fcm) \II(\Xi_{m_0}) \|^2  & \leq & \| \hbvm - \bvm \|^2  
+ A^2  2^{-2J s^*} +  A^2 (2 C_{LU})^{\lkr 1- 2/p \rkr_+}\  2^{-2\mo s'} 2^{-2s^*(m-\mo)}\\
& + & \sum_{j'=m}^{J-1} \sum_{k \in \Kjmop} (\hbjkp - \bjkp)^2 \II(|\hbjkp - \bjkp|> 0.5 d 2^{j \alpha/2} |k - k_{0j'}|^{-\alpha/2}).
\eeqns
Now, by Lemma   \ref{lem:large_dev1} with $w=\ph$ and $w = \psi$,   obtain 
\beqns 
{\cal P}_{c,j,m} & \leq & \PP \lkr \| \hbvm - \bvm \| >  0.25\, \lam \, 2^{j\af/2}\, \rho_n - A^2  2^{-2J s^*} +  A^2 
(2 C_{LU})^{\lkr 1-2/p\rkr_+}\  
2^{-2\mo s'} 2^{-2s^*(m-\mo)} \rkr \\
& + & \sum_{j' =m}^{J-1} \sum_{k \in \Kjmop} \PP (|\hbjkp  - \bjkp |> 0.5 d 2^{j \alpha/2} |k - k_{0j'}|^{-\alpha/2})\\
& = &  O \lkr n^{- (C_\tau C_{\lam 0})^{-1} \lambda } \rkr + O \lkr n^{\frac{1}{\alpha +1} -\frac{d}{2C_d} }  \rkr,  
\eeqns
which completes the proof.


\subsection{Proofs of the statements in Section  \ref{est_fc_f0}: the zero-affected 
part of the wavelet thresholding estimator }
\label{proofs_f0}


{\bf Proof of Lemma  \ref{lem:error_sys_sol}.   }
Note that  
$$
\DD_1   =  \| \EE \hfo^{(m)}  - f_0^{(m)} \|^2 =   \sum_{j=m}^\infty \sumkomh \bjk^2,\ \ \ 
\DD_2  =  \EE \| \hfo^{(m)} - \EE \hfo^{(m)} \|^2 = \sumkomh \EE (\hamk - \amk)^2, 
$$ 
where      $\hamk = \hbumk$ for  $k \in \Komh$. 
From  the characterization \fr{bpqs} of Besov spaces, it follows that, for any $k$, one has
$ \bjk^2 \leq A 2^{-2js'}$, and, therefore, since the number of indices in the set $\Komh$ is finite,
\beqn 
\DD_1 & = & O \lkr  \sum_{j=m}^\infty   2^{-2 j {s'}} \rkr = O \lkr 2^{-2 m {s'}} \rkr. 
\label{dd1l}
\eeqn

Now, consider $\DD_2$.  Let, as in Lemma \ref{lem:inv_matr}, $\bAm$ be the matrix with the entries given  by  
\fr{matrA},  $\bD^{(m)}  = \sqrt{\diag(\bAm)}$  and $\bQ^{(m)} = (\bD^{(m)})^{-1} \bA^{(m)}(\bD^{(m)})^{-1}$. 
In the following proof, for the sake of clarity, we shall  suppress the    index $m$.
Rewrite   the systems of equations \fr{system} and \fr{estsys}, respectively, as
\be  \label{modsys}
\bQ\,  \bD\, \bu = \bD^{-1}\, \bc + \bD^{-1}\,\beps - \bD^{-1}\, \bB \bv,\ \ \ \
\bQ\,  \bD\,  \hbu = \bD^{-1}\, \hbc   - \bD^{-1}\, \bB \hbv,
\ee
so that 
\be  \label{differ}
\hbu - \bu = \bD^{-1} \bQ^{-1} \bD^{-1} (\hbc - \bc) - \bD^{-1} \bQ^{-1} \bD^{-1} \bB (\hbv - \bv) 
+ \bD^{-1} \bQ^{-1} \bD^{-1}  \beps.
\ee
Therefore, 
\beqn  \label{dd2l}
\DD_2 & = & \EE \|\hbu - \bu\|^2 = O \lkr \DD_{21} + \DD_{22} + \DD_{23} \rkr,  
\eeqn
with 
\be \label{dd2er}
 \DD_{21} = \EE \|\bD^{-1} \bQ^{-1} \bD^{-1} (\hbc - \bc)\|^2,\ \ \ 
 \DD_{22} = \EE \|\bD^{-1} \bQ^{-1} \bD^{-1} \bB (\hbv - \bv)\|^2,\ \ \ 
 \DD_{23} =  \|\bD^{-1} \bQ^{-1} \bD^{-1}  \beps \|^2.\ \ \ 
\ee 
By Lemma \ref{lem:sys_matr}, one has
$$
\bD_{ii} \geq C 2^{-m\af/2} \exp(-0.5\, b 2^{\beta(m+1)}),
$$
and since $\bD$ is the  finite-dimensional diagonal matrix, the latter implies 
\be \label{D_inv_upper}
\| \bD^{-1}\| = O \lkr 2^{m \af/2}  \exp(0.5\, b 2^{\beta(m+1)}) \rkr.
\ee
Therefore, since the set $\Komh$ is finite,  by Lemma \ref{lem:sys_matr}, one has
$$
\EE  \|\bD^{-1} (\hbc - \bc) \|^2  
= \sumkomh \Var(\hbcmk)/\bAmkk = O \lkr n^{-1} \rkr, 
$$ 
so that we derive 
\be  \label{dd21er}
\DD_{21} = O \lkr \|\bD^{-1}\|^2   \|\bQ^{-1}\|^2\  \EE  \|\bD^{-1} (\hbc - \bc) \|^2  \rkr  
=  O \lkr  n^{-1}\ 2^{m \af}\ \exp(b 2^{\beta(m+1)}) \rkr.
\ee

In order to derive an upper bound for $\DD_{22}$, note that from \fr{Amkk}, \fr{dd2er}  
and considerations above, it follows that
$$
 \DD_{22} =  O \lkr \|\bD^{-1}\|^4   \|\bQ^{-1}\|^2\  \EE  \|\bB (\hbv - \bv)\|^2 \rkr
= O \lkr 2^{2m \af} \exp(2b\, 2^{\beta(m+1)})\  \EE  \|\bB (\hbv - \bv)\|^2  \rkr.
$$
Since $\exp (- b 2^{m\beta} |z|^{-\beta})$ is an increasing function of $|z|$ and, 
for $\Lh \leq z + \kom -k \leq \Uh$ and $k \in \Koms$, one has $|z| \leq 2(\Uh - \Lh)$,
for $k  \in \Koms$, we derive
\beqns
C_{kk} =   \int \phimk^2 (x)   g(x)\II(2^m x -l \in \Om_\delta) dx 
 &= &O \lkr  2^{-m\af}\ \int \ph^2(z + \kom -k)\, |z|^\af\, \exp \lfi - b 2^{m\beta} |z|^{-\beta} \rfi dz \rkr  \\
&= &   O \lkr 2^{-m\af}\ \exp \lfi - b 2^{m(\beta-1)} (\Uh - \Lh)^{-\beta} \rfi \rkr.   
\eeqns
Hence, since  the  sets $\Komh$ and $\Koms$ are finite, by definition of vector  $\hbv$, 
Lemmas \ref{lem:coef_moments} and \ref{lem:sys_matr} and the Cauchy-Schwarz inequality,  we  obtain
\beqns
\EE  \|\bB (\hbv - \bv)\|^2 &=& \sum_{h \in \Komh}\ \sum_{k,l \in \Koms}
B_{hk} B_{hl} \Cov(\hat{a}_k, \hat{a}_l)  
  \leq   n^{-1}\ \sum_{h \in \Komh}\ \sum_{k,l \in \Koms} \Jmkl A_{hh} \sqrt{C_{kk} C_{ll}}\\
& = & O \lkr n^{-1}  2^{-m\af} \exp \lfi b 2^{m \beta}  - b 2^{m(\beta-1)} (\Uh - \Lh)^{-\beta} 
-  b 2^{m \beta}\, M_\ph^{-\beta}  \rfi \rkr,
\eeqns	
where $M_\ph$ is defined in Lemma \ref{lem:sys_matr}. Since $U_\ph - L_\ph \geq 4$,   we  finally obtain
\be  \label{dd22er}
\DD_{22} =  O \lkr  n^{-1}\ 2^{m \af}\ \exp( b\, 2^{m \beta} [2^{\beta +1} + 1] ) \rkr.
\ee

Now, for the  function $\eps_m(x)$ defined in \fr{fmepsm}, one has
\be \label{dd2f1}
 \DD_{23} =  O \lkr \|\bD^{-1}\|^2\, \|\bQ^{-1}\|^2\, \|\bD^{-1} \beps\|^2 \rkr,
\ee 
where 
\be \label{dd2f2}
\|\bD^{-1} \beps\|^2 = \sumkomh A_{kk}^{-2} \lkv \int \eps_m (x)\, \phimks (x)\,  g(x) dx \rkv^2.
\ee
If $b=0$, by Cauchy-Schwarz inequality,  one obtains
$\|\bD^{-1} \beps\|^2 \leq \sumkomh A_{kk}^{-2} \|\eps_{m 0}\|^2 \ \|\phimk \, g \|^2$,
where $\eps_{m 0} (x) = \eps_m (x) \II(|x - x_0| \leq C 2^{-m})$.
By calculations similar to proof of Lemma \ref{lem:sys_matr}
in the case of $b=0$, one can show that 
$\|\phimk \, g \|^2 = \int \phimk^2 (x) g(x) dx = O \lkr 2^{-2m\af} \rkr$.  
Also, since $\bjk^2 \leq A 2^{-2js'}$, one has
\beqns 
\|\eps_{m 0}\|^2  & = & O \lkr \sum_{j=m}^\infty \sum_{|k - k_{0j}| \leq C 2^{j-m}} \rkr \\
& = & O \lkr \sum_{j=m}^\infty 2^{-2js'} 2^{(j-m)(1 -2/p)} \rkr = O \lkr 2^{-2ms'} \rkr.
\eeqns
Recalling \fr{Amkk},  we  obtain  in the case of $b=0$,
\be \label{dd23er}
 \DD_{23} =  O \lkr 2^{-2m s'} \rkr.
\ee


Now, let us consider the case of $b>0$.
Denote $\phimks (x) = \phimk (x) \II(2^m x - k \in \Omega_\dd)$,
$\Ijmkl = \int \phimks(x) \psijl(x) g(x) dx$ 
and let
\be  \label{zmaxmix}
\zmax(\phimks, \psijl) = \arg\max_x [\phimks(x) \psijl(x) g(x)].
\ee
Observe that since $\phimks(\zmax) \neq 0$, we have 
$\Ijmkl / \bAmkk = O(1)$.  Consider the collection of indices  
$$
\Lmjk = \lfi l:\ 0 \leq l \leq 2^j -1,\  \supp (\phimks) \cap \supp (\psijl) \neq \emptyset \rfi.
$$
It is easy to see that $\Lmjk \subseteq [2^{j-m} (\Lh + \dd_b +k) - \Us,\ 2^{j-m} (\Uh + \dd_b +k) - \Ls]$,
so, for each $k$, there are $O(2^{j-m})$ terms such that $l \in \Lmjk$. 
Note that $|\zmax(\phimks, \psijl)| \leq |\zmax(\phimks)|$ and, for each $k$, there is only finite number of
terms such that  $|\zmax((\phimks, \psijl)| = |\zmax((\phimks)|$.  Indeed, straightforward calculation shows that
$$
\zmax(\phimks, \psijl) = \min [(\Uh - \dd_b +k - \kom), 2^{m-j}(\Us + l - \koj)] \quad \text{if} \quad  \kom - k < 0.5\, (\Uh + \Lh),
$$
and 
$$
\zmax (\phimks, \psijl) = \max [(\Lh + \dd_b +k - \kom), 2^{m-j}(\Ls + l - \koj)] \quad \text{if} \quad\kom - k \geq 0.5\, (\Uh + \Lh).
$$ 
Hence,  $|\zmax (\phimks, \psijl)| = |\zmax (\phimks)|$ if $l \geq 2^{j-m} (\Uh - \dd_b + k) - \Us$ or
$l \leq 2^{j-m} (\Lh + \dd_b + k) - \Ls$. Since we also need $l \in \Lmjk$, we obtain that 
 $|\zmax((\phimks, \psijl)| = |\zmax((\phimks)|$ if $l \in \Lmjks$
where 
$$
\Lmjks \subseteq  [2^{j-m} (\Lh + \dd_b +k) - \Us, 2^{j-m} (\Lh + \dd_b +k) - \Us] \cup 
[2^{j-m} (\Uh + \dd_b +k) - \Us, 2^{j-m} (\Uh + \dd_b +k) - \Ls]
$$
and, thus, $\Lmjks$ contains at most $2(\Us - \Ls)$ values of $l$ for each $k$.
If $l \in \Lmjk \setminus \Lmjks = \Lmjkc$, then 
\be \label{LmjkIneq}
2^{j-m} (\Lh + \dd_b + k) - \Ls < l < 2^{j-m}(\Uh - \dd_b + k) - \Us.
\ee
Then, by \fr{gen_asymp},
\beqns
\Ijmkl & \sim & 2^{-\frac{j-m}{2}} \int \ph^*(2^{m-j}t + \kom -k) \, 
\psi(t+ \koj -l) 2^{-j\af} \, |t|^\af \exp(-b |t|^{-\beta} 2^{j \beta}) dt\\
& = & O \lkr 2^{(m-j)/2} 2^{-j \af} 2^{-j\beta } 2^{(j-m)\af} 
\exp \lfi - b 2^{j\beta} |\tmaxkl|^{-\beta} \rfi \rkr,
\eeqns
where
$$
\tmaxkl = \Us+l-\koj  \quad \text{if} \quad \kom - k < (\Uh + \Lh)/2
$$ 
and 
$$
\tmaxkl = \Ls+l-\koj \quad \text{if} \quad \kom - k \geq (\Uh + \Lh)/2.
$$ 
Using formula \fr{AmllAsymp}, we derive that
\be \label{IoverA}
(\bAmkk)^{-1} \; |\Ijmkl| =  O \lkr 2^{(j-m)(\beta + 1/2)} 
\exp \lfi - b 2^{j\beta} \lkv |\tmaxkl|^{-\beta} -  2^{-(j-m)\beta} |\zmaxkk|^{-\beta} \rkv \rfi  \rkr,
\ee
where $\zmaxkk$ is defined in \fr{zmaxll}.

Denote $\hjmkl = |\tmaxkl|  -  2^{(j-m)} |\zmaxkk|$ and observe that 
$$
\hjmkl = 2^{(j-m)}(\Uh - \dd_b + k) - \Us - l \quad \text{if} \quad \kom - k < (\Uh + \Lh)/2,
$$ 
and 
$$
\hjmkl = l- 2^{(j-m)}(\Lh + \dd_b + k) + \Ls \quad \text{if} \quad \kom - k \geq (\Uh + \Lh)/2.
$$ 
Comparing the latter formulae with definition of $\Lmjkc$, we derive that, for $l \in \Lmjkc$,
$0< \hjmkl < C_h 2^{j-m}$ for every value of $k$, where $C_h>0$ is a constant which depends only
on the choice of the wavelet basis. 
Now,   for any $0<x<y$ and $\beta >0$ one has, for some $0 < \xi < y-x$,
$$
x^{-\beta} - y^{-\beta} = \beta (y-x) (y-\xi)^{-\beta} \geq \beta (y-x) y^{-(\beta+1)}.
$$
Applying the  above  inequality with $x=|\tmaxkl|$ and $y =  2^{(j-m)} |\zmaxkk|$, we obtain 
that 
$$
 |\tmaxkl|^{-\beta} -  2^{-(j-m)\beta} |\zmaxkk|^{-\beta} \geq \beta \hjmkl 2^{-(j-m)(\beta +1)} |\zmaxkk|^{-(\beta+1)},
$$
and,  thus, for $l \in \Lmjkc$,  we have
\be  \label{IAFin}
(\bAmkk)^{-1}\, |\Ijmkl|  = O \lkr 2^{(j-m)(\beta + 1/2)} 
\exp \lfi - b \beta |\zmaxkk|^{-\beta}  2^{m\beta} 2^{-(j-m)} \hjmkl \rfi  \rkr.
\ee 

Now, it follows from \fr{dd2f1} that $ \DD_{23} =  \DD_{231} +  \DD_{232}$,  where
\beqns
 \DD_{231} & = &  O \lkr \sumkomh \lkv \sum_{j=m}^\infty \sum_{l \in \Lmjks} 
(\bAmkk)^{-1}\, |\Ijmkl| |\bjl| \rkv^2 \rkr,\\ 
 \DD_{232}  & = &  O \lkr \sumkomh \lkv \sum_{j=m}^\infty \sum_{l \in \Lmjkc} 
(\bAmkk)^{-1}\, |\Ijmkl| |\bjl| \rkv^2 \rkr.
\eeqns
Using the facts thta the set $\Komh$ is finite, $|\bjl| = O(2^{-j{s'}}$ and $(\bAmkk)^{-1}\, |\Ijmkl| = O(1)$,
we derive that,  as $m \rightarrow \infty$, 
\be \label{DD231}
 \DD_{231} = O \lkr 2^{-2m{s'}} \rkr.
\ee 
For $ \DD_{232}$, using \fr{IAFin} and taking into account that $\hjmkl$ changes by unit increments,  we obtain
\beqn \label{DD232}
 \DD_{232} & = &  O \lkr \sumkomh \lkv \sum_{j=m}^\infty \sum_{\hjmkl=0}^{C_h 2^{j-m}} 
2^{-j{s'}} 2^{(j-m)(\beta + 1/2)} \exp \lfi - b \beta |\zmaxkk|^{- (\beta+1)}  2^{m\beta} 2^{-(j-m)} \hjmkl \rfi \rkv^2  \rkr \nonumber\\
 & = & O \lkr -2m{s'} \lkv \sum_{j=m}^\infty 2^{-(j-m)(\beta-1/2+s')} \rkv^2 \rkr = 
 O \lkr 2^{-2m{s'}} \rkr  
\eeqn
since  $s' \geq 1/2$ and $\beta >0$.
Finally, combining expressions \fr{dd2l}, \fr{dd21er}, \fr{dd22er}, \fr{DD231} and \fr{DD232},  we  obtain 
\be \label{DD2Er}
\DD_2 = O \lkr n^{-1}\ 2^{m \af}\ \exp(2b\, 2^{\beta(m+1)}) +  2^{-2m{s'}} \rkr.
\ee
To complete the proof of \fr{linest_error}, set $m=\mo$, where $\mo$ is defined in \fr{mvalue} and combine \fr{DD2Er} with 
 \fr{dd1l}, \fr{dd2l}, \fr{dd21er} and \fr{dd22er}.


Now, we need to show that  $\EE \| \hfo^{(m)} - \EE \hfo^{(m)} \|^4 = o(1)$.
Note that  it follows from  \fr{modsys}--\fr{dd2er}  that 
$\DD^* = O \lkr \DD^*_1 + \DD^*_2 + \DD^*_3 \rkr$ where, similarly to the case of squared difference, 
\beqns
\DD^*_1  & = & O \lkr  \|\bD^{-1}\|^8   \|\bQ^{-1}\|^4\  \EE  \| \hbc - \bc  \|^4  \rkr,\\
\DD^*_2  & = & O \lkr  \|\bD^{-1}\|^8\  \|\bB \|^4   \|\bQ^{-1}\|^4\  \EE  \| \hbv - \bv \|^4 \rkr, \\
\DD^*_3  & = & O \lkr \|\bD^{-1}\|^4\, \|\bQ^{-1}\|^4\, \|\bD^{-1} \beps\|^4 \rkr.
\eeqns
Applying Lemma \ref{lem:inv_matr} and using  \fr{Bc_bounnds} and \fr{D_inv_upper} with $b=0$, we obtain
$\DD^*_1 =  O \lkr  n^{-2} 2^{2 m \af}    \rkr = o(1)$ and  
$\DD^*_2 =  O \lkr  2^{2 m \af}    \EE  \| \hbv - \bv \|^4 \rkr$. 
Also, similarly to \fr{dd2f1} and \fr{dd23er}, 
$\DD^*_3 =   O \lkr 2^{-4m s'} \rkr$.
To complete the proof  of the lemma, recall the definitions of $\hbv$ and $\bv$, 
apply \fr{4thmoment} with $k \in K_{0m}^*$, and note that,  for $k \in K_{0m}^*$,
one has $|k -\kom| = O(1)$.
\\


\subsection{Proofs of the statements in Section  \ref{est_fc_f0}: the zero-free part of the wavelet threholding estimator  }
\label{sec:zero-free}



\noindent
{\bf  Proof of Lemma  \ref{lem:exponential_zero} }. 
Let $R = \EE \| \hfcmo - \fcmo \|^2 = R_1 + R_2 + R_3$,  where
\beqns
R_1 & = &   \sumkomhc \Var ( \hat{a}_{m_0 k}),\ \ \ 
R_2 =  \sum_{j=J}^\infty \sumkojsc \bjk^2,\ \ \ 
R_3 =   \sum_{j=m_0}^{J-1} \sumkojsc  \EE   (\hbjk - \bjk)^2. 
\eeqns
By Lemma \ref{lem:coef_moments} we derive that, as $n \rightarrow \infty$,  
\beqns  
R_1 & = &   O \lkr n^{-1}\, 2^{\mo \af} \sum_{k \in K_{0 m_0 c}^\ph}  \lkv |k - k_{0 \mo}|^{-\af}  \exp(b 2^{\mo \beta} |k -  k_{0 \mo}|^{-\beta}) \rkv \rkr\\
& = &   O \lkr n^{-1} 2^{\mo (1+\af)} \exp (2^{-(\beta + 1)} \ln n) \rkr = o \lkr (\ln n)^{-\frac{2s'}{\beta}} \rkr.
\eeqns 
Using \fr{Jvalueo} and  \fr{besball}, we derive that
\beqns  
R_2 & = &   O \lkr 2^{-2Js^*} \rkr = O \lkr (\ln n)^{-\frac{4s^*}{\beta}} \rkr =  O \lkr (\ln n)^{-\frac{2s'}{\beta}} \rkr,
\eeqns
since $s^*=s'$ for  $1 \leq p \leq 2$  and $s^* = s \geq (s+1/2-1/p)/2$ for  $2 <p \leq \infty$  due to $s \geq 1/2$.
For $R_3$, we have 
\beqns 
R_3 & = &    \sum_{j=m_0}^{J-1}  \sum_{ |k - \koj| \leq 2^{j-\mo}}  \bjk^2 +  \sum_{j=m_0}^{J-1}  \sum_{ |k - \koj| > 2^{j-\mo}} \Var ( \tbjk ) \\
& = &   O \lkr \sum_{j=m_0}^{J-1} \lkv 2^{-2js'}  (2^{j-m_0})^{1-2/p} + n^{-1}\, 2^j\, 2^{\af \mo }\, \exp \lkr b 2^{\beta \mo }\rkr
\rkv \rkr\\
& = &   O \lkr 2^{-m_0 (s+2/p-1)} + (\ln n)^{(2+\af)/\beta} n^{-1 + 2^{-(\beta+1)}} \rkr =  O \lkr (\ln n)^{-\frac{2s'}{\beta}} \rkr.
\eeqns
To  complete  the proof  of the lemma, note that the  upper bounds are uniform for $f \in  B_{p,q}^s (A)$.
\\


\noindent
{\bf Proof of Lemma  \ref{lem:polynomial_zero}. }   Note  that 
\be \label{Rtotal}
R = \EE \| \hfcm - f_c \|^2 = R_1 + R_2 + R_3 + R_4,
\ee
where 
\beqns
R_1 & = &   \sumkomhc \EE (\hamk - \amk)^2,\ \ \ 
R_2 =  \sum_{j=J}^\infty \sumkojsc \bjk^2,\\
R_3 & = &   \summJ \sumkojsc \EE \lkv (\tbjk - \bjk)^2 \ 
 \II( \tbjk^2 >  d^2\, \ron \, 2^{j\af}\, |k - \koj|^{-\af}) \rkv,\\
R_4 & = &  \summJ \sumkojsc \bjk^2 \ \PP  (\tbjk^2 \leq d^2 \, \ron \,  2^{j\af}\, |k - \koj|^{-\af}),
\eeqns
with $\ron$ defined in \fr{ron}. 
Using Lemma \ref{lem:coef_moments}, we  obtain
\be  \label{R1}
R_1 = O \lkr n^{-1}\ 2^{m \af} \sumkomhc\ |k - \kom|^{-\af} \rkr =
O \lkr n^{-1}\ 2^{m \af} (\ln n)^{\II (\af =1)} \rkr,
\ee
since the set $\Komhc$ contains $O(\ln n)$ terms and the sum $\sumkomhc\ |k - \kom|^{-\af}$
is uniformly bounded if $\af >1$.

It is well-known (see, e.g., Johnstone (2002), Lemma 19.1) 
that if $f \in \Bpqsa$, then for some  constant
$c^{\star}>0$, dependent on $p$, $q$, $s$ and $A$ only, one has 
\be \label{besball}
\sumjk \bjk^2 \leq c^{\star} 2^{-2 j {s^*}}\ \ \mbox{with}\ \ s^*=\min(s,s'). 
\ee
Therefore, an upper bound for $R_2$ is of the form
$$
R_2 = \sum_{j = J}^\infty \sumjk\bjk^2 = O \lkr 2^{-2 J {s'}} \rkr. 
$$
If $1 \leq p \leq 2$, then $s^*=s'$ and 
$R_2 = O \lkr n^{-\frac{2{s'}}{2{s'} + \af}}\ (\ln n)^{ \frac{2{s'}}{2{s'} + \af}} \rkr$.
If  $2 \leq p \leq \infty$, then $s^* = s$ and, since $s \geq 1/2$, one has 
$p > (4s - 2\af -2)/(4s^2 - \af -1)$. Hence, 
$$
2s/(\af +1) > 2s'/(2s' + \af),
$$
so that, for  $1 \leq p \leq \infty$, one has
\be  \label{R2}
R_2 =   O \lkr n^{-\frac{2{s'}}{2{s'} + \af}}\ (\ln n)^{ \frac{2{s'}}{2{s'} + \af}} \rkr.
\ee

In oder to obtain an upper bound for $R_3$ and $R_4$, note that 
\be \label{r3r4ineq}
R_3 \leq R_{31} + R_{32},\ \ \ R_4 \leq R_{41} + R_{42},
\ee
 where
\beqn 
R_{31} & = & \summJ \sumkojsc \EE \lkv (\tbjk - \bjk)^2 \ 
 \II( (\tbjk - \bjk)^2 > 0.25\,  d^2\, \ron \,  2^{j\af}\, |k - \koj|^{-\af}) \rkv, \nonumber \\
R_{32} & = & \summJ \sumkojsc \EE \lkv (\tbjk - \bjk)^2 \ 
 \II( \bjk^2 > 0.25\,  d^2 \, \ron^2 \,  2^{j\af}\, |k - \koj|^{-\af}) \rkv,\nonumber \\
R_{41} & = & \summJ \sumkojsc \bjk^2 \ 
\PP  (  (\tbjk - \bjk)^2 > 0.25\, d^2 \, \ron^2 \,  2^{j\af}\, |k - \koj|^{-\af}),  \label{partition}\\
R_{42} & = & \summJ \sumkojsc \bjk^2 \ 
\II  ( \bjk^2  \leq 2.25\ d^2 \, \ron^2 \,  2^{j\af}\, |k - \koj|^{-\af}). \nonumber 
\eeqn

Applying Lemma \ref{lem:large_dev1} with $w (\cdot) = \psi(\cdot)$ and $\tau = 0.5 d$,  we  obtain 
$$
\PP  (  (\tbjk - \bjk)^2 > 0.25\, d^2 \, \ron^2 \,  2^{j\af}\, |k - \koj|^{-\af}) = 
O \lkr n^{-0.5 d/ C_d} \rkr,
$$ 
where $C_d$ is given by \fr{Cdvalue}.
Hence, by Lemma \ref{lem:coef_moments} and inequality $\sqrt{a+b} \leq \sqrt{a} + \sqrt{b}$, for $d > 4 C_d$,
as $n \rightarrow \infty$,  we  obtain
\beqn 
R_{31} & \leq & \summJ \sumkojsc \lkv \EE  (\tbjk - \bjk)^4\ \cdot \ 
\PP  (  (\tbjk - \bjk)^2 > 0.25\, d^2 \, \ron^2 \,  2^{j\af}\, |k - \koj|^{-\af}) \rkv ^{1/2} \nonumber \\
& = &  O \lkr n^{-\frac{d}{4 C_d}} \lkv n^{-\frac{3}{2}} 2^{\frac{j(3\af +1)}{2}} \sumkojsc |k - \koj|^{-\frac{3\af}{2}}
+ n^{-1} 2^{j\af} \sumkojsc |k - \koj|^{-\af} \rkv \rkr\nonumber \\
& = &  O \lkr n^{-\frac{d}{4 C_d}} \rkr = o \lkr n^{-\frac{2{s'}}{2{s'} + \af}}\ (\ln n)^{ \frac{2{s'}}{2{s'} + \af}} \rkr.
\label{R31} \eeqn 
Similarly, by \fr{besball}, 
\beqn \label{R41}
R_{41} & = &  O \lkr n^{-\frac{d}{2 C_d}} \rkr\  \summJ \sumkojsc \bjk^2  = o(n^{-1}).  
 \eeqn 
Now, consider $R_{32}$ and $R_{42}$. Note that it follows from Lemma \ref{lem:coef_moments} that
\beqns
R_{32} & = & O \lkr \summJ \sumkojsc \lkv n^{-1} 2^{j\af} |k - \koj|^{-\af} 
 \II( \bjk^2 > 0.5\,  d^2 \, n^{-1} \ln n \,  2^{j\af}\, |k - \koj|^{-\af}) \rkv \rkr \\
& = & O  \lkr \summJ \sumkojsc \min \lkv (\ln n)^{-1} \bjk^2, n^{-1}  2^{j\af} |k - \koj|^{-\af} \rkv \rkr
\eeqns
and, similarly, 
\beqns
R_{42} & = &  O  \lkr \summJ \sumkojsc \min \lkv   \bjk^2, n^{-1} \ln n\,   2^{j\af} |k - \koj|^{-\af} \rkv \rkr.
\eeqns
Hence, 
\be \label{R3242}
R_{32} = O \lkr (\ln n)^{-1} R_{42} \rkr =  O \lkr   R_{42} \rkr
\ee
 so that one needs to study only $R_{42}$.
Partition $R_{42}$ as 
$R_{42}= R_{421}+R_{422}+R_{423}$,  where
\beqns
R_{421} & = & \sum_{j=m}^{j_1} \sumkojsc \lkv n^{-1}\, \ln n\,  2^{j\af} |k - \koj|^{-\af} \rkv,\ \ 
R_{422} = \sum_{j=j_2}^{J-1} \sumkojsc    \bjk^2,\\
R_{423} & = & \sum_{ j=j_1+1 }^{j_2-1} \lkv \sum_{|k - \koj| > N_j} n^{-1} \, \ln n\,  2^{j\af} |k - \koj|^{-\af} + 
\sum_{|k - \koj| \leq  N_j}   \bjk^2 \rkv,
\eeqns
and the values of $j_1$,  $j_2$ and $N_j$ will be defined later.
It is easy to see that, by \fr{besball}, 
\beqn \label{R421422}
R_{421}  & = & O \lkr n^{-1}\, \ln n\,  2^{j_1 \af} (\ln n)^{\II(\af =1)} \rkr,\ \ \ 
R_{422}  = O \lkr   2^{-2 j_2 s^*} \rkr, \\
\label{R423o}
R_{423} & = & O \lkr \sum_{ j=j_1+1}^{j_2-1} \lkv n^{-1}\, \ln n\,  2^{j\af} N_j^{1 -\af} (\ln n)^{\II(\af =1)} +  
  2^{-2 j s'}  N_j^{1-2/p} \rkv \rkr.
\eeqn
If $\af \neq 1$, the two terms in \fr{R423o} are equal to each other when 
$$
N_j = \lkr n^{-1} \ln n 2^{j(2s' + \af)} \rkr^{1/(\af - 2/p)},
$$
and, for this value of $N_j$, one has 
\be \label{R423}
R_{423} =  O \lkr \sum_{j=j_1+1}^{j_2-1} (n/\ln n)^{\frac{2/p-1}{\af - 2/p}}  
2^{\frac{ 2j(s' - \af s)}{\af - 2/p}} \rkr.
\ee
Therefore, $R_{423}$ behave differently when $\af s \geq s'$ and $\af s < s'$, and we consider those cases separately.

First, consider the case when $\af s = s'$. Then 
$$
R_{423} = O \lkr (j_2- j_1)  (n/\ln n)^{\frac{2/p-1}{\af - 2/p}} \rkr = O \lkr (\ln n/n)^{ \frac{2s'}{2s' + \af} } \, \ln n  \rkr
 \quad \mbox{if}\ \   \af s = s'.
$$
If $\af >1$, $\af s > s'$, choose $j_1$ and $j_2$ such that
$$
2^{j_1} = (n/\ln n)^\frac{1}{2s' + \af},\ \ \ 
2^{j_2} = (n/\ln n)^\frac{s'}{s^* (2s' + \af)}.
$$
Note that if  $1 \leq p \leq 2$, one has $s^* = s \geq s'$, so that $j_2 \leq j_1$ and $R_{423}=0$.
If  $2 < p \leq \infty$, then $j_2>j_1$. Also, it follows from \fr{R421422} and \fr{R423} that
$R_{423} =  O \lkr n^{\frac{2/p-1}{\af - 2/p}} (\ln n)^{\frac{1 - \af}{\af - 2/p}} 
2^{\frac{ 2j_1(s' - \af s)}{\af - 2/p}} \rkr$. Hence,  
$R_{421} = O \lkr (n/\ln n)^{-\frac{2s'}{2s' + \af}}  \rkr,$
$R_{422} = O \lkr (n/\ln n)^{-\frac{2s'}{2s' + \af}}  \rkr$ and 
$R_{423} = O \lkr (n/\ln n)^{-\frac{2s'}{2s' + \af}}  \rkr$,
so that
\be \label{R42case1af}
R_{42} =  O \lkr (n/\ln n)^{-\frac{2s'}{2s' + \af}}  \rkr\ \ 
\mbox{if}\ \   \af s \geq s',\, \af >1.
\ee
Similarly, if $\af >1$, $\af s < s'$, choose $j_1$ and $j_2$ such that
$$
2^{j_1} =  (n/\ln n)^\frac{1}{\af(2s+1)},\ \ \ 
2^{j_2} = (n/\ln n)^\frac{1}{2s+1}.
$$
In this case, $R_{423} =  O \lkr n^{\frac{2/p-1}{\af - 2/p}} (\ln n)^{\frac{1 - \af}{\af - 2/p}} 
2^{\frac{ 2j_2(s' - \af s)}{\af - 2/p}} \rkr$, and direct calculations yield 
$R_{421} = O \lkr (n/\ln n)^{-\frac{2s}{2s+1}}  \rkr,$
$R_{422} = O \lkr (n/\ln n)^{-\frac{2s}{2s+1}}  \rkr$ and 
$R_{423} = O \lkr (n/\ln n)^{-\frac{2s}{2s+1}}  \rkr$,
so that
\be \label{R42case2af}
R_{42} = O \lkr (n/\ln n)^{-\frac{2s}{2s+1}}  \rkr\ \ 
\mbox{if}\ \   \af s < s',\, \af >1.
\ee
Finally, if $\af=1$, set $j_1=j_2$ such that
\beqns
2^{j_1} =  (n/\ln^2 n)^\frac{1}{2s^* +1}
\eeqns
and obtain 
\be \label{R42cases}
R_{42} = O \lkr n^{-\frac{2s^*}{2s^* + 1}}\ (\ln n)^{\frac{4s^* -1}{2s^* + 1}} \rkr  \quad
\mbox{if}\ \   \af =1. 
\ee
Now, to complete the proof of  \fr{risk_hfc}, one just need to combine \fr{Rtotal}, \fr{R1}, \fr{R2}, \fr{R31}, \fr{R41},  \fr{R3242}
and \fr{R42case1af}--\fr{R42cases}, and to  note that all upper bounds are uniform for $ f \in B_{p,q}^s (A)$.
\\

In order to prove \fr{risk_hfc_4}, note that  
$$
R^* = \EE \| \hfcm - f_c \|^4 \leq R_1^* + R_2^* + R_3^*,
$$
where 
\beqns
R_1^* & = & O \lkr \EE  \| \sumkomhc (\hamk - \amk) \phimk (x) \|^4 \rkr,\ \ \ 
R_2^* =   O \lkr  \| \sum_{j=m}^\infty \sumkojsc \bjk  \psijk(x) \|^4  \rkr,\\
R_3^* & = & O \lkr   \EE  \lkv  \summJ \sumkojsc   (\tbjk - \bjk)^2 \ 
 \II( \tbjk^2 >  d^2 \, \ron \, 2^{j\af}\, |k - \koj|^{-\af}) \rkv^2  \rkr.\\
\eeqns
Observe that, by Lemma \ref{lem:coef_moments}, since $2^{m(\af +1)}  = o(n/\ln n)$, as $n \rightarrow \infty$,
\beqns
R_1^* & = & O \lkr 2^m \sumkomhc \EE (\hamk - \amk)^4 \rkr 
=  O \lkr n^{-3}\, 2^{m(3\af +2)} + n^{-2}\, 2^{m(2\af +1)} \rkr \\
& = & O \lkr  n^{-2}\, 2^{m(2\af +1)} \rkr = o(1).
\eeqns
For $R_2^*$, by \fr{besball},  we have
\beqns
R_2^* & = & O \lkr \lkv \sum_{j=m}^\infty \sumkojsc \bjk^2 \rkv^2 \rkr 
=  O \lkr \lkv 2^{-2 m {s'}} \rkv^2 \rkr = o(1).
\eeqns
Finally,  similarly to \fr{partition},  partition 
$R_3^*$ as $R_3^* = R_{31}^*   + R_{32}^*$  with $R_{31}^*$ and $R_{32}^*$
corresponding to $\II( |\tbjk - \bjk|^2 > 0.25\, d^2 \, \ron \,  2^{j\af}\, |k - \koj|^{-\af})$
and $\II( \bjk^2 > 0.25\,  d^2 \, \ron \,  2^{j\af}\, |k - \koj|^{-\af})$,
respectively.  
For $R_{31}^*$,  applying Lemmas \ref{lem:coef_moments} and  \ref{lem:large_dev1} with  $w= \psi$  and 
$C_d$ given by \fr{Cdvalue},  and also noting that 
$\sumkojsc |k - \koj|^{-l} = O(1)$ for $l>1$,  we derive  
\beqns
R_{31}^* & = & O \lkr n\  \summJ \sumkojsc  \EE \lkv |\tbjk - \bjk|^4 \ 
\II ( |\tbjk - \bjk|^2 > 0.25\, d^2 \, \ron \, 2^{j\af}\, |k - \koj|^{-\af}) \rkv   \rkr \\
 & = & O \lkr n  \summJ \sumkojsc  \lkv \EE |\tbjk - \bjk|^6  \rkv^{2/3}\   
\lkv \PP ( |\tbjk - \bjk|^2 > 0.25\, d^2 \, \ron \,  2^{j\af}\, |k - \koj|^{-\af}) \rkv^{1/3} \rkr  \\
 & = & O \lkr  n\  \summJ n^{- d/(3 C_d) } \ \lkv n^{-10/3}\, 2^{j(10\af +4)/3} + n^{-8/3}\, 2^{j(8\af +2)/3} +  n^{-2}\, 2^{2 j \af} \rkv \rkr\\ 
 & = & o \lkr n^{1- d/(3 C_d) } \rkr = o(1), \ \ \ n \rightarrow \infty,
\eeqns
since $d > 3 C_d$.
For $R_{32}^*$,  using Lemma \ref{lem:coef_moments} and \fr{besball}, we derive that
\beqns
R_{32}^* & = & O \lkr n\  \summJ \sumkojsc  \EE \lkv |\tbjk - \bjk|^4 \ 
\II( \bjk^2 > 0.25\,  d^2 \, \ron \,  2^{j\af}\, |k - \koj|^{-\af})\rkv  \rkr \\
& = & O \lkr n\  \summJ \sumkojsc [ 2^j\, \ln^{-3} n\, \bjk^6 + \ln^{-2}n\,  \bjk^4 ] \rkr
= o \lkr n\  \summJ [ 2^{j(1-6 {s'})} + 2^{-4j {s'}} ] \rkr
\eeqns
since $n^{-1}\, 2^{j \af} |k -\koj|^{-\af} < 0.25\, \bjk^2/(d^2 \ln n)$. Note that
$m \geq \mo$ implies $2^m \geq n^{1/(2 {s'} + \af)}$, so that
\beqns
R_{32}^* & = & o \lkr n^{-\frac{6 {s'} -1 }{2 {s'} + \af}} + n^{-\frac{4 {s'}} {2 {s'} + \af}} \rkr
= o(1),
\eeqns
which completes the proof  of the lemma.
\\


\subsection{Proof of the asymptotic minimax upper bounds for the $L^2$-risk in Section \ref{minimax_risk}  }
\label{sec:minimax_proof}


{\bf Proof of Theorem \ref{cor:upper_bound}. }  
Since $\hm = \mo$ for $b>0$,  the  validity of Theorem \ref{cor:upper_bound} for $b>0$  follows directly from 
Lemma \ref{lem:exponential_zero}. 
 For $b=0$, observe that
\beqns
\DD    =   \EE [ \| \hfm - f \|^2 &=& \sum_{m=m_1}^{m_0} \EE [ \| \hfm - f \|^2\ \ \II(\hm = m \leq m_0) ] + 
\EE [  \| \hfm - f \|^2\ \ \II(\hm = m > \mo) ]  \\
&\equiv& \DD_{1} + \DD_{2}, 
\eeqns 
and consider terms $\DD_{1}$ and $\DD_{2}$  separately.

Denote
\be \label{Rn}
 R(n) = \lfi
\begin{array}{ll} 
O \lkr n^{-\frac{2s}{2s + 1}}\ (\ln n)^{\mu_1} \rkr  
& {\rm if}\ \ \  b=0,\ \af s < s',\\ 
O \lkr   n^{-\frac{2s'}{2s' + \af}}\ (\ln n)^{\mu_2} \rkr    
& {\rm if}\ \ \  b=0,\  \af s \geq s',
\end{array} \right. 
\ee
and note that, for any $m \geq m_1$,  
\beqns
 \EE  \| \hfm - f \|^2 & \leq &  2 [\, \EE  \| \hfmo - f \|^2 +  \EE  \| (\hfm - \hfmo) \II(x \in \Xi_m) \|^2 +  
\EE  \| (\hfm - \hfmo) \II(x \in \Xi_m^c) \|^2 ]
\eeqns
where $m_1$ is defined in \fr{Jvalueo} and   set $\Xi_m$ is defined in \fr{Xi_neighborhood}.
By Lemmas \ref{lem:error_sys_sol}  and \ref{lem:polynomial_zero}, we obtain
$$
\EE  \| \hfmo - f \|^2 = O \lkr n^{-\frac{2s'}{2s' + \alpha}} +  R(n) \rkr. 
$$
If  $\hm = m \leq m_0$, then by definition of $\hm$, we derive that 
$$
 \EE  \| (\hfm - \hfmo) \II(x \in \Xi_m) \|^2  \leq   \lam^2\, 2^{m_0 \af}\, n^{-1} \ln n  = 
O \lkr n^{-\frac{2s'}{2s' + \alpha}} \rkr.
$$
Now, recall that $\Xi_m$ is defined in such a way that $\supp (\fom) \in \Xi_m$ for any $m$,
and that $\Xi_{j1} \subset \Xi_{j2}$ for $j_1 > j_2$, so that, for $m \leq m_0$,
\beqns
\EE  \| (\hfm - f) \II(x \in \Xi_m^c) \|^2 & = &  \EE  \| (\hfom + \hfcm - \fom - \fcm) \II(x \in \Xi_m^c) \|^2  \\
& = & \EE  \| (\hfcm - \fcm) \II(x \in \Xi_m^c) \|^2 \leq \EE  \| \hfcm - \fcm  \|^2 = O \lkr R(n) \rkr
\eeqns
 as $n \rightarrow \infty$. Noting that 
$$
\EE  \| (\hfm - \hfmo) \II(x \in \Xi_m^c) \|^2 \leq 2 \lkv \EE  \| (\hfm - f) \II(x \in \Xi_m^c) \|^2
+ \EE  \| (\hfmo - f) \II(x \in \Xi_m^c) \|^2 \rkv
$$
 and combining all formulae above, we obtain that $\DD_1 = O \lkr R(n) \rkr$ as $n \rightarrow \infty$.

By Lemmas \ref{lem:error_sys_sol} and \ref{lem:polynomial_zero}, one has
$\EE \| \hfom - \fom \|^4 = o \lkr  1 \rkr$ and $\EE \| \hfcm - \fcm \|^4 = o \lkr  1 \rkr$.  
Then, Lemma   \ref{lem:mopt_dev} yields
\beqns
\DD_{2}  & \leq  &  \sqrt{\EE [ \| \hfm - f  \|^4}\ 
\sqrt{\PP (\hm = m > \mo)} 
= O \lkr  n^{- \frac{\lam}{2 C_\lam}  } + n^{\frac{1}{2(\alpha+1)} - \frac{d}{4 C_d} } \rkr =O(n^{-1}),
\eeqns
provided $\lam \geq   \max \lkr  C_{\lam 1}, C_{\lam 2}, 2 C_\lam \rkr$ 
and $d > 2 (\alpha +1)^{-1} (2 \alpha +3)  C_d $, 
which   completes the proof  of   Theorem \ref{cor:upper_bound}.

\subsection{Proof of the asymptotic minimax upper bounds for the $L^2$-risk in Section \ref{sec:small_alpha}  }
\label{sec:proofs_small_alpha}

The proof of Theorem \ref{th:frisk_af_small} is based on the following lemma.

\begin{lemma} \label{lem:moments}
Let Assumption A hold with $b=0$ and $0<\af <1$. Then, 
\beqns
\Var (\tbjk) & = & O \lkr n^{-1} 2^{j \af} \, \min(1, |k-\koj|^{-\af}) \rkr, \\
\EE  |\tbjk - \bjk|^\frac{\af+3}{\af+1}  & = & O \lkr n^{-\frac{2}{\af +1}} 2^{j \frac{(\af+3)}{2(\af+1)}} + n^{-\frac{\af +3}{2(\af +2)}} 2^j \rkr.
\eeqns
\end{lemma}

\noindent
{\bf Proof of Lemma \ref{lem:moments}}. Proof of the first statement is very similar to the proof of validity of formula \fr{lemmom1}.
Proof of the second statement is based on Lemma 3.1. in Chesneau (2007) which states that whenever $\int [g(x)]^{1-\nu} dx < \infty$
for some $\nu>2$, one has
\be \label{chesn2007}
\EE  |\tbjk - \bjk|^\nu = O \lkr n^{1-\nu} \int |\psijk(x)|^\nu  \, [g(x)]^{1-\nu} dx + n^{-\nu/2} \int \psijk^2(x)\,  [g(x)]^{-\nu/2} dx \rkr. 
\ee  
To complete the proof, note that for $\nu = 1 + 2/(\af +1)>2$  one has 
$\int [g(x)]^{1-\nu} dx < \infty$  and  apply   \fr{chesn2007}. 
\\

\noindent
{\bf Proof of Theorem \ref{th:frisk_af_small}}. The proof of this statement is  similar to the proof of Lemma 
\ref{lem:polynomial_zero}. Indeed, similarly to the proof of Lemma \ref{lem:polynomial_zero}, 
partition the risk as 
$R = \EE \| \hf  - f \|^2 = R_1 + R_2 + R_3 + R_4$
where, similarly to the proof of Lemma \ref{lem:polynomial_zero},
\beqns
R_1 & = &     \sum_{k=0}^{2^{m_1}-1} \EE (\hat{a}_{m_1 k}- a_{m_1 k})^2,\ \ \ 
R_2 =  \sum_{j=J}^\infty \sumjk  \bjk^2,\\
R_3 & = &   \sumJ \sumjk  \EE \lkv (\tbjk - \bjk)^2 \ 
 \II( \tbjk^2 >  d^2\, \ron \, 2^{j\af}\, |k - \koj|^{-\af}) \rkv,\\
R_4 & = &  \sumJ \sumjk  \bjk^2 \ \PP  (\tbjk^2 \leq d^2 \, \ron \,  2^{j\af}\, |k - \koj|^{-\af})
\eeqns
with $\ron$ defined in \fr{ron}. 
Since $1/g$ is integrable and $m_1$ in \fr{Jvalueo} is finite, it is easy to show that $R_1 = O \lkr n^{-1} \rkr$.
Also, same as before,
$R_2 =   O \lkr 2^{-2 J {s^*}} \rkr.$
If $p>2$, then $\af +1< 2s+1$ since $s \geq \max(1/2,1/p)$ and  $\af <1$, so that
$R_2 = O \lkr n^{- 2s/(2s+1)} \rkr$. If $1 \leq p \leq 2$, then $s^* = s'$ and 
$2s'/(1 +\af) > \max\lfi 2s'/(2s' + \af), 2s/(2s+1) \rfi$,
so that  
$$
R_2 = O \lkr \max \lfi n^{- 2s/(2s+1)}, n^{-2s'/(2s' + \af)} \rfi \rkr.
$$
Now, similarly to the proof of Lemma \ref{lem:polynomial_zero}, partition $R_3$ and $R_4$ as  
$R_3 \leq R_{31} + R_{32}$ and $R_4 \leq R_{41} + R_{42}$. Using Lemma \ref{lem:moments}, as $n \rightarrow \infty$, obtain upper bounds
\beqns 
R_{31} & \leq &   \sumJ \sumjk  
\lkv \PP  (  (\tbjk - \bjk)^2 > 0.25\, d^2 \, \ron^2 \,  2^{j\af}\, |k - \koj|^{-\af}) \rkv ^{1 -2/\nu} \ 
\lkv  \EE  |\tbjk - \bjk|^\nu \rkv^{2/\nu} \\
& = &  O \lkr  \sumJ 2^j \, n^{-d(1-2/\nu)/(2 C_d)}\ \lkv n^{1-\nu} 2^{j\nu/2} +   n^{-\nu/2}\ 2^j \rkv^{2/\nu}  \rkr \\
& = &  O \lkr  2^j \, n^{-d(1-2/\nu)/(2 C_d)}\ \lkv   n^{-\nu/2}\ 2^J \rkv^{2/\nu}  \rkr = O \lkr n^{-1} \rkr,
 \eeqns 
provided \fr{d_value} holds, and also
\beqns
R_{41} & = &  O \lkr n^{-\frac{d}{2 C_d}} \rkr\  \summJ \sumkojsc \bjk^2  = o(n^{-1}).  
 \eeqns 

Now, same as before, 
$R_{32} = O \lkr (\ln n)^{-1} R_{42} \rkr =  O \lkr   R_{42} \rkr$,
so that we need to construct upper bounds for    $R_{42}$  only.
Partition $R_{42}$ as 
$R_{42}= R_{421}+R_{422}+R_{423}$ where
\beqns
R_{421} & = & \sum_{j=0}^{j_1} \sumjk \lkv n^{-1}\, \ln n\,  2^{j\af} |k - \koj|^{-\af} \rkv,\ \ \ \ 
R_{422} = \sum_{j=j_2}^{J-1} \sumjk    \bjk^2,\\
R_{423} & = & \sum_{j=j_1+1}^{j_2-1} \lfi \sum_{|k - \koj| > N_j} |\bjk|^p \lkv  n^{-1} \, \ln n\,  2^{j\af} N_j^{-\af} \rkv^{1-p/2} + 
\sum_{|k - \koj| \leq  N_j}      n^{-1} \, \ln n\,  2^{j\af} N_j^{1 -\af}   \rfi,
\eeqns
and the values of $j_1$,  $j_2$ and $N_j$ will be defined later.
It is easy to see that, same as before, 
$R_{421}  = O \lkr n^{-1}\, \ln n\,  2^{j_1 \af}   \rkr$  and $R_{422}  = O \lkr   2^{-2 j_2 s^*} \rkr$.
For $R_{423}$ we can write the following expression
\beqns 
R_{423} & = & \sum_{j=j_1+1}^{j_2-1} \lkv 2^{-js'p} \lkr  \frac{\ln n}{n}\,  2^{j\af} N_j^{-\af} \rkr^{1-p/2}
+    \frac{\ln n}{n}\,    2^{j\af} N_j^{1 -\af}  \rkv.
\eeqns
If $p \geq 2$, we choose $j_1=j_2$ such that $2^{j_1} = (\ln n /n)^{1/(2s+1)}$ and obtain
$R_{42}= O \lkr (\ln n/n)^{2s/(2s + 1)} \rkr$.    
If $1 \leq p <2$, we choose   $N_j$ which equalize the two terms in $R_{423}$ and  obtain, similarly to \fr{R423},
\be \label{R423new}
R_{423} =  \lfi \begin{array}{ll} 
O \lkr   (n/\ln n)^{\frac{2/p-1}{\af - 2/p}}  
2^{\frac{ 2j_2 (s' - \af s)}{\af - 2/p}} \rkr & {\rm if}\;\;\;    \af s < s'\\
O \lkr  (n/\ln n)^{\frac{2/p-1}{\af - 2/p}}  
2^{\frac{ 2j_1 (s' - \af s)}{\af - 2/p}} \rkr &  {\rm if}\;\;\;   \af s > s'\\
O \lkr (j_2 - j_1)\, (n/\ln n)^{\frac{2/p-1}{\af - 2/p}}  
 \rkr &  {\rm if}\;\;\;   \af s = s'
\end{array} \right.
\ee
%

If $\af s < s'$, then choose 
$$
2^{j_1} = (n/ \ln n)^\frac{1}{2 s+1},\ \ \ 2^{j_2} = (n/ \ln n)^\frac{s}{s' (2s +1)},
$$
so that $j_1 < j_2$. Direct calculations show that in this case
$$
R_{42} = O \lkr   (n/ \ln n)^{\zeta_1} \rkr \  \   {\mbox with }\ \ \  \zeta_1 = \frac{2/p-1}{2/p - \af} + \frac{2(s' - \af s)}{(2s+1)(2/p-\af)}
= \frac{2s}{2s+1}
$$
and $R_{42} = O \lkr (\ln n/n)^{2s/(2s + 1)} \rkr$.  
If $\af s > s'$, then set 
$$
2^{j_1} = (n/ \ln n)^\frac{\af}{2 s' + \af},\ \ \ 2^{j_2} = (n/ \ln n)^\frac{1}{2 s' + \af},
$$
so that  again $j_1 < j_2$. Here we have
$$
R_{42} = O \lkr   (n/\ln n)^{\zeta_2}\rkr \  \   {\mbox with }\ \ \  \zeta_2 = \frac{2/p-1}{2/p - \af} - \frac{2(\af s - s')}{(2s'+\af)(2/p-\af)}
= \frac{2s'}{2s' + \af}
$$
and $R_{42} = O \lkr (\ln n/n)^{2s'/(2s'+\af)} \rkr$.   
If $\af s = s'$, then  note that $j_2 - j_1 = O (\ln n)$, so that
$R_{42} = O \lkr (\ln n/n)^{2s'/(2s'+\af)} \rkr = O \lkr (\ln n/n)^{2s/(2s + 1)} \rkr$.
Now, to complete the proof, just combine the expressions for $R_1$, $R_2$, $R_{31}$, $R_{41}$, $R_{32}$ and $R_{42}$.


\end{document}